\documentclass[aps,prd,preprint,superscriptaddress,showpacs]{revtex4}
\pdfoutput=1
\usepackage{slashed}
\usepackage{graphicx}
\usepackage{verbatim}
\usepackage{multirow}
\usepackage{ulem}
\usepackage{color}
\definecolor{My_red}        {cmyk}{0.00,1.00,1.00,0.20}

\usepackage[centertags]{amsmath}
\usepackage{amssymb}

\begin{document}

\title{The Renormalizable Three-Term Polynomial Inflation with Large Tensor-to-Scalar Ratio}

\author{Tianjun Li}

\email{tli@itp.ac.cn}

\affiliation{State Key Laboratory of Theoretical Physics
and Kavli Institute for Theoretical Physics China (KITPC),
Institute of Theoretical Physics, Chinese Academy of Sciences,
Beijing 100190, P. R. China}

\affiliation{School of Physical Electronics,
University of Electronic Science and Technology of China,
Chengdu 610054, P. R. China}

\author{Zheng Sun}

\email{sun_ctp@scu.edu.cn}

\affiliation{State Key Laboratory of Theoretical Physics
and Kavli Institute for Theoretical Physics China (KITPC),
Institute of Theoretical Physics, Chinese Academy of Sciences,
Beijing 100190, P. R. China}

\affiliation{Center for theoretical physics,
College of Physical Science and Technology,
Sichuan University, Chengdu, 610064, P. R. China}

\author{Chi Tian}

\email{rectaflex@gmail.com}

\affiliation{State Key Laboratory of Theoretical Physics
and Kavli Institute for Theoretical Physics China (KITPC),
Institute of Theoretical Physics, Chinese Academy of Sciences,
Beijing 100190, P. R. China}

\affiliation{School of Physical Electronics,
University of Electronic Science and Technology of China,
Chengdu 610054, P. R. China}

\author{Lina Wu}

\email{wulina@std.uestc.edu.cn}

\affiliation{State Key Laboratory of Theoretical Physics
and Kavli Institute for Theoretical Physics China (KITPC),
Institute of Theoretical Physics, Chinese Academy of Sciences,
Beijing 100190, P. R. China}

\affiliation{School of Physical Electronics,
University of Electronic Science and Technology of China,
Chengdu 610054, P. R. China}

\begin{abstract}

We systematically study the renormalizable three-term polynomial inflation
in the supersymmetric and non-supersymmetric models. The supersymmetric
inflaton potentials can be realized in supergravity theory, and only have
two independent parameters. We show that the general renormalizable
supergravity model is equivalent to one kind of
our supersymmetric models. We find
that the spectral index and tensor-to-scalar ratio can be consistent
with the Planck and BICEP2 results, but the running of spectral
index is always out of the $2\sigma$ range. If we do not consider the
BICEP2 experiment, these inflationary models can be highly consistent
with the Planck observations and saturate its upper bound on the
tensor-to-scalar ratio ($r \le 0.11$). Thus, our models can be tested
at the future Planck and QUBIC experiments.

\end{abstract}

\maketitle

\section{Introduction }

It is well-known that the standard big bang cosmology has some problems,
for instance, the flatness, horizon, and monopole problems, etc, which can be
solved naturally by inflation~\cite{Staro, Guth:1980zm, Linde:1981mu, Albrecht:1982wi}.
Also, the observed temperature fluctuations in the cosmic microwave background radiation (CMB)
strongly suggests an accelerated expansion at a very early stage of
our Universe evolution, {\it i.e.}, inflation. Moreover,
the inflationary models predict the cosmological perturbations in the matter density and
spatial curvature from the vacuum fluctuations of the inflaton, which can explain the primordial
power spectrum elegantly. Besides the scalar perturbation, the tensor perturbation is produced as well,
which has special features in the B-mode of the CMB polarization data
as a signature of the primordial inflation.

The Planck satellite measured the CMB temperature anisotropy with
an unprecedented accuracy. From its first-year observational data~\cite{Ade:2013ktc}
in combination with the nine years of Wilkinson Microwave Anisotropy Probe (WMAP) polarization
low-multipole likelihood data~\cite{Hinshaw:2012aka} and the high-multipole spectra data from the Atacama
Cosmology Telescope (ACT)~\cite{Das:2013zf} and the South Pole Telescope (SPT)~\cite{Keisler:2011aw}
(Planck+WP+highL), the scalar spectral index $n_s$,
the running of the scalar spectral index $n_s' \equiv d n_s/d\ln k$,
the tensor-to-scalar ratio $r$, and the scalar amplitude $A_s$
for the power spectrum of the curvature perturbations  are respectively
constrained to be~\cite{Ade:2013zuv, Ade:2013uln}
\begin{eqnarray}
&& n_s = 0.9603 \pm 0.0073~,~~n_s'=-0.0134\pm 0.0090 ~,~~\nonumber \\
&& r \le 0.11~,~~A^{1/2}_s = 4.6856^{+0.0566}_{-0.0628} \times 10^{-5}~.~\,
\label{Planck-A}
\end{eqnarray}
As given by the Planck Collaboration, we also quote 68\% errors on the measured parameters and 95\%
upper limits on the other parameters.

Recently, the BICEP2 experiment announced the discovery of the gravitational waves or primordial tensor perturbations
in the B-mode power spectrum around $\ell \sim 80$~\cite{Ade:2014xna}. If confirmed by future experiments,
it will definitely be a huge progress in fundamental physics. The measured tensor-to-scalar ratio is
\begin{eqnarray}
 r ~=~ 0.20^{+0.07}_{-0.05}~.~\,
\end{eqnarray}
Subtracting the various dust models and re-deriving the $r$
constraint still results in high significance of detection, we have
\begin{eqnarray}
 r ~=~ 0.16^{+0.06}_{-0.05}~.~\,
\end{eqnarray}
Thus, the BICEP2 results are in tension with the Planck results.
To be consistent with both experiments,
one can consider the running of the spectral index.
With it, we have the following results from the { Planck}+WP+highL data~\cite{Ade:2013zuv}
\begin{eqnarray}
n_s=0.9570\pm 0.0075~,~~~n'_s=-0.022\pm 0.010~,~~r ~< ~0.26 ~(95\%~{\rm C.L.})~.~\,
\label{Planck-B}
\end{eqnarray}
And the combined Planck+WP+highL+BICEP2 data give
\begin{eqnarray}
n_s= 0.9574^{+0.0073}_{-0.0074}~,~~~n'_s= -0.0292\pm 0.0096 ~,~~r ~= ~0.21^{+0.05}_{-0.06} ~.~\,
\label{Planck-C}
\end{eqnarray}
Therefore,  we must at least require the running of the spectral index $n'_s$ to be  smaller than
0.0004 at  3$\sigma$ level for any viable inflationary model. However, there might exist the
foreground subtleties in the BICEP2 experiment such as dust effects, etc.
As we know, the recent observations from the Planck and BICEP2/Keck Array Collaborations provided
strong constraints on the primordial tensor
fluctuations~\cite{Ade:2015tva, Planck:2015xua, Ade:2015lrj}, $r < 0.11$
($r < 0.12$ from BICEP2/Keck Array) at 95\% Confidence Level (C.L.). Because
these results were announced seven months after we submitted our paper to arXiv,
we will not consider them here.

Obviously, such a large tensor-to-scalar ratio $r$ from the BICEP2 measurement
does impose a strong constraint on the inflationary models. For example,
most inflationary models from string theory predict small $r$ far below $0.01$
and then contradict with the BICEP2 results~\cite{Burgess:2013sla}.
With $r=0.16$ or 0.20, we obtain that the Hubble scale during inflation
is about $1.0\times 10^{14}$~GeV, and the inflaton potential is
around the Grand Unified Theory (GUT) scale $2\times 10^{16}$~GeV which might
have some connections with GUTs.
From the naive analysis of Lyth bound~\cite{Lyth:1996im}, we will have
large field inflation, and then the effective field theory might not
be valid since the high-dimensional operators are suppressed by the
reduced Planck scale. The inflationary models, which can realize $n_s\simeq 0.96$ and
$r\simeq 0.16/0.20$, have been studied extensively~\cite{Anchordoqui:2014uua, Czerny:2014wua,
Hamada:2014iga, Kobayashi:2014jga, Ferrara:2014ima, Choudhury:2014kma, Gong:2014cqa, Ashoorioon:2014nta,
Okada:2014lxa, Ellis:2014rxa, Hamaguchi:2014mza, DiBari:2014oja, Kawai:2014doa,
Antusch:2014cpa, Freivogel:2014hca, Bousso:2014jca, Kaloper:2014zba, Choudhury:2014wsa, Choi:2014aca,
Murayama:2014saa, McDonald:2014oza, Gao:2014fha, Gao:2014yra, Li:2014owa, Chialva:2014rla,
Li:2014xna, Kallosh:2014xwa,
Gao:2014pca, Gao:2014uha, Nakayama:2014hga, Choudhury:2013iaa,
Li:2014lpa, Ben-Dayan:2014lca, Okada:2014nia}.
Especially, the simple chaotic and natural inflation models are favoured.

From the particle physics point of view, supersymmetry is
the most promising new physics beyond the Standard Model (SM).
Especially, it can stabilize the scalar masses, and has a non-renormalized
 superpotential.  Moreover, gravity is very important in the
early Universe. Thus, a natural framework for inflationary model building
is supergravity theory~\cite{SUGRA}. However,
 supersymmetry breaking scalar masses in a generic supergravity theory
are of the same order as gravitino mass, giving rise to
the so-called $\eta$ problem~\cite{eta},
where all the scalar masses are at the order of the Hubble scale
due to the large vacuum energy density during inflation~\cite{glv}.
Two elegant solutions were proposed: no-scale supergravity~\cite{Cremmer:1983bf,
Ellis:1984bf, Enqvist:1985yc, Ellis:2013xoa, Ellis:2013nxa, Li:2013moa, Ellis:2013nka},
and shift symmetry in the K\"ahler potential~\cite{Kawasaki:2000yn, Yamaguchi:2000vm,
Yamaguchi:2001pw, Kawasaki:2001as, Kallosh:2010ug, Kallosh:2010xz, Nakayama:2013jka,
Nakayama:2013txa, Takahashi:2013cxa, Li:2013nfa}.

The Planck satellite experiment  might measure the tensor-to-scalar ratio $r$ down to
0.03-0.05 in one or two years. And the target of future QUBIC experiment is to constrain
 the tensor-to-scalar ratio of 0.01 at the 90\% Confidence Level (C.L.) with one year
of data taking from the Concordia
Station at ${\rm D\hat{o}me}$ C, Antarctica~\cite{Battistelli:2010aa}.
Thus, even if the BICEP2 results on tensor-to-scalar ratio $r$ were too large,
as long as $r$ is not smaller than 0.01, for example, $r=0.05$ or $0.1$,
how to construct the inflationary models which highly agree
 with the Planck results and have large tensor-to-scalar ratio is still
a very important question since these models can be tested in the near future.

The simple inflationary models have one parameter, for example, the monomial inflaton potentials.
So the next to the simple inflationary models have two parameters. In the supergravity models
with two parameters, we will generically have three terms due to the square of the F-term.
In particular, we show that the general renormalizable
supergravity model is equivalent to one kind of our supersymmetric models.
Thus, in this paper,
 we will classify the renormalizable three-term polynomial inflationary models
for both supersymmetric and non-supersymmetric models. The supersymmetric
inflaton potentials can be obtained from supergravity theory. We find that their
spectral indices and tensor-to-scalar ratios
can be consistent with the Planck and BICEP2 experiments. However, $n'_s$ is always out
of the $2\sigma$ range. In addition, even if we do not consider the BICEP2 results,
we find that the three-term polynomial inflationary models can be
 consistent with the Planck observations. Especially,
the tensor-to-scalar ratio can not only be larger than $0.01$ in the $1\sigma$ region,
above the well-known Lyth bound~\cite{Lyth:1996im}, but also saturate
the Planck upper bound $0.11$ in the $1\sigma$ region. Thus, these models produce
the typical large field inflation,
and can be tested at the future Planck and QUBIC experiments.

This paper is organized as follows. In Section II, we briefly review
the slow-roll inflation. In Section III, we construct the supersymmetric
models from the supergravity theory. In Section IV, we systematically
study the three-term polynomial inflation. Our conclusion is given
in Section V.

\section{Brief Review of Slow-Roll Inflation}

In the inflation, the slow-roll parameters are defined as
\begin{gather}
\label{slow1}
\epsilon=\frac{M_{\rm Pl}^2V_\phi^2}{2V^2},\\
\label{slow2}
\eta=\frac{M_{\rm Pl}^2V_{\phi\phi}}{V},\\
\label{slow3}
\xi^2=\frac{M_{\rm Pl}^4V_\phi V_{\phi\phi\phi}}{V^2},
\end{gather}
where $M^2_{\rm Pl}=(8\pi G)^{-1}$ is the reduced Planck scale,
$V_\phi \equiv \partial V(\phi)/\partial \phi$, $V_{\phi\phi} \equiv \partial^2V(\phi)/\partial \phi^2$,
and $V_{\phi\phi\phi} \equiv \partial^3V(\phi)/\partial \phi^3$. Also,
the scalar power spectrum in the single field inflation is
\begin{equation}
\label{power}
\mathcal{P}_{\mathcal{R}}=A_s\left(\frac{k}{k_*}\right)^{n_s-1+n_s'\ln(k/k_*)/2},
\end{equation}
where the subscript ``*" means the value at the horizon crossing, the scalar amplitude is
\begin{equation}
\label{power1}
A_s\approx \frac{1}{24\pi^2M^4_{\rm Pl}}\frac{\Lambda^4}{\epsilon}~,~\,
\end{equation}
and the scalar spectral index as well as its running at the second order are~\cite{Lyth:1998xn,Stewart:1993bc}
\begin{gather}
\label{nsdef}
\begin{split}
n_s~=~1+2\eta-6\epsilon+2\left[\frac{1}{3}\eta^2+(8C-1)\epsilon \eta\right.\\
\left.-\left(\frac{5}{3}+12C\right)\epsilon^2-\left(C-\frac{1}{3}\right)\xi^2\right],
\end{split}\\
\label{rundef}
n_s'~=~16\epsilon\eta-24\epsilon^2-2\xi^2,
\end{gather}
where $C=-2+\ln 2+\gamma \simeq -0.73$ with $\gamma$ the Euler-Mascheroni constant.
Moreover, the tensor power spectrum is
\begin{equation}
\label{power}
\mathcal{P}_{T}=A_T\left(\frac{k}{k_*}\right)^{n_t}~,
\end{equation}
where the tensor spectral index is~\cite{Lyth:1998xn,Stewart:1993bc}
\begin{equation}
\label{ntdef}
n_t=-2\epsilon\left[1+\left(4C+\frac{11}{3}\right)\epsilon-2\left(\frac{2}{3}+C\right)\eta\right]~.~
\end{equation}
Thus, the tensor-to-scalar ratio is given by~\cite{Lyth:1998xn,Stewart:1993bc}
\begin{equation}
\label{rdef}
r\equiv \frac{A_T}{A_s}=16 \epsilon \left[ 1+8\left(C+\frac{2}{3}\right)(2\epsilon-\eta) \right]~.
\end{equation}
Because $8(C+\frac{2}{3})\simeq -0.506667$, we can safely neglect the term $8 (C+\frac{2}{3})(2\epsilon-\eta)$
at the next leading order in the above equation. Thus, we will take the next leading order
approximation $r=16 \epsilon $ for simplicity.  Therefore, with the BICEP2 result $r=0.16/0.20$,
we obtain  the inflation scale about $2\times 10^{16}$ GeV and the Hubble scale
around $1.0\times 10^{14}$ GeV.

The number of e-folding before the end of inflation is
\begin{equation}
\label{efolddef}
N(\phi)~=~
\int_{t_i}^{t_e}Hdt\approx \frac{1}{M_{\rm Pl}^2}\int_{\phi_e}^{\phi_i}\frac{V(\phi)}{V_\phi(\phi)}d\phi=\frac{1}{\sqrt{2}
\,M_{\rm Pl}}\int_{\phi_e}^{\phi_i}\frac{d\phi}{\sqrt{\epsilon(\phi)}},
\end{equation}
where the value $\phi_i$ of the inflaton at the beginning of the inflation
is the value at the horizon crossing, and
the value $\phi_e$ of the inflaton at the end of inflation is defined by either $\epsilon(\phi_e)=1$
or $\eta(\phi_e)=1$.
From the above equation, we get the Lyth bound~\cite{Lyth:1996im}
\begin{equation}
\Delta \phi \equiv |\phi_i-\phi_e| > {\sqrt{2\epsilon_{\rm min}}} N(\phi) M_{\rm Pl}~,~\,
\end{equation}
where $\epsilon_{\rm min}$ is the minimal $\epsilon$ during inflation.
If $\epsilon(\phi)$ is a monotonic function of $\phi$, we have
$\epsilon_{\rm min}=\epsilon(\phi_i)\equiv\epsilon$. Thus,
for $r=0.01$, 0.05, 0.1, 0.16, and 0.21,
we obtain the large field inflation due to $\Delta\phi > 1.77 ~M_{\rm Pl}$, $4.0 ~M_{\rm Pl}$, $5.6 ~M_{\rm Pl}$,
$7.1 ~M_{\rm Pl}$, and $8.1 ~M_{\rm Pl}$ for $N(\phi)=50$, respectively.
Moreover, to violate the Lyth bound and have the magnitude of $\phi$ smaller than the reduced Planck scale
during inflation, we require that $\epsilon(\phi)$ be not a monotonic function and
have a minimum between $\phi_i$ and $\phi_e$.

In this paper, we will consider the renormalizable three-term polynomial inflation with large tensor-to-scalar
ratio. With slow-roll condition, each term in the polynomial potential will be around $10^{-8}M^4_{\rm Pl}$ or
smaller. However, without slow-roll condition and with fine-tuning, each term could be much larger than
$10^{-8}M^4_{\rm Pl}$ and there exist large cancellations among three terms. Thus, the quantum corrections can
be very large and then out of control during large field inflation.

\section{Supergravity Model Building}

In this paper, to simplify the discussions,
we take $M_{\rm Pl}=1$.  In the non-supersymmetric inflationary models,
we will consider the following polynomial potentials
at the renormalizable level
\begin{eqnarray}
V &=& a_0 + a_1 \phi + a_2 \phi^2 + a_3 \phi^3 + a_4 \phi^4 ~,~
\label{NS-potential}
\end{eqnarray}
where $\phi$ is the inflaton, and $a_i$ are couplings.
In the supersymmetric inflationary models from the supergravity theory,
there are some relations among $a_i$.
Before we construct the concrete models,
let us briefly review the supergravity model building.

In the supergravity theory with a K\"ahler potential $K$ and a superpotential $W$,
the scalar potential is
\begin{equation}
V=e^K\left((K^{-1})^{i}_{\bar{j}}D_i W D^{\bar{j}} \overline{W}-3|W|^2 \right)~,~
\label{sgp}
\end{equation}
where $(K^{-1})^{i}_{\bar{j}}$ is the inverse of the K\"ahler metric
$K_{i}^{\bar{j}}=\partial^2 K/\partial \Phi^i\partial{\bar{\Phi}}_{\bar{j}}$, and $D_iW=W_i+K_iW$.
Moreover, the kinetic term for a scalar field is
\begin{equation}
{\cal L} ~=~ K_{i}^{\bar{j}} \partial_{\mu} \Phi^i \partial^{\mu} {\bar \Phi}_{\bar{j}}~.~\,
\end{equation}

We first briefly review the generic model building. Introducing two superfields
$\Phi$ and $X$, we consider the  K\"ahler potential and superpotential as below
\begin{eqnarray}
K=-\frac{1}{2}(\Phi-{\bar \Phi})^2+X{\bar X}-\delta(X{\bar X})^2~,~
\label{KP-A}
\end{eqnarray}
\begin{eqnarray}
W~=~Xf(\Phi)~.~\,
\label{SP-A}
\end{eqnarray}
Thus, the above K\"ahler potential $K$ is invariant under the following shift
symmetry~\cite{Kawasaki:2000yn, Yamaguchi:2000vm,
Yamaguchi:2001pw, Kawasaki:2001as, Kallosh:2010ug, Kallosh:2010xz, Nakayama:2013jka,
Nakayama:2013txa, Takahashi:2013cxa, Li:2013nfa}
\begin{eqnarray}
\Phi\rightarrow\Phi+CM_{\rm Pl}~,~\,
\label{SSymmetry-A}
\end{eqnarray}
with $C$ a dimensionless real parameter. In general,
the K\"ahler potential $K$ is a function of $\Phi-{\bar \Phi}$ and independent on
the real part of $\Phi$. Before further discussions, we shall present a few comments
on the K\"ahler potential and superpotential

\begin{itemize}

\item {If shift symmetry is a global symmetry, it will be violated by quantum gravity
effects, {\it i.e.}, one might add high-dimensional operators suppressed by the
reduced Planck scale. To solve this problem, one can consider gauged discrete symmetry
from anomalous $U(1)_X$ gauge symmetry inspired from string models, and then
quantum gravity violating effects can be forbidden. }

\item {Shift symmetry is violated by the superpotential in Eq.~(\ref{SP-A}). In principle,
we can break the shift symmetry spontaneuously by introducing a spurion field $S$ and
extending the shift symmetry as follows~\cite{Kawasaki:2000ws}
\begin{eqnarray}
\Phi\rightarrow\Phi+CM_{\rm Pl}~,~~ S \rightarrow \frac{S\Phi}{\Phi+CM_{\rm Pl}}\,~.~
\end{eqnarray}
And we consider the following superpotential
\begin{eqnarray}
W~=~Xf(S\Phi/M_{\rm Pl})~,~\,
\label{SP-A}
\end{eqnarray}
which is clearly invariant under the extended shift symmetry.
After $S$ obtains a non-zero vacuum expectation values, we obtain the
superpotential in Eq.~(\ref{SP-A}). The effects from spontaneous shift symmetry
breaking have been studied in Ref.~\cite{Mazumdar:2014bna}.}

\item {In a supersymmetric theory, the superpotential is non-renormalized, while
there indeed exist quantum corrections to the K\"ahler potential in general. In the renormalizable
 three-term polynomial inflation which we shall study in the following, the inflaton
value is about $10M_{\rm Pl}$, and each term in the scalar potential is about
$10^{-8}M^4_{\rm Pl}$ or smaller during inflation. The  K\"ahler potential for $\Phi$ in
Eq.~(\ref{KP-A})  is about
$100M^2_{\rm Pl}$, and the quantum corrections will be around $10^{-6}M^2_{\rm Pl}$ from the naive dimensional
annalyses with loop factor. Thus, such quantum corrections are under control and negligible.

In addition, supersymmetry is violated during inflation. Thus, the masses for the scalar and fermionic components
of any superfield may be splitted. And then we might have additional one-loop effective scalar potential, which may
affect the inflation and is beyond the scope of our current paper.   }

\end{itemize}

From the above K\"ahler potential and superpotential, the scalar potential is given by
\begin{eqnarray}
V&=& e^K\left[ |(\Phi - {\bar \Phi})Xf(\Phi)+X \frac{\partial f(\Phi)}{\partial \Phi}|^2+|({\bar X}-2\delta X{\bar X}^2)Xf(\Phi)+f(\Phi)|^2
\right. \nonumber \\ && \left.
-3|Xf(\Phi)|^2\right]~.~\,
\end{eqnarray}
Because there is no real component ${\rm Re} [\Phi]$ of $\Phi$ in the K\"ahler potential
due to the shift symmetry, this scalar potential along ${\rm Re} [\Phi]$ is very flat and then
${\rm Re} [\Phi]$ is a natural inflaton candidate.
From the previous studies~\cite{Kallosh:2010ug, Kallosh:2010xz, Li:2013nfa},
we can stabilize the imaginary component ${\rm Im} [\Phi]$ of $\Phi$ and $X$
 at the origin during inflation, {\it i.e.}, ${\rm Im} [\Phi]=0$ and $X=0$.
Therefore, with ${\rm Re} [\Phi]=\phi/{\sqrt 2}$, we get the inflaton potential
\begin{eqnarray}
V~=~|f(\phi/{\sqrt 2})|^2~.~\,
\end{eqnarray}

For a renormalizable superpotential, we have
\begin{eqnarray}
f(\Phi) ~=~a'_0 + a'_1 {\sqrt 2} \Phi +  2 a'_2 \Phi^2 ~,~\,
\end{eqnarray}
where we choose $a_i$ as real numbers.
And then the polynomial inflaton potential is
\begin{eqnarray}
V &=& |a'_0+a'_1 \phi+ a'_2 \phi^2|^2~.~\,
\end{eqnarray}

The polynomial inflations from supergravity model building have been considered before.
At the renormalizable level, only the case with $a_1'\not=0$ and $a_2'\not=0$ has been
studied in the literatures~\cite{Kallosh:2014xwa, Nakayama:2013jka, Nakayama:2013txa}.
In this paper, we also consider the following three cases with $a_0'\not=0$:
(1) $a_0'\not=0$ and $a_1'\not=0$; (2) $a_0'\not=0$ and $a_2'\not=0$; (3) The most general case with
$a_0'\not=0$, $a_1'\not=0$, and $a_2'\not=0$. Moreover, we study the three-term polynomial
inflations whose coefficients for the lowest and highest order terms in the inflaton potential
can be negative.
These inflations cannot be realized in supergravity model building where the
coefficients for the lowest and highest order terms must be positive.

\section{The Renormalizable Three-Term Polynomial Inflation}

To classify the three-term polynomial inflation at renormalizable level,
we consider the following inflaton potential
\begin{eqnarray}
V &=&  a_j \phi^j + a_k \phi^k + a_l \phi^l ~,~
\label{NS-potential-T}
\end{eqnarray}
where $0\le j<k<l\le 4$. With $(j,~k,~l)$, we will study
all the renormalizable
non-supersymmetric and supersymmetric three-term polynomial inflation
with large tensor-to-scalar ratio $r$, which can be consistent with
the Planck and/or BICEP2 experiments. For simplicity, we denote
the maximum and minimum of the inflaton potential as $\phi_M$ and $\phi_m$,
respectively. Because we shall consider the super-Planckian inflation,
our inflation around the maxima and minima of inflaton potentials
is similar to the inflection point inflation~\cite{Allahverdi:2006iq, Allahverdi:2006we,
Allahverdi:2008bt, Enqvist:2010vd}.

\subsection{Inflaton Potential with $(j,~k,~l)=(0, ~1, ~2)$}

\begin{figure}[h]
\centering
\includegraphics[height=5cm]{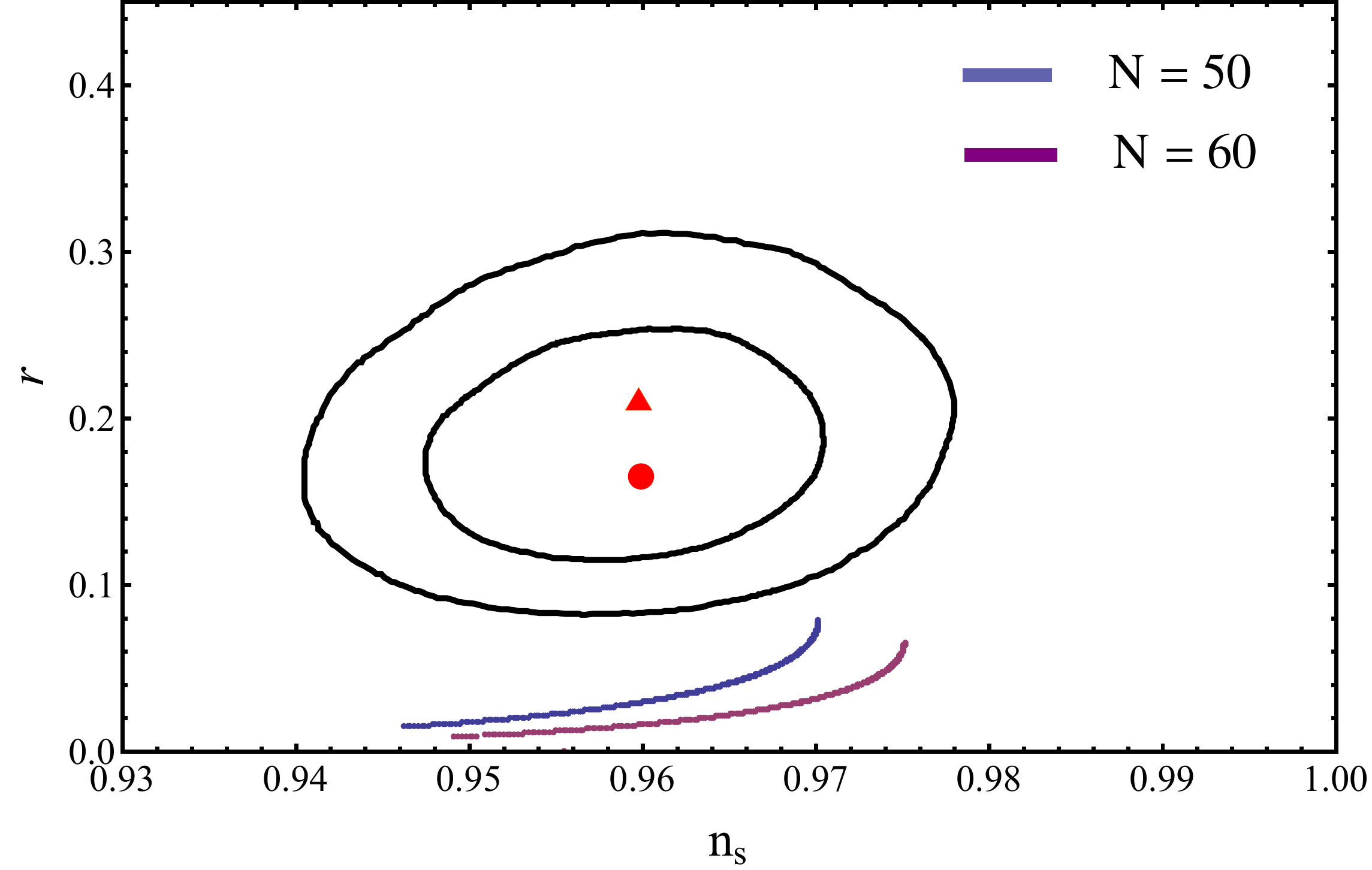}
\caption {$r$ versus $n_s$ for the
inflaton potential with $(j,~k,~l)=(0, ~1, ~2)$ and $a_2<0$.
The inner and outer circles are $1\sigma$ and $2\sigma$ regions, respectively.
} \label{012(a2<0)}
\end{figure}

 First, we consider the non-supersymmetric models with
 the inflaton potential $V=a_0+a_1 \phi + a_2 \phi^2$.
For $a_2<0$, there exists a maximum at $\phi_M=-\frac{a_1}{2a_2}$.
No matter the slow-roll inflation occurs at the right or left
of this maximum (which is the same because of symmetry), we cannot find any $r$
within the $2\sigma$ range of the BICEP2 data. And the numerical results
for $r$ versus $n_s$ is given in Fig.~\ref{012(a2<0)}.
When $n_s$ is within the $1\sigma$ range $0.9603\pm0.0073$, the range of $r$ is $[0.0132, ~0.0534]$.

\begin{figure}[h]
\centering
\includegraphics[height=5cm]{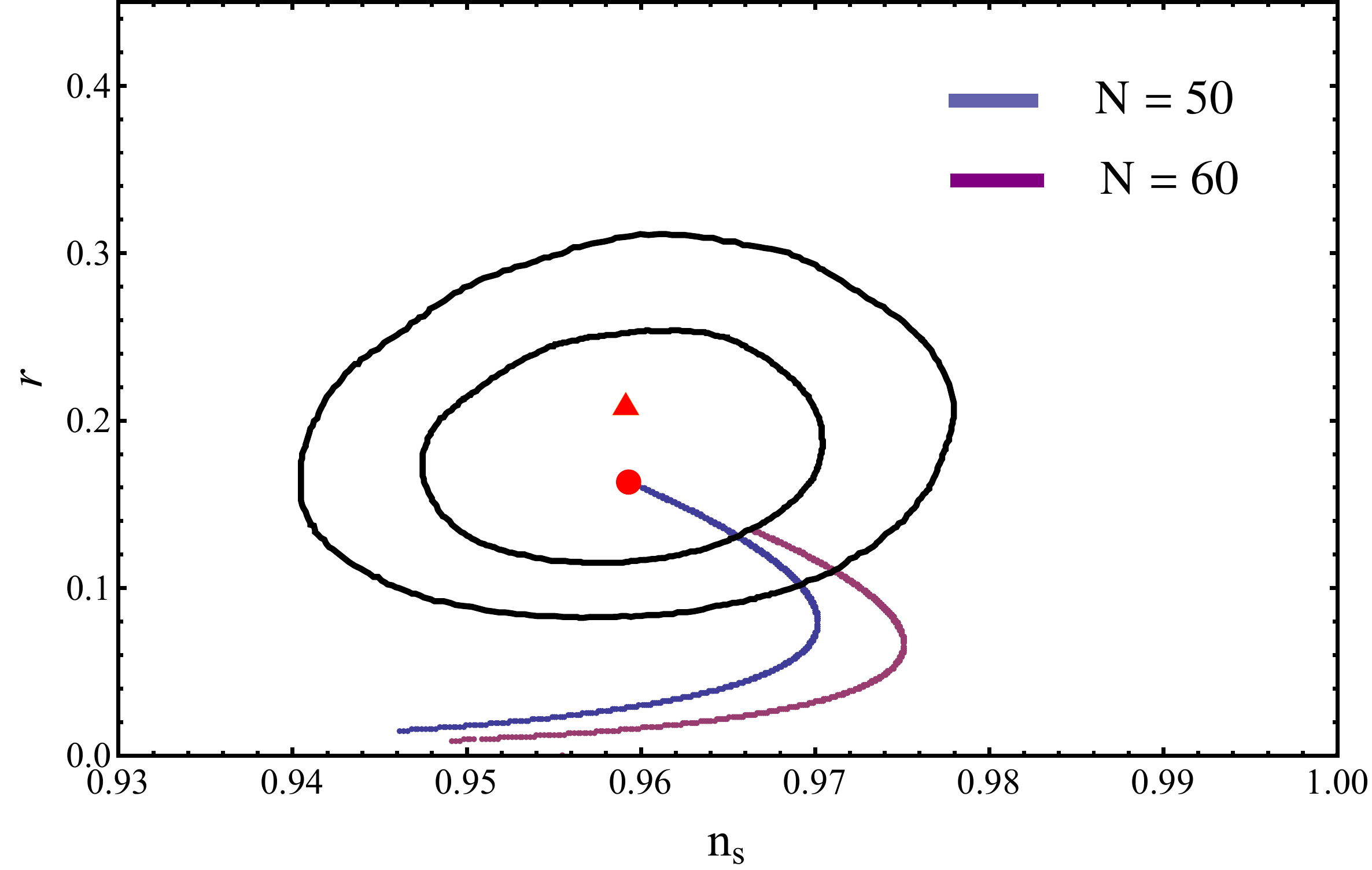}
\caption {$r$ versus $n_s$ for the
inflaton potential with $(j,~k,~l)=(0, ~1, ~2)$, $a_1<0$, and $a_2>0$.} \label{012(a1<0)}
\end{figure}

Moreover, for $a_2>0$ and $a_1<0$, we have a minimum at $\phi_m=-\frac{a_1}{2a_2}$.
We present the numerical results for $r$ versus $n_s$ in Fig.~\ref{012(a1<0)},
where the inner and outer circles are $1\sigma$ and $2\sigma$ regions, respectively.
For $n_s$ in the $1\sigma$ range $0.9603\pm0.0073$,
the range of $r$ is $[0.0132, ~0.16160]$, which can be consistent with the BICEP2 experiment.
In addition, for the number of e-folding
$N_e=50$, $n_s$ and $r$ are within $1\sigma$ and $2\sigma$ regions
of the BICEP2 experiment for $a_1>-30a_2$ and $a_0<\frac{a_1^2+2a_2^2}{4a_2}$ and for
$a_1>-10a_2$ and $a_0<\frac{a_1^2+2a_2^2}{4a_2}$, respectively.
Also, for $N_e=60$, $n_s$ and $r$ are within $2\sigma$ region for
$a_1>-10a_2$ and $a_0<\frac{a_1^2+2a_2^2}{4a_2}$, but no viable parameter space
for $1\sigma$ region. In particular, the best fit point with $n_s=0.96$ and $r=0.16$
for the BICEP2 data can be obtained for $N_e=50$, $a_2\approx -a_1$ and
 $a_2 \approx -3 a_0$. For example, $a_0=3, ~a_1=-10$, and $a_2=10$, and the corresponding
$\phi_i, ~\phi_e$, and $\phi_{m}$ respectively are $-13.621, ~0.464$, and $0.5$.
Thus, we obtain $\Delta \phi=14.085$, which satisfies the Lyth bound.
In the following discussions, we will not comment on $\Delta \phi$ since
the Lyth bound is always satisfied in our models.

Second, we consider the supersymmetric model with inflaton potential
 $V=a^2+2ab\phi + b^2 \phi^2$, which
has a minimum at $-a/b$. We obtain that for $\phi = -a/b\pm \sqrt{2}$, both $\epsilon$ and
$\eta$ are equal to 1, and then the slow-roll inflation ends. Also,
 we find that no matter the slow-roll inflation occurs at the left or right of the minimum,
$n_s$ and $r$ can be written as functions of the e-folding number $N_e$
 \begin{eqnarray}
 n_s~=~1-\frac{8}{4N_e+2}~,~~~
 r~=~\frac{32}{4N_e+2}
 \end{eqnarray}
Thus, for $N_e=50$, we get $n_s=0.9604$ and $r=0.1584$. And for $N_e=60$,
we get $n_s=0.9669$ and $r=0.1322$. In fact, this is similar to the
chaotic inflation with inflaton potential $\phi^2$.

\subsection{Inflaton Potential with $(j,~k,~l)=(0, ~1, ~3)$}

\begin{figure}[h]
\centering
\includegraphics[height=5cm]{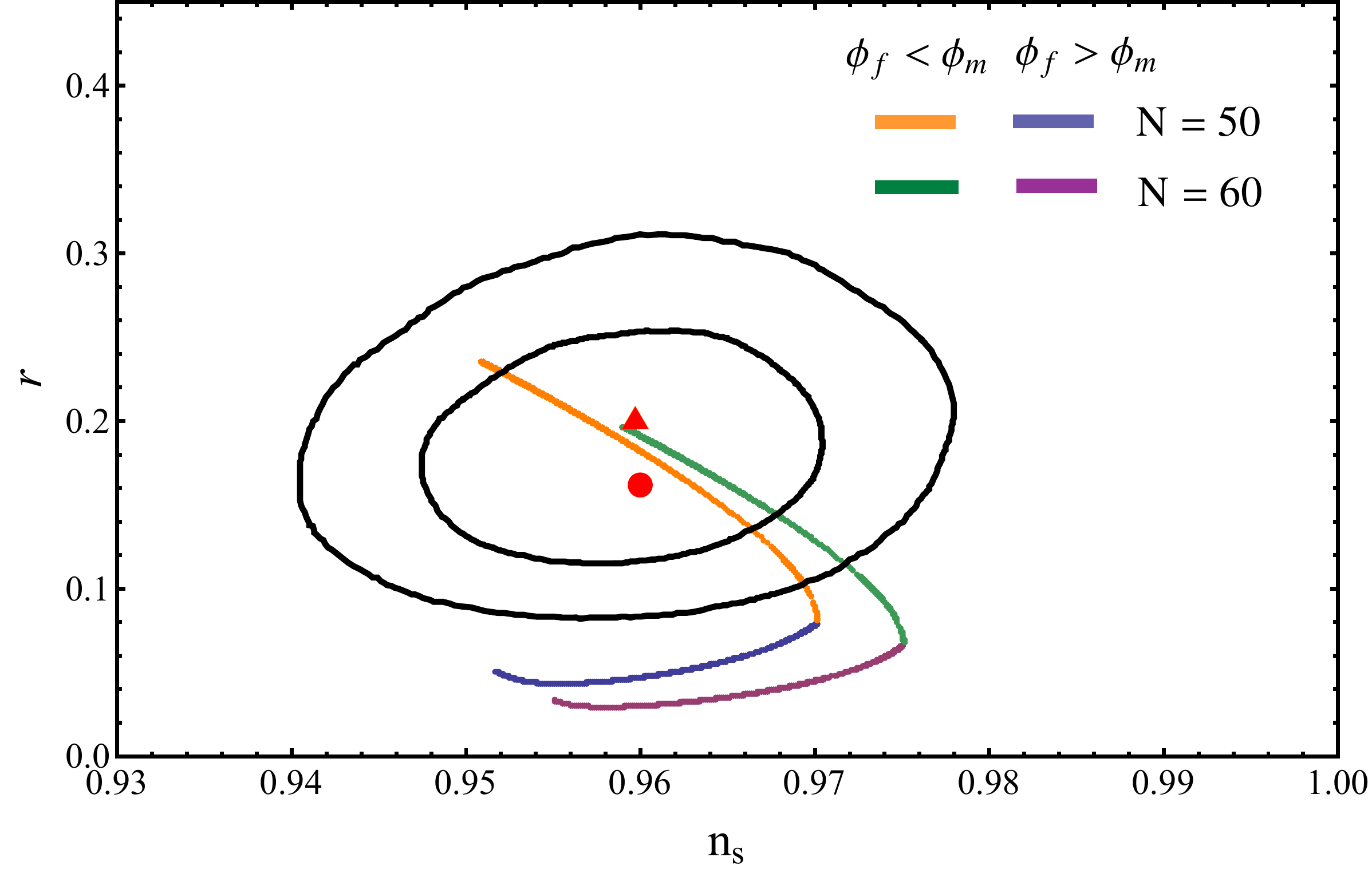}
\caption {
$r$ versus $n_s$ for the
inflaton potential with $(j,~k,~l)=(0, ~1, ~3)$, $a_1>0$, and $a_3<0$,
where the inflationary trajectories are
at the left and right of the minimum.
}
\label{013_1}
\end{figure}

The inflaton potential is $V=a_0+a_1 \phi + a_3 \phi^3$. First,
we consider $a_1>0$ and $a_3<0$. Because there is
a minimum at $\phi_m=-\sqrt{-\frac{a_1}{3a_3}}$ and a maximum at
$\phi_M=\sqrt{-\frac{a_1}{3a_3}}$, we have three inflationary trajectories,
and let us discuss them one by one.
When the slow-roll inflation occurs at the left of
the minimum,  the numerical results for $r$ versus $n_s$ is given in Fig.~\ref{013_1}.
The range of $r$ is about  $[0.1231,~0.2237]$ for $n_s$ within its
$1\sigma$ range $0.9603\pm0.0073$, which is consistent with the BICEP2 results.
In the viable parameter space, we generically have $a_0 < a_1$.
For the number of e-folding $N_e=50$, $n_s$ and $r$ are within $1\sigma$ and $2\sigma$ regions
of the BICEP2 experiment for $-11a_3 < a_1<-1000a_3$ and $-11a_3 < a_1<-5000a_3$, respectively.
For the number of e-folding $N_e=60$, $n_s$ and $r$ are within $1\sigma$ and $2\sigma$ regions
of the BICEP2 experiment for $ a_1<-125a_3$ and $ a_1<-600a_3$, respectively.
To be concrete, we will present two best fit points for the BICEP2 data.
The best fit point with $n_s=0.963$ and $r=0.16$ can be realized for
 $N_e=50$, $a_0\thickapprox a_1 \thickapprox -230 a_3$, for example, $a_0=1, a_1=1$ and
$a_3=-0.00436$, and the corresponding  $\phi_i, ~\phi_f, ~\phi_{m}$, and $\phi_{M}$ are
respectively $-27.1459, -15.3793, -8.74372$, and $8.74372$.
Another best fit point with $n_s=0.959$ and $r=0.196$ can be obtained
for $N_e=60$, $a_0\thickapprox a_1 \thickapprox -2 a_3$,  for example,
 $a_0=1, ~a_1=1$, and $a_3=-0.5$, and the corresponding  $\phi_i, ~\phi_f, ~\phi_{m}$, and
$\phi_{M}$ are $-19.2289, ~-2.35496, ~-0.816497$, and $0.816497$, respectively.
In addition, when slow-roll inflation occurs at the right of the minimum,
we also present the numerical results for $r$ versus $n_s$ in Fig.~\ref{013_1}.
The range of $r$ is about [0.0337, 0.0669]  for $n_s$ within its
$1\sigma$ range $0.9603\pm0.0073$. Although we can not fit
the BICEP2 data, we still have large enough tensor-to-scalar ratio,
which can be tested at the future Planck and QUBIT experiments.

\begin{figure}[h]
\centering
\includegraphics[height=5cm]{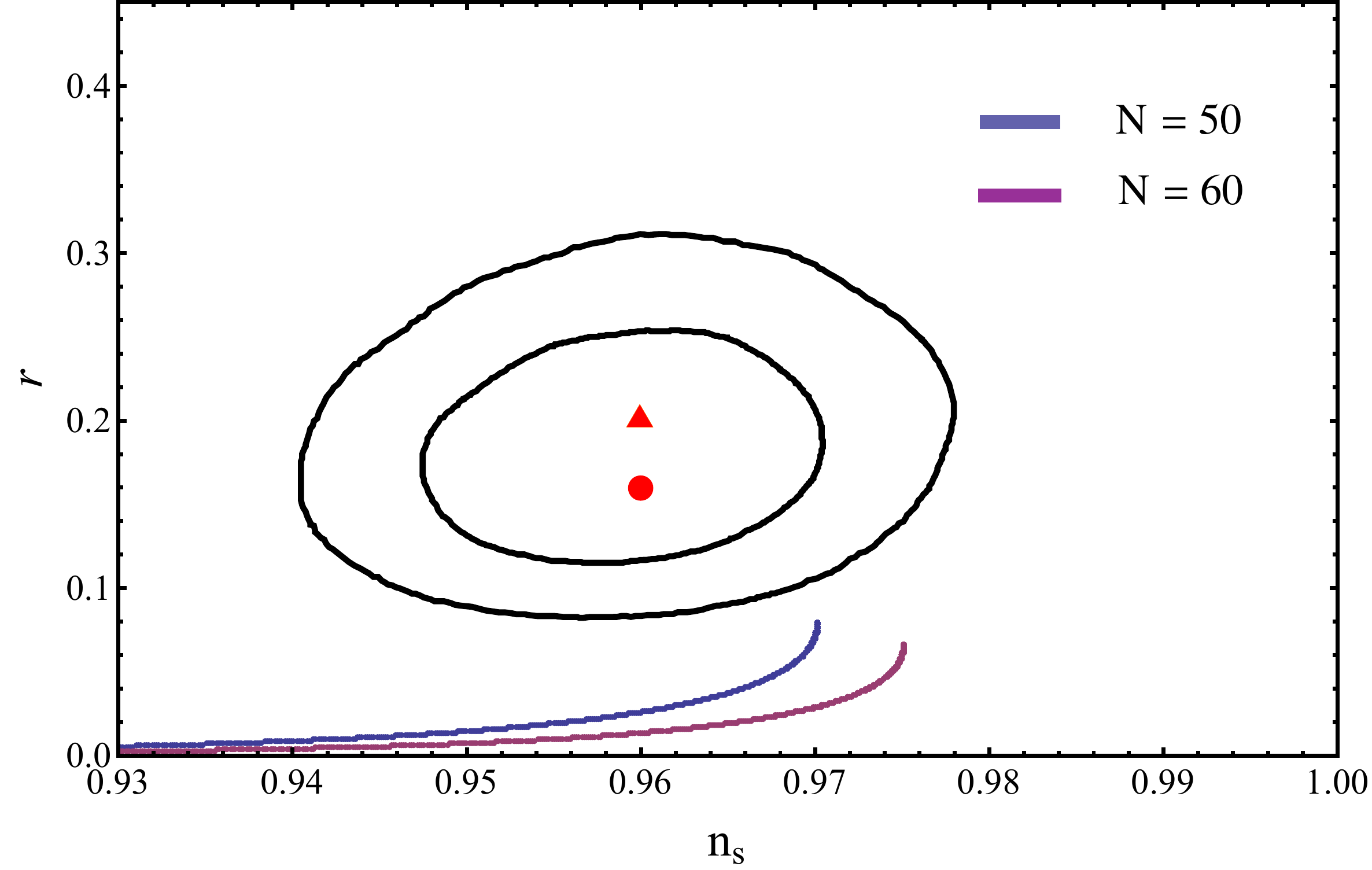}
\caption{
$r$ versus $n_s$ for the
inflaton potential with $(j,~k,~l)=(0, ~1, ~3)$, $a_1>0$, and $a_3<0$,
 where the inflationary trajectory is
at the right of the maximum.
}
\label{013_2}
\end{figure}

Furthermore, for the slow-roll inflation at the right of the maximum,
 the numerical results for $r$ versus $n_s$ is given in Fig.~\ref{013_2}.
For $n_s$ in the $1\sigma$ range $0.9603\pm0.0073$,
the range of $r$ is $[0.0085, ~ 0.0482]$, which is within the reach of
the future Planck and QUBIT experiments.

\begin{figure}[h]
\centering
\includegraphics[height=5cm]{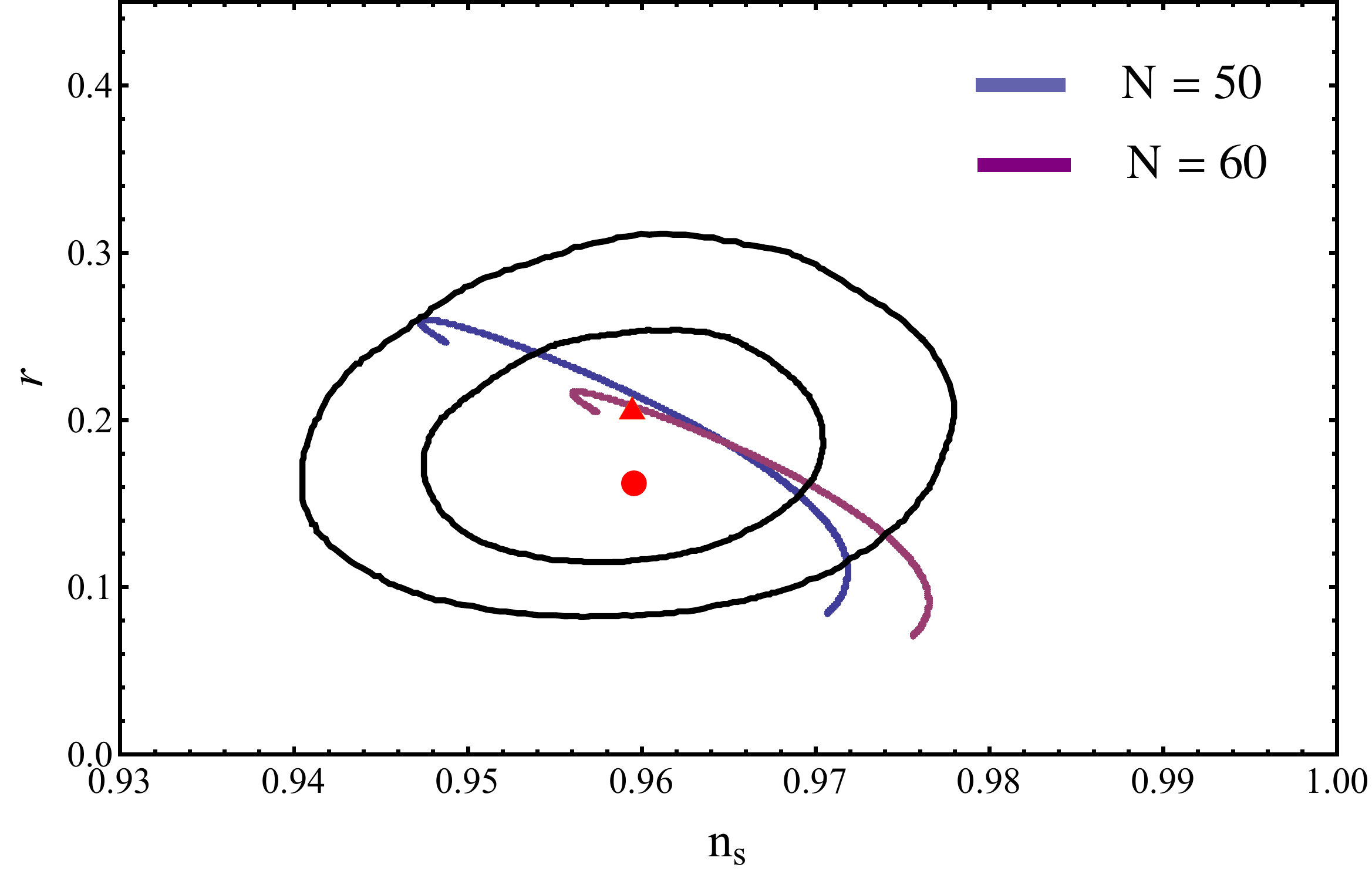}
\caption{
$r$ versus $n_s$ for the
inflaton potential with $(j,~k,~l)=(0, ~1, ~3)$, $a_1<0$, and $a_3<0$.
}
\label{013_3}
\end{figure}

Second, we consider $a_1<0$ and $a_3<0$, the potential will decrease monotonically, and
the curves for $r$ versus $n_s$ are given in Fig.~\ref{013_3}.
The range of $r$ is about  $[0.1670,~0.2427]$ for $n_s$
within its $1\sigma$ range $0.9603\pm0.0073$, which is consistent with the BICEP2 results.
In the viable parameter space, we generically have $a_0\thickapprox  1$.
For the number of e-folding $N_e=50$, $n_s$ and $r$ are within $1\sigma$ and $2\sigma$ regions
of the BICEP2 experiment for $-90a_3 < -a_1<-300a_3$ and $-a_1<-1000a_3$, respectively.
For the number of e-folding $N_e=60$, $n_s$ and $r$ are within $1\sigma$ and $2\sigma$ regions
of the BICEP2 experiment for $ -a_1<-210a_3$ and $ -a_1<-300a_3$, respectively.
The best fit point with $n_s=0.96$ and $r=0.206$ can be realized for
 $N_e=60$, $a_0\thickapprox 1 $ and $ - a_1\thickapprox -90 a_3$, for example, $a_0=1, a_1=-90$ and
$a_3=-1$, and the corresponding  $\phi_i$ and $\phi_f$ are
respectively $-15.2061$ and $-0.7038$.

\subsection{Inflaton Potential with $(j,~k,~l)=(0, ~1, ~4)$}

\begin{figure}[h]
\centering
\includegraphics[height=5cm]{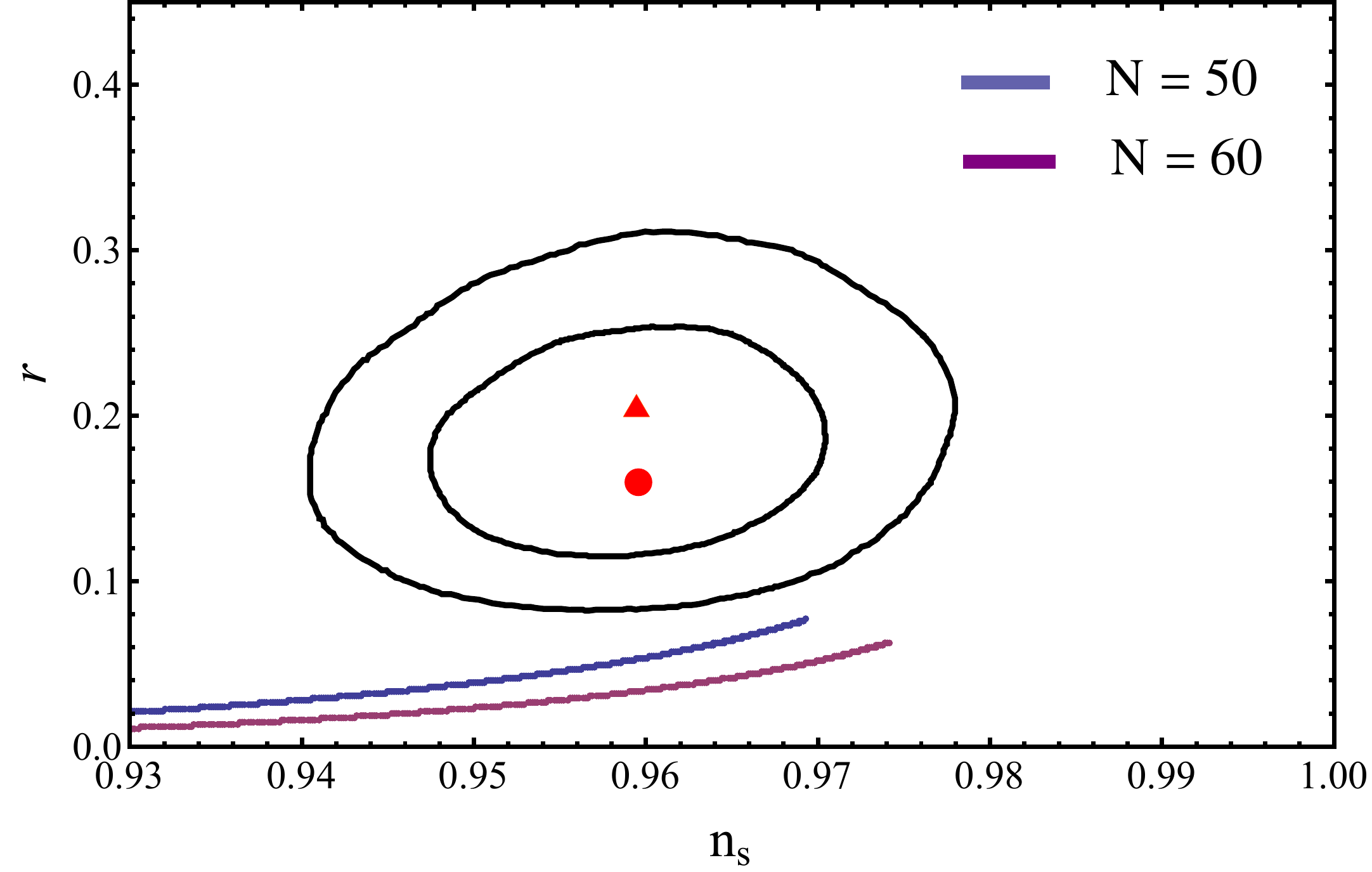}
\caption {
$r$ versus $n_s$ for the
inflaton potential with $(j,~k,~l)=(0, ~1, ~4)$, $a_1>0$, and $a_4<0$ for inflation
at the left of the maximum.
}
\label{014(a4>0)}
\end{figure}

\begin{figure}[h]
\centering
\includegraphics[height=5cm]{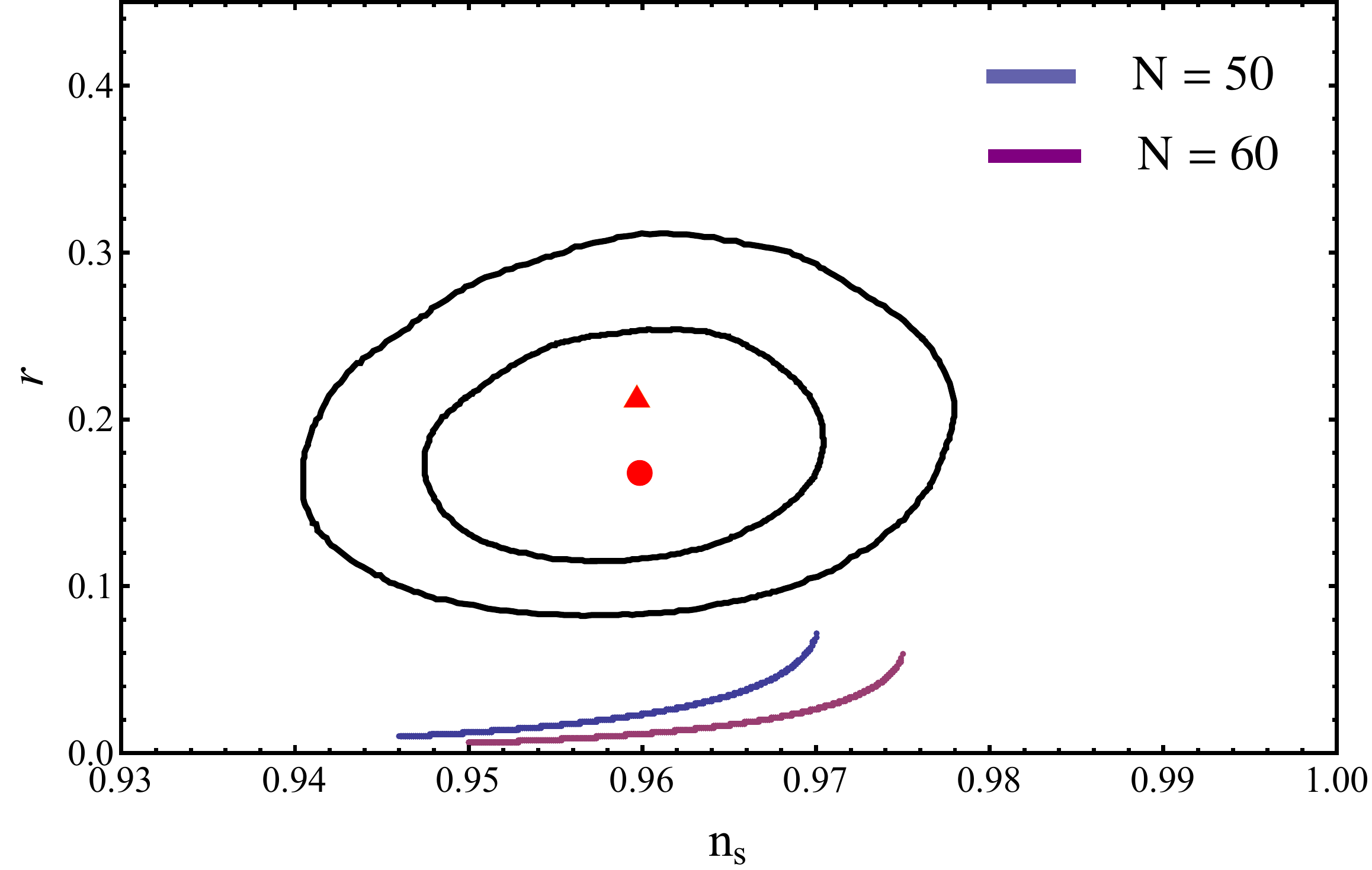}
\caption {
$r$ versus $n_s$ for the
inflaton potential with $(j,~k,~l)=(0, ~1, ~4)$, $a_1>0$, and $a_4<0$ for inflation
at the right of the maximum.
}
\label{014_2}
\end{figure}

We consider the non-supersymmetric inflation models with $V=a_0+a_1 \phi + a_4 \phi^4$.
First, we consider $a_1>0$ and $a_4<0$. There is a maximum at $\phi_M=\left(-\frac{a_1}{4a_4}\right)^{1/3}$.
When slow-roll inflation occurs at the left and right of the maximum,
we present the numerical results for $r$ versus $n_s$ in Figs.~\ref{014(a4>0)}
and \ref{014_2}, respectively. For $n_s$ in the $1\sigma$ range $0.9603\pm0.0073$,
the corresponding ranges of $r$ are $[0.0250,~0.0732]$ and $[0.0077,~0.0459]$,
respectively, which is large enough to be tested
at the future Planck and QUBIT experiments.

\begin{figure}[h]
\centering
\includegraphics[height=5cm]{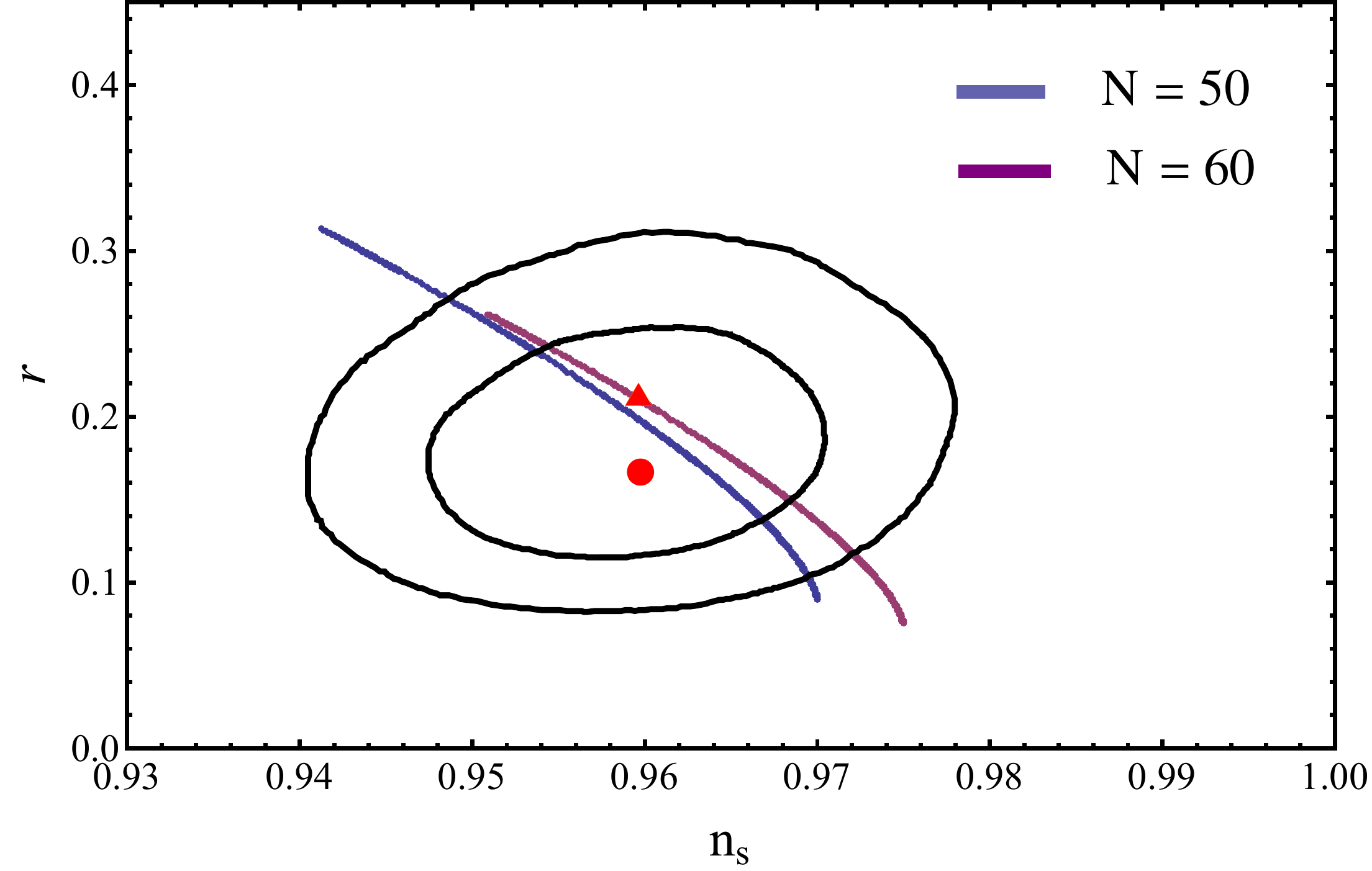}
\caption {
$r$ versus $n_s$ for the
inflaton potential with $(j,~k,~l)=(0, ~1, ~4)$, $a_1<0$, and $a_4>0$.
} \label{014_4}
\end{figure}

Second, we consider $a_1<0$ and $a_4>0$. There exists a minimum
at $\phi_m=\left(-\frac{a_1}{4a_4}\right)^{1/3}$.
 If the slow-roll inflation occurs at the left of the minimum,
we obtain $n_s \leq 0.94$ and $r>0.3$, which is not consistent with
the Planck and BICEP2 data.
When the slow-roll inflation occurs at the right of the minimum,
the numerical results for $r$ versus $n_s$ is given in Fig.~\ref{014_4}.
With $n_s$ in the $1\sigma$ range $0.9603\pm0.0073$,
 the range of $r$ is about $[0.1288,~0.2498]$,
which agrees with the BICEP2 experiment.
Moreover, for the number of e-folding
$N_e=50$, $n_s$ and $r$ are within $1\sigma$ and $2\sigma$ regions
of the BICEP2 experiment  for $-5 \times 10^4 a_4<a_1<-1000a_4$
and $-1 \times 10^6a_4<a_1<-3000a_4$, respectively.
And for $N_e=60$, $n_s$ and $r$ are within $1\sigma$ and $2\sigma$ regions
of the BICEP2 experiment for $-1\times 10^4 a_4<a_1<-100 a_4$
and $-5\times10^4 a_4<a_1$, respectively.
To be concrete, we will present the best fit point for the BICEP2 data.
The best fit point with $n_s=0.9607$ and $r=0.2035$ can be realized for
$N_e=60$, $a_1<-100a_0$, and $a_1\thickapprox -1000 a_4$,
for example, $a_0=1, ~a_1=-1000$, and $a_4=1$, and the corresponding
$\phi_i, ~\phi_f$, and $\phi_{m}$ are respectively $26.1887, ~10.8134$, and $6.29961$.

\subsection{Inflaton Potential with $(j,~k,~l)=(0, ~2, ~3)$}

\begin{figure}[h]
\centering
\includegraphics[height=5cm]{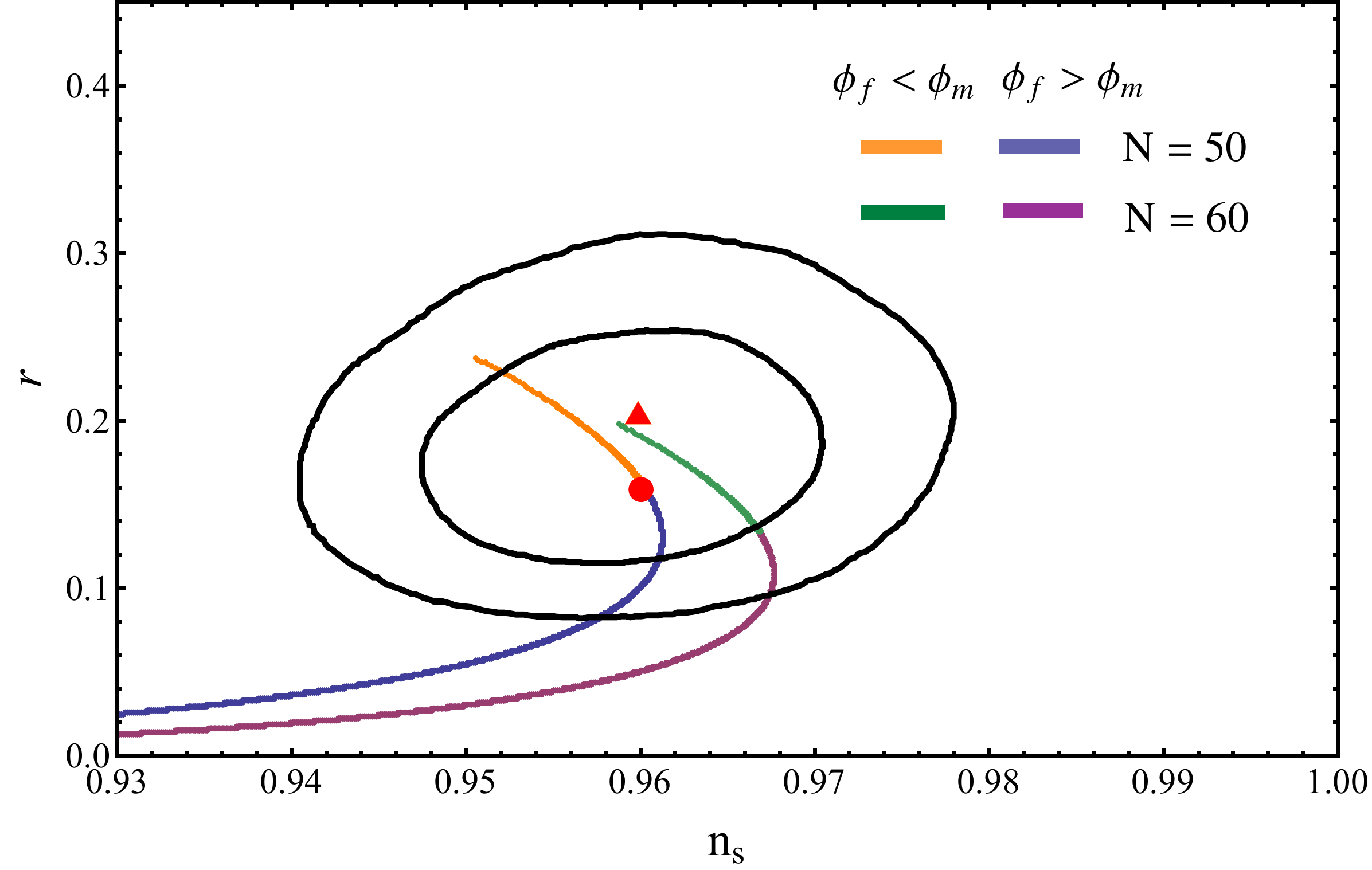}
\caption {
$r$ versus $n_s$ for the
inflaton potential with $(j,~k,~l)=(0, ~2, ~3)$ where the inflationary trajectories are
at the left and right of the minimum.
}
\label{023_1}
\end{figure}

We consider the inflationary model with potential $V=a_0+a_2\phi^2+a_3\phi^3$.
First, for $a_2>0$ and $a_3<0$, there exist
 a minimum at $\phi_m=0$ and a maximum at $\phi_M=-\frac{2a_2}{3a_3}$.
So we have three inflationary trajectories,
and let us discuss them one by one.
When the slow-roll inflation occurs at the left of
the minimum,  we present the numerical results for $r$ versus $n_s$ in Fig.~\ref{023_1}.
For $n_s$ within its $1\sigma$ range $0.9603\pm0.0073$,
the range of $r$ is about  $[0.1363,~0.2206]$, which agree with the BICEP2 results.
For the number of e-folding $N_e=50$, $n_s$ and $r$ are within $1\sigma$ and $2\sigma$ regions
of the BICEP2 experiment for $a_2>-5a_3$ and $a_0<{\rm Max}(a_2/2,-2a_3)$
and for $a_0<{\rm Max}(a_2/2,-2a_3)$, respectively.
Also, for $N_e=60$, $n_s$ and $r$ are within $1\sigma$ and $2\sigma$ regions
of the BICEP2 experiment for $a_0<{\rm Max}(a_2/2,-2a_3)$.
To be concrete, we will present two best fit points for the BICEP2 data.
The best fit point with $n_s=0.96$ and $r=0.16$ can be realized for
$N_e=50$, $a_2>10a_0$ and $a_2\thickapprox10^3 a_3$, for instance,
 $a_0=1, ~a_2=10$, and $a_3=-0.01$, and the corresponding
$\phi_i, ~\phi_f$, and $\phi_{m}$ are respectively $-14.2222, ~-1.34067$, and $0$.
Another best fit point with $n_s=0.958$ and $r=0.199$ can be obtained
for $N_e=59$, $a_2\approx40a_0$, and $a_2\thickapprox-2 a_3$,  for example,
 $a_0=1, a_2=10$ and $a_3=-0.01$,
and the corresponding  $\phi_i, ~\phi_f$, and $\phi_{m}$ are
respectively $-18.3869, ~-1.73496$, and $0$.

In addition, when slow-roll inflation occurs at the right of the minimum,
 the numerical results for $r$ versus $n_s$ are given in Fig.~\ref{023_1} as well.
The range of $r$ is about $[0.0645,~0.160]$  for $n_s$ within its
$1\sigma$ range $0.9603\pm0.0073$. In the viable parameter space,
we have $a_0<a_2/2$ in general.
For the number of e-folding $N_e=50$, $n_s$ and $r$ are within $1\sigma$ and $2\sigma$ regions
of the BICEP2 experiment for $a_2>-50a_3$ and $a_2>-30a_3$, respectively.
And for the number of e-folding $N_e=60$, $n_s$ and $r$ are out of the  $1\sigma$ region
of the BICEP2 experiment and are  within $2\sigma$ region
for $a_2>-50a_3$.
The best fit point with $n_s=0.96$ and $r=0.158$ for the BICEP2 data can be realized for
 $N_e=50$, $a_2>10a_0$, and $a_2>-10^4 a_3$, for instance,
 $a_0=0.1, ~a_2=1$, and $a_3=-10^{-4}$,  and the corresponding
$\phi_i, ~\phi_f, ~\phi_{m}$, and $\phi_M$ are respectively
$14.1854, ~1.33945, ~0.0$, and $6666.67$.

\begin{figure}[h]
\centering
\includegraphics[height=5cm]{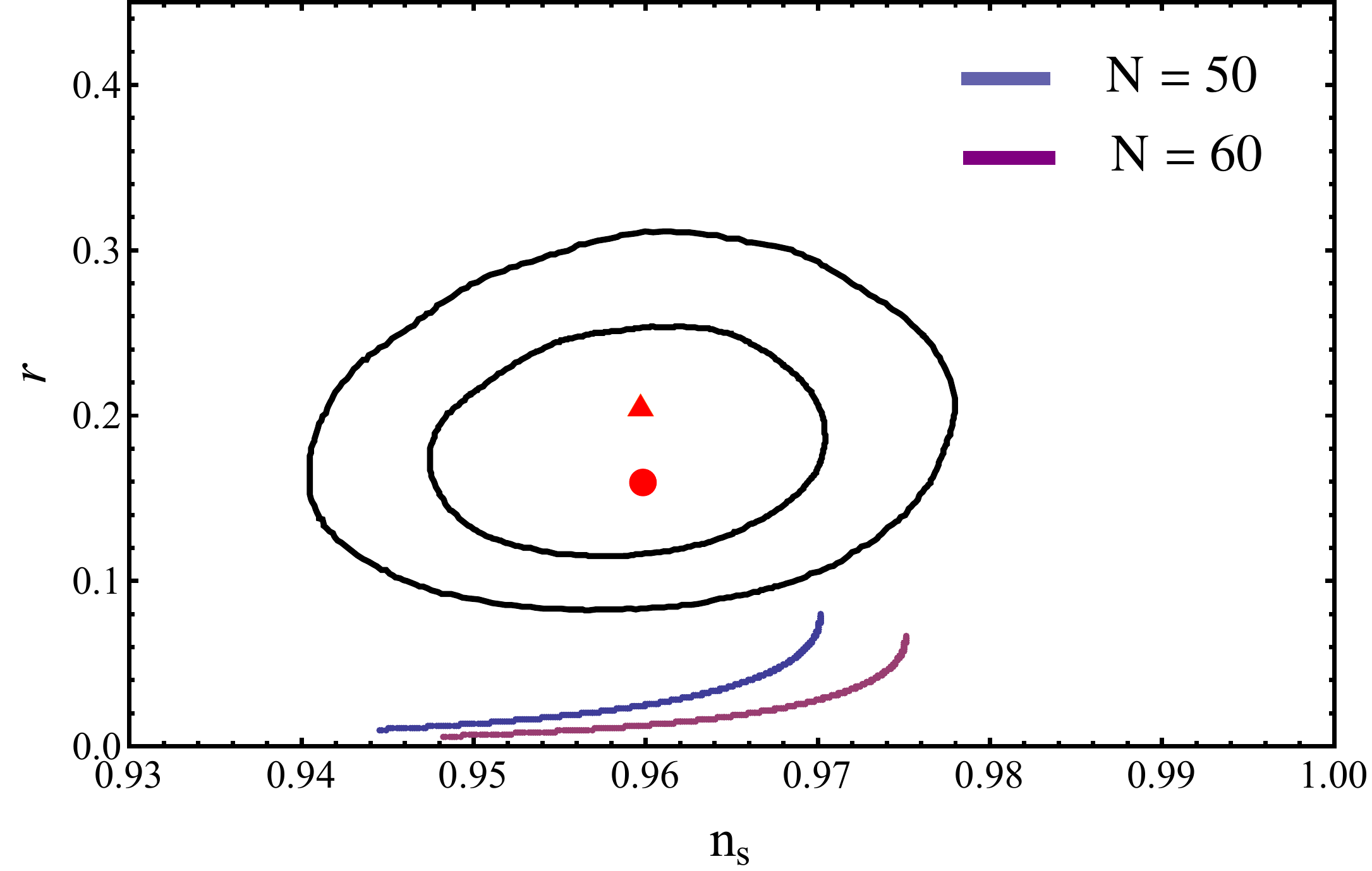}
\caption {
$r$ versus $n_s$ for the
inflaton potential with $(j,~k,~l)=(0, ~2, ~3)$ where the inflationary trajectory is at the right of the maximum.
} \label{023_2}
\end{figure}

Furthermore, for the slow-roll inflation at the right of the maximum,
 the numerical results for $r$ versus $n_s$ are given in Fig.~\ref{023_2}.
For $n_s$ in the $1\sigma$ range $0.9603\pm0.0073$,
the range of $r$ is $[0.0097, ~ 0.0431]$,
which can be tested at the future Planck and QUBIT experiments.

\begin{figure}[h]
\centering
\includegraphics[height=5cm]{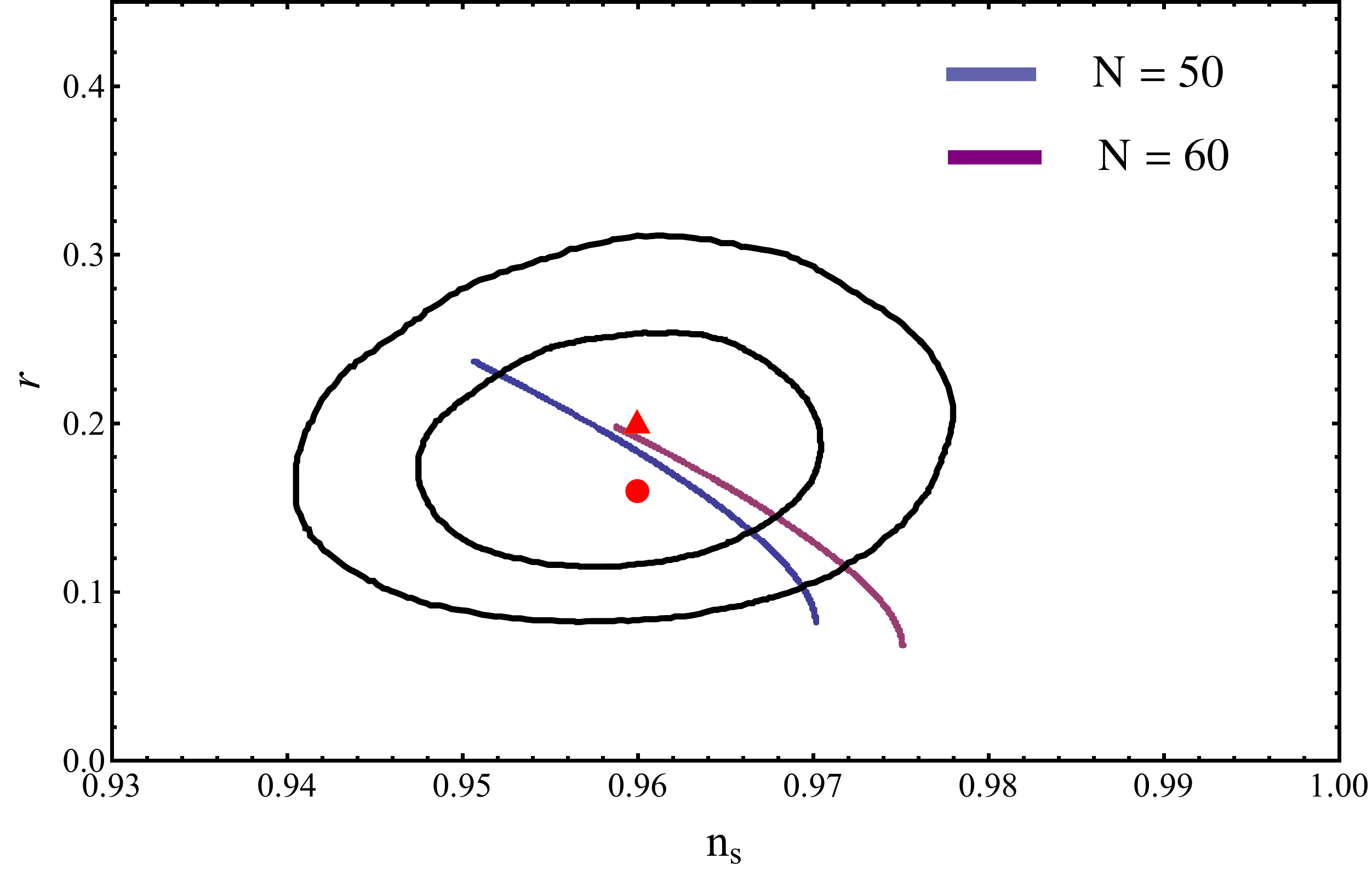}
\caption{
$r$ versus $n_s$ for the inflaton potential with $(j,~k,~l)=(0,~2,~3)$
where the inflationary trajectory is at the left of the minimum.
}
\label{0(-2)(-3)_1}
\end{figure}

Second, for $a_2<0$ and $a_3<0$, there exist a minimum at $\phi_m=-\frac{2a_2}{3a_3}$ and a maximum at $\phi_M=0$.
Similar to the above discussions, there exist three inflationary trajectories,
and we will discuss them one by one.
When the slow-roll inflation occurs at the left of the minimum,
we present the numerical results for $r$ versus $n_s$ in Fig.~\ref{0(-2)(-3)_1}.
For $n_s$ within its $1\sigma$ range $0.9603\pm0.0073$,
the range of $r$ is about $[0.1249,~0.2242]$, which can be consistent with the BICEP2 experiment.
Generically, we have  $a_0\thickapprox 1$. For the number of e-folding $N_e=50$,
 $n_s$ and $r$ are within $1\sigma$ and $2\sigma$ regions
of the BICEP2 experiment for $-a_2<-30a_3$ and $-a_2<-100a_3$, respectively.
Also, for $N_e=60$, $n_s$ and $r$ are within $1\sigma$ and $2\sigma$ regions
of the BICEP2 experiment respectively for $-a_2<-15a_3$ and $-a_2<-35a_3$.
To be concrete, we will present two best fit points for the BICEP2 data.
The best fit point with $n_s=0.959$ and $r=0.196$ can be realized for
$N_e=60$, $a_0=1$, and $-a_2\thickapprox -2 a_3$, for instance,
 $a_0=1, ~a_2=-2$, and $a_3=-1$, and the corresponding
$\phi_i, ~\phi_f$, and $\phi_{m}$ are respectively $-19.8863, ~-3.10761$, and $-1.33333$.

\begin{figure}[h]
\centering
\includegraphics[height=5cm]{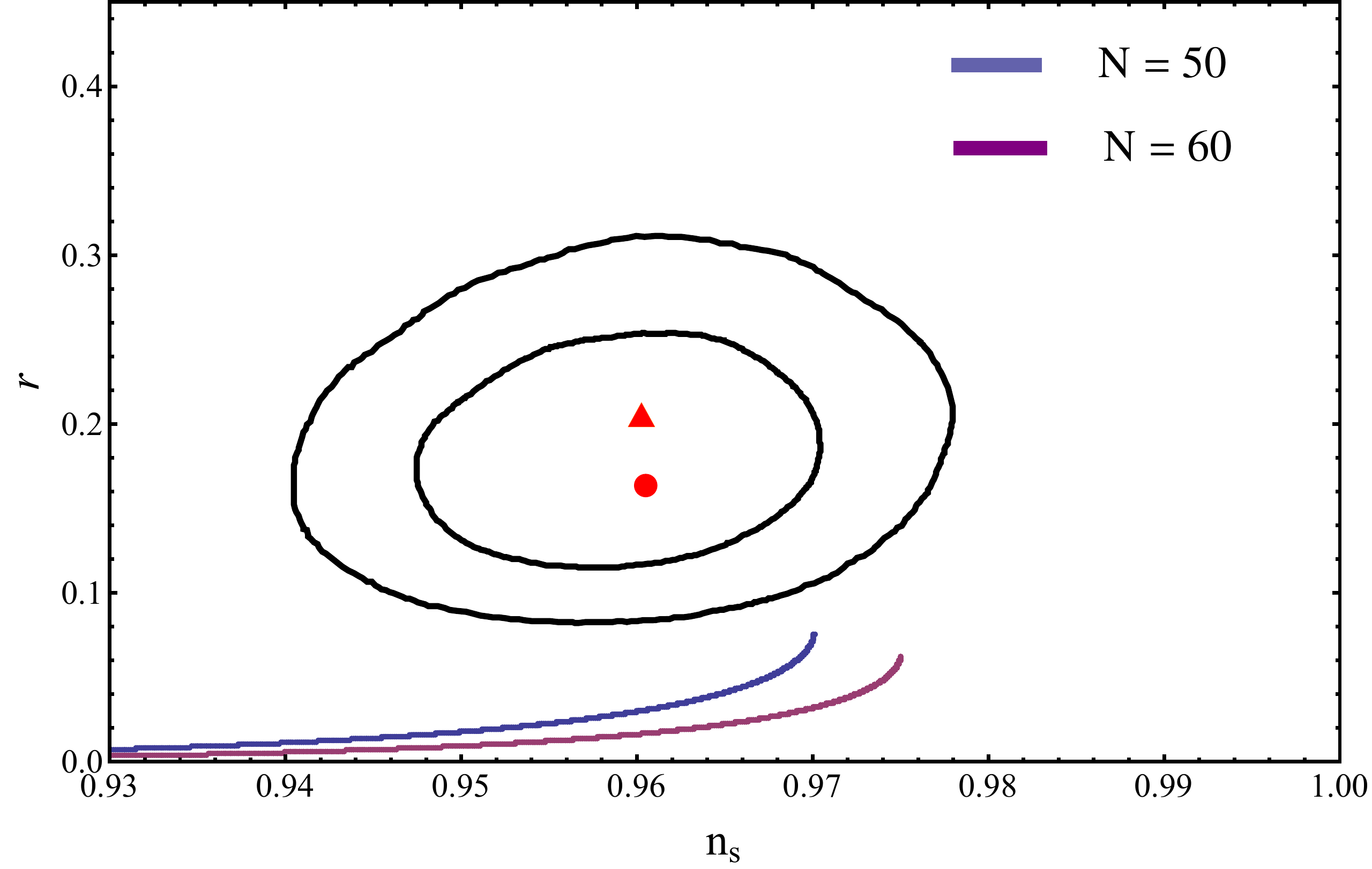}
\caption{
$r$ versus $n_s$ for the inflaton potential with $(j,~k,~l)=(0,~2,~3)$
where the inflationary trajectory is at the right of the minimum.
}
\label{0(-2)(-3)_2}
\end{figure}

\begin{figure}[h]
\centering
\includegraphics[height=5cm]{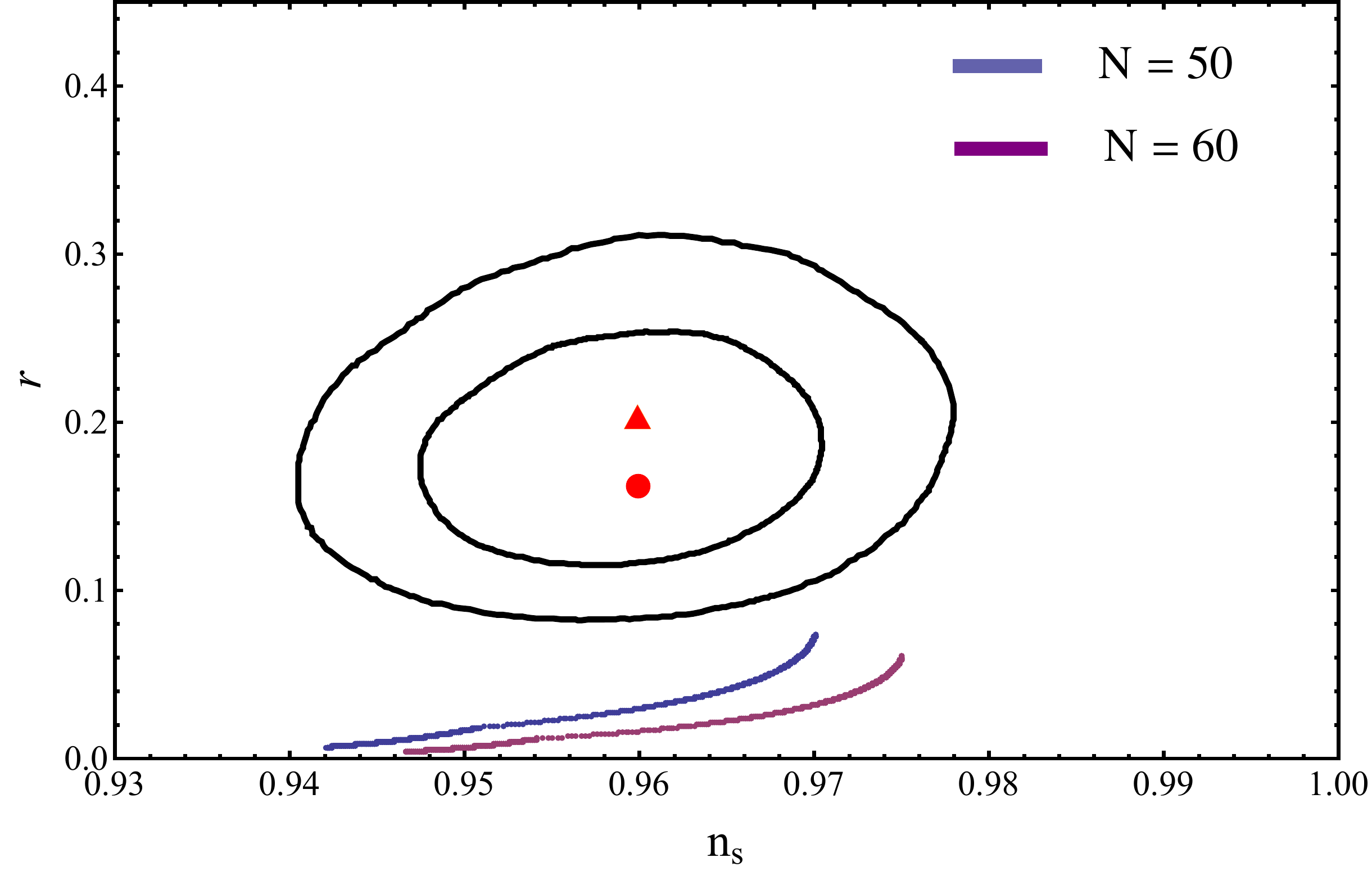}
\caption{
$r$ versus $n_s$ for the inflaton potential with $(j,~k,~l)=(0,~2,~3)$ where
the inflationary trajectory is at the right of the maximum.
}
\label{0(-2)(-3)_3}
\end{figure}

In addition, when the slow-roll inflations occur at the right of the minimum and
maximum, we present the numerical results
for $r$ versus $n_s$ in Figs.~\ref{0(-2)(-3)_2} and \ref{0(-2)(-3)_3}.
For $n_s$ within its $1\sigma$ range $0.9603\pm0.0073$,
the corresponding ranges of $r$ are respectively $[0.0104,~0.0512]$
and $[0.0099,~0.0505]$, which are within the reach of the
future Planck and BICEP2 experiments.

\begin{figure}[h]
\centering
\includegraphics[height=5cm]{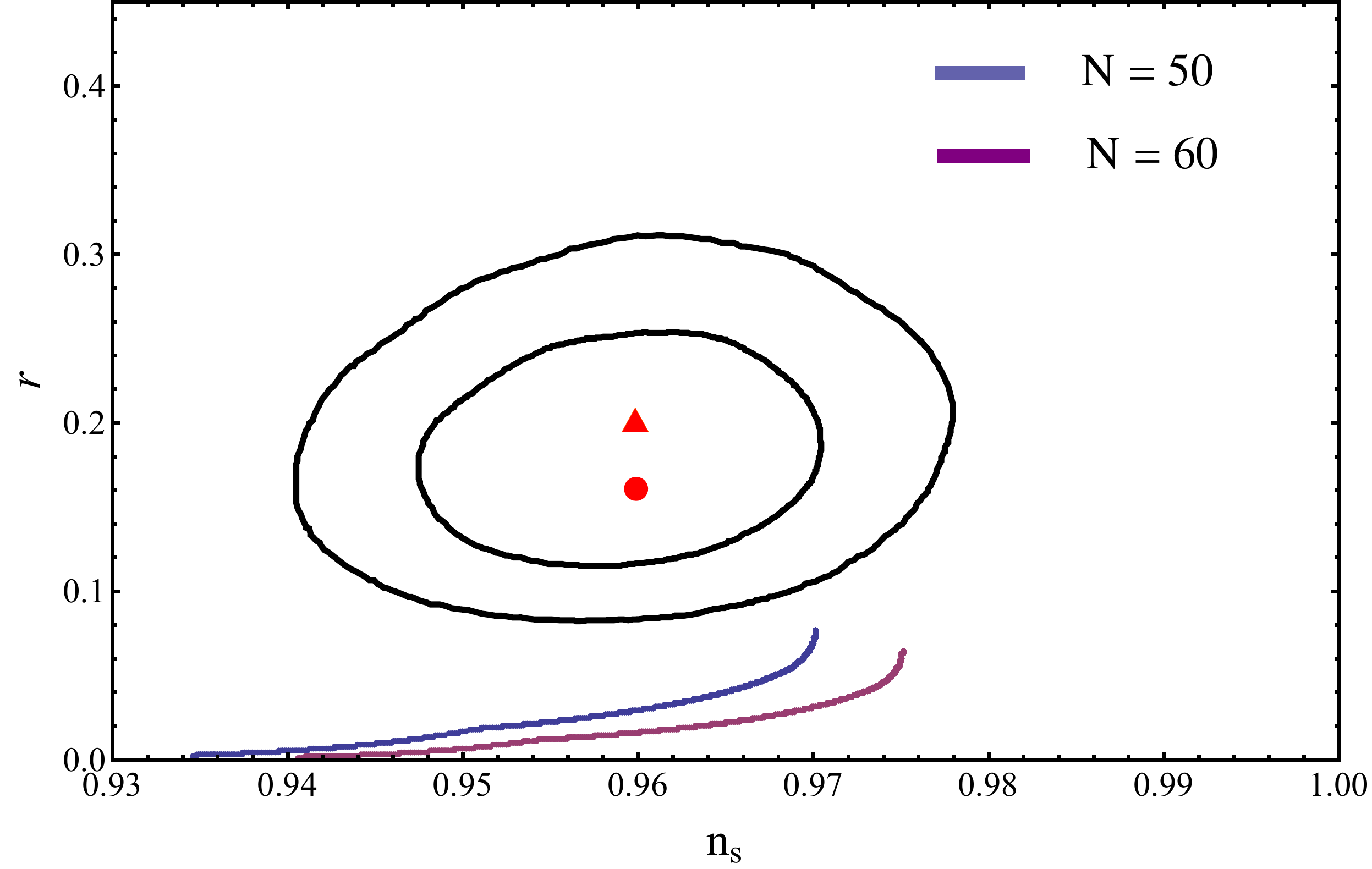}
\caption{
$r$ versus $n_s$ for the inflaton potential with $(j,~k,~l)=(0,~2,~3)$ where
the inflationary trajectory is at the left of the maximum.
}
\label{0(-2)3_1}
\end{figure}

\begin{figure}[h]
\centering
\includegraphics[height=5cm]{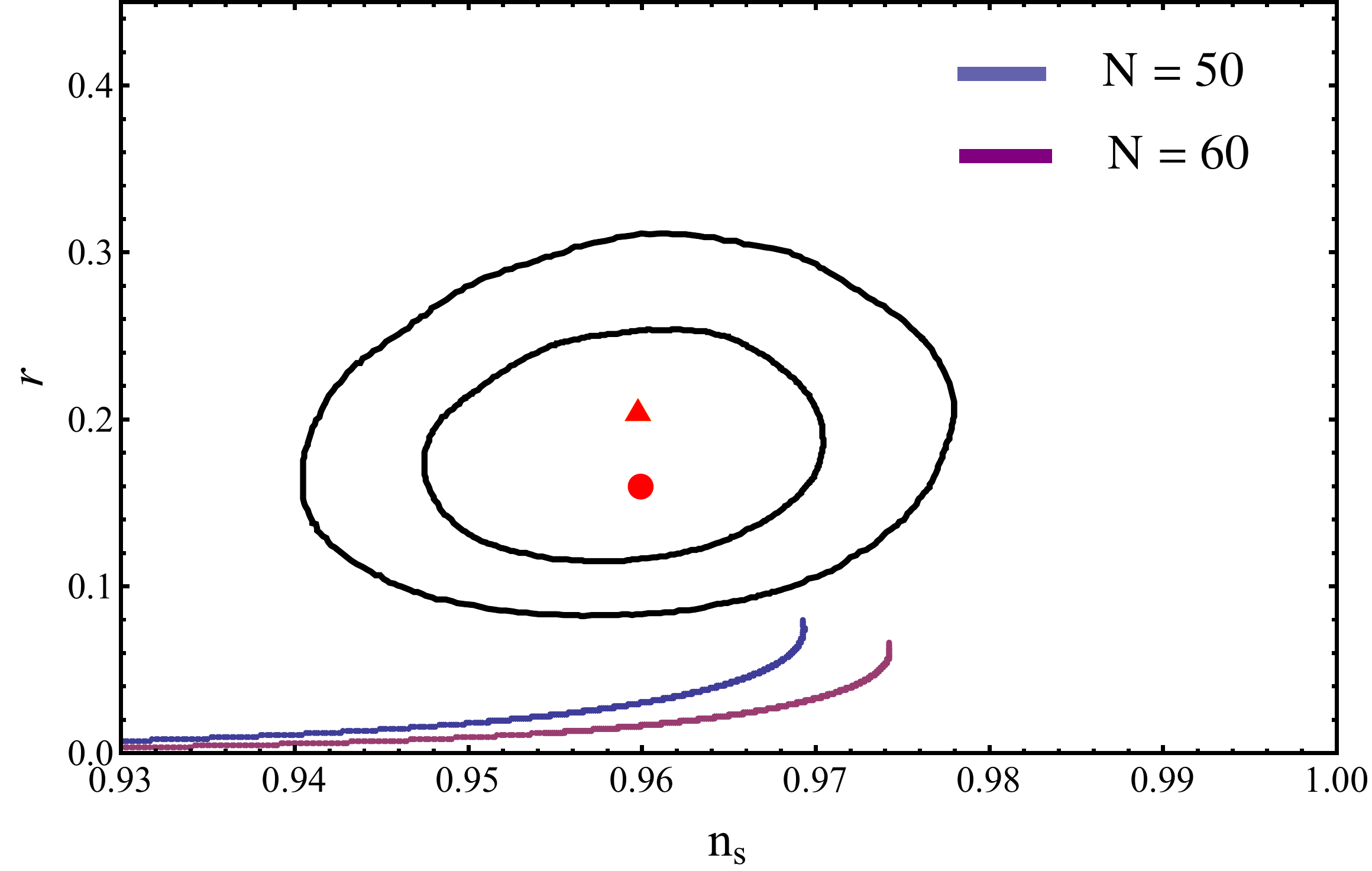}
\caption{
$r$ versus $n_s$ for the inflaton potential with $(j,~k,~l)=(0,~2,~3)$
where the inflationary trajectory is at the right of the maximum.
}
\label{0(-2)3_2}
\end{figure}

Third, for $a_2<0$ and $a_3>0$, there exist a maximum at $\phi_M=0$ and a minimum
at $\phi_m=-\frac{2a_2}{3a_3}$.
Similarly, we have three inflationary trajectories, and will discuss them one by one as well.
When the slow-roll inflations occur at the left and right of the maximum, we present
the numerical results for $r$ versus $n_s$ in Figs.~\ref{0(-2)3_1} and \ref{0(-2)3_2},
respectively. For $n_s$ within its $1\sigma$ range $0.9603\pm0.0073$,
the corresponding ranges of $r$ are $[0.0099,~0.0485]$ and $[0.0097,~0.0515]$,
which can be tested at the future Planck and BICEP2 experiments.

\begin{figure}[h]
\centering
\includegraphics[height=5cm]{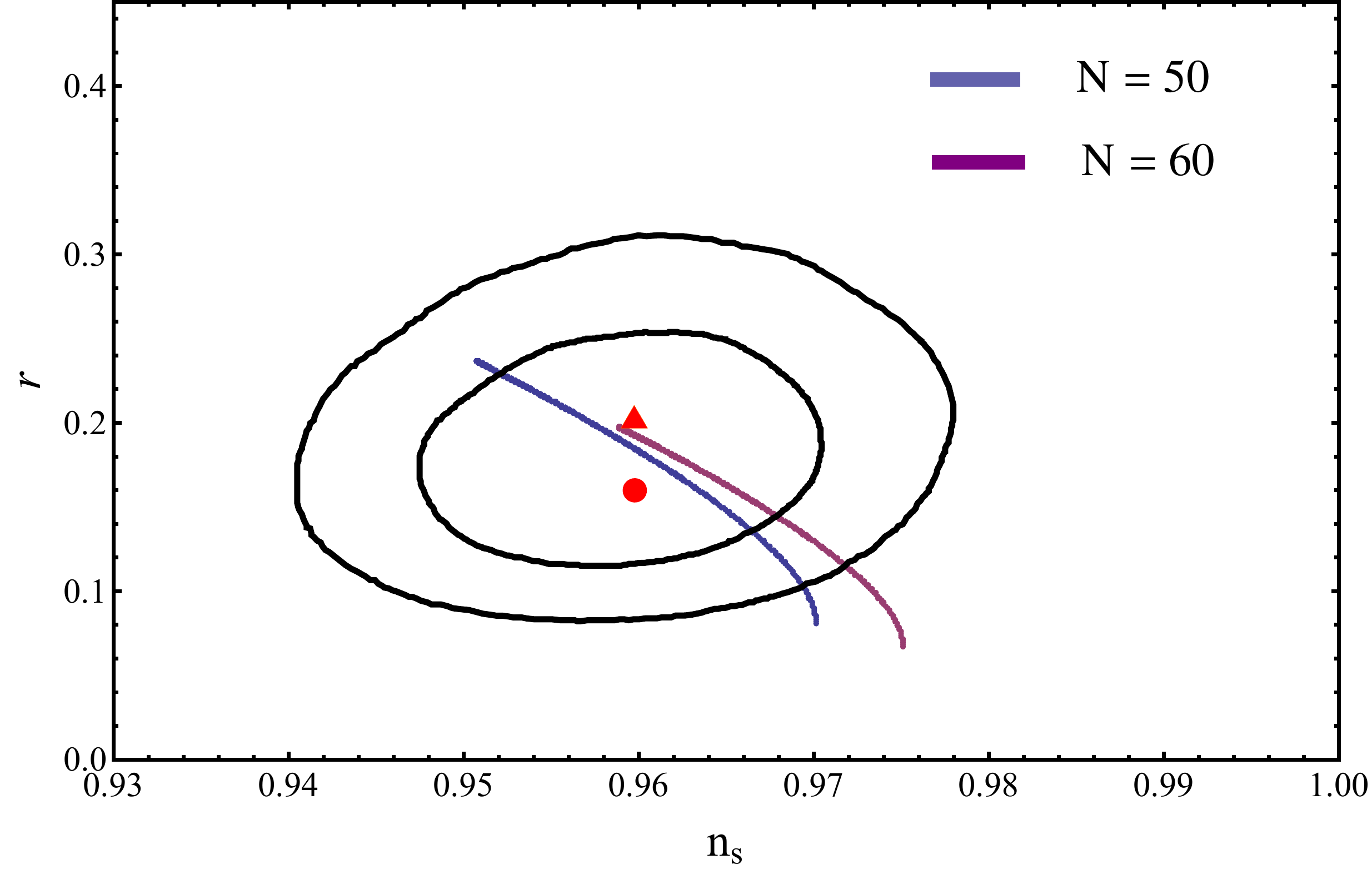}
\caption{
$r$ versus $n_s$ for the inflaton potential with $(j,~k,~l)=(0,~2,~3)$
where the inflationary trajectory is at the right of the minimum.
}
\label{0(-2)3_3}
\end{figure}

Furthermore, for the slow-roll inflation at the right of the minimum, the numerical results
for $r$ versus $n_s$ are given in Fig.~\ref{0(-2)3_3}. For $n_s$ in the $1\sigma$ range
$0.9603\pm0.0073$, the range of $r$ is $[0.1232,~0.2253]$, which can be consistent with
the BICEP2 experiment. In general, we can take $a_0\thickapprox 1$.
For the number of e-folding $N_e=50$,  $n_s$ and $r$ are within $1\sigma$ and $2\sigma$ regions
of the BICEP2 experiment for $-a_2<30a_3$ and $-a_2<100a_3$ respectively.
Also, for $N_e=60$, $n_s$ and $r$ are within $1\sigma$ and $2\sigma$ regions
of the BICEP2 experiment respectively for $-a_2<15a_3$ and $-a_2<35a_3$.
The best fit point with $n_s=0.959$ and $r=0.196$ can be realized for
$N_e=60$, $a_0=1$ and $-a_2\thickapprox -2 a_3$, for instance,
 $a_0=1, ~a_2=-2$, and $a_3=1$, and the corresponding
$\phi_i, ~\phi_f$, and $\phi_{m}$ are respectively $19.8863, ~3.10761$, and $1.33333$.

\subsection{Inflaton Potential with $(j,~k,~l)=(0, ~2, ~4)$}

\begin{figure}[h]
\centering
\includegraphics[height=5cm]{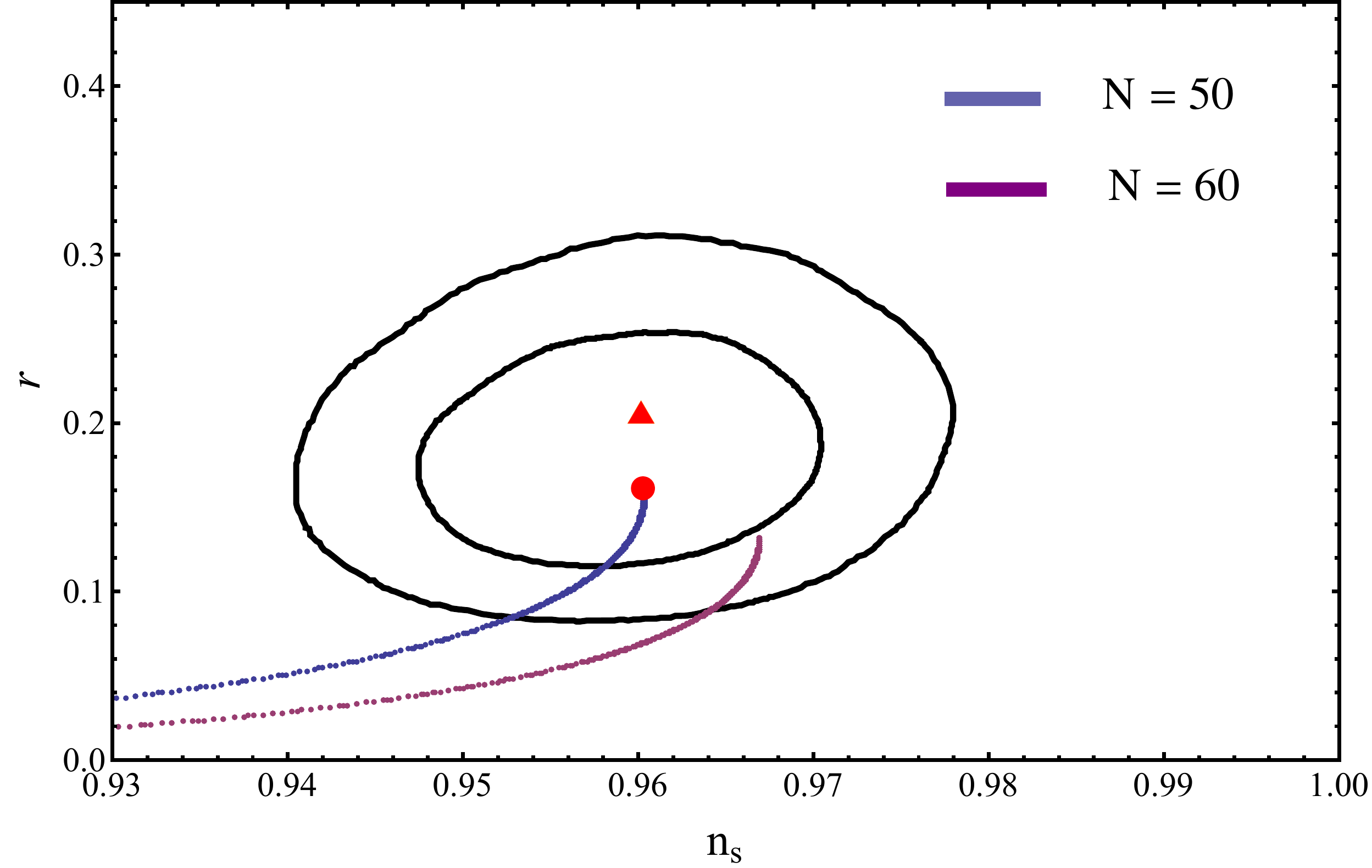}
\caption {
$r$ versus $n_s$ for the non-supersymmetric
inflaton potential with $(j,~k,~l)=(0, ~2, ~4)$, $a_0>0$, $a_2>0$, and $a_4<0$,
where the inflationary trajectory is at the left of the maximum. }
\label{024_1}
\end{figure}

First, we consider the non-supersymmetric inflation models with potential
 $V=a_0+a_2\phi^2+a_4\phi^4$. For simplicity, we only study
 the hill-top scenario with $a_0>0$, $a_2>0$, and $a_4<0$.
Thus, there is a maximum at $\phi=\phi_M=\sqrt{-\frac{a_2}{2a_4}}$.
For the slow-roll inflation occurs at the left of the maximum with
$ 0<\phi_f<\phi_i<\phi_M$, to achieve a  proper $r$,
we require $|a_4|\ll a_2$ to get a relatively large $\phi_M$, and thus,
 the $\phi^2$ term dominates the potential.
We present the numerical results for $r$ versus $n_s$  in Fig.~\ref{024_1}.
For $n_s$ in the $1\sigma$ range $0.9603\pm0.0073$,
the range of $r$ is $[0.0480, ~ 0.1565]$, which can be consistent with the BICEP2 experiment.
In the viable parameter space, we always have $a_2>10a_0$.
Moreover, for the number of e-folding
$N_e=50$, $n_s$ and $r$ are within $1\sigma$ and $2\sigma$ regions
of the BICEP2 experiment for $a_2>-1000a_4$ and $a_2>-700a_4$, respectively.
Also, for $N_e=60$, $n_s$ and $r$ are within $2\sigma$ region for
$a_2>-1200a_4$, but no viable parameter space for $1\sigma$ region.
The best fit point with $n_s=0.96$ and $r=0.158$
for the BICEP2 data can be obtained for $N_e=50$,  $a_2>10^5 a_4$, and $a_2>10a_0$.
For example, $a_0=1$, $a_2=10$, and $a_4=-10^{-4}$, and the corresponding
$\phi_i$, $\phi_f$, and $\phi_M$ are respectively $14.1817, 1.33953$, and $223.607$.

\begin{figure}[h]
\centering
\includegraphics[height=5cm]{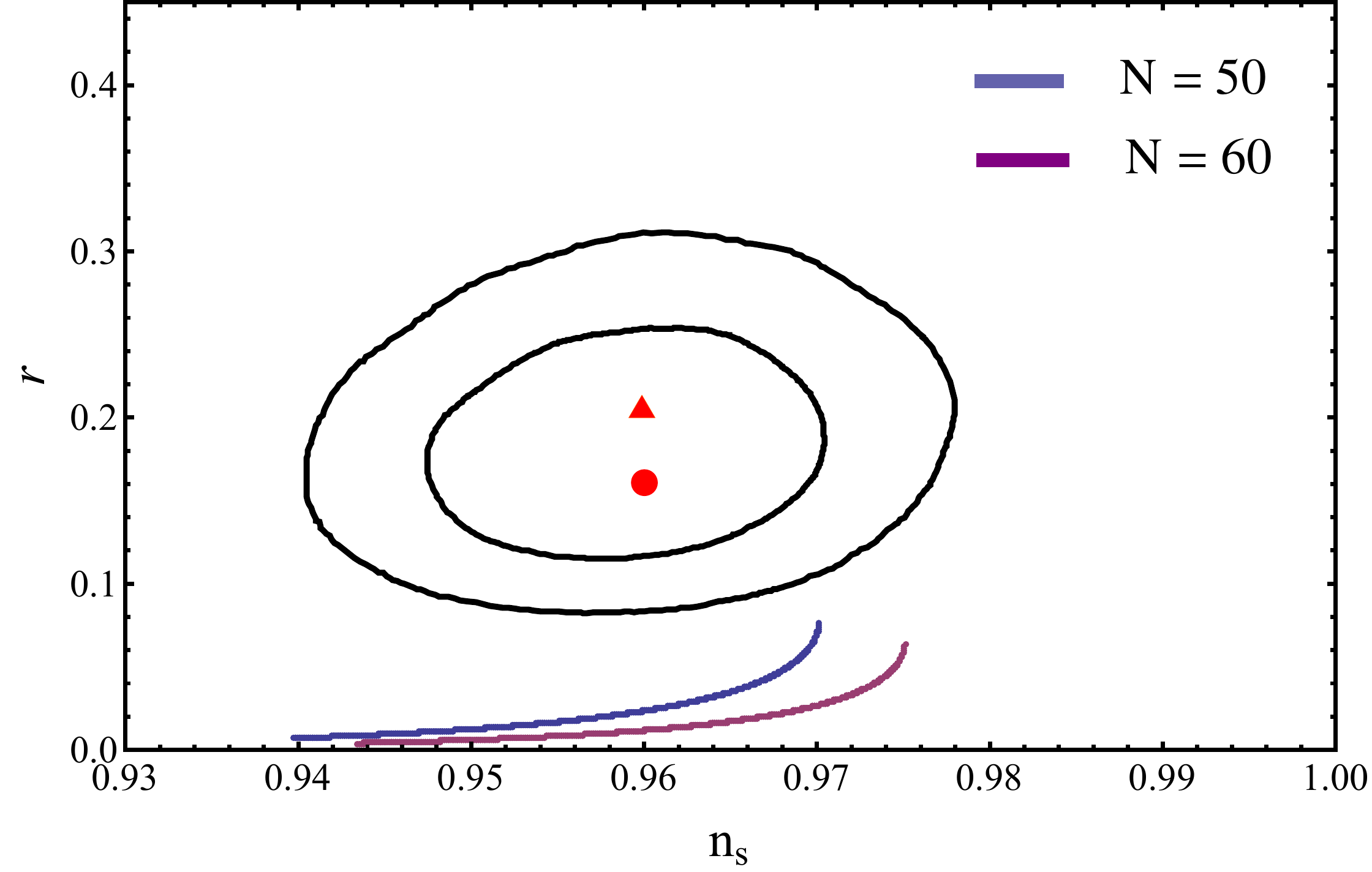}
\caption {
$r$ versus $n_s$ for the non-supersymmetric
inflaton potential with $(j,~k,~l)=(0, ~2, ~4)$, $a_0>0$, $a_2>0$, and $a_4<0$,
where the inflationary trajectory is at the right of the maximum.
}
\label{024_2}
\end{figure}

In addition, when slow-roll inflation occurs at the right of the maximum, {\it i.e.},
$ \phi_M<\phi_i<\phi_f$,
 the numerical results for $r$ versus $n_s$ are given in Fig.~\ref{024_2}.
 For $n_s$ within its $1\sigma$ range $0.9603\pm0.0073$,
the range of $r$ is about $[0.0072,~0.0444]$, which is within the reach of
the future Planck and QUBIT experiments.

\begin{figure}[h]
\centering
\includegraphics[height=5cm]{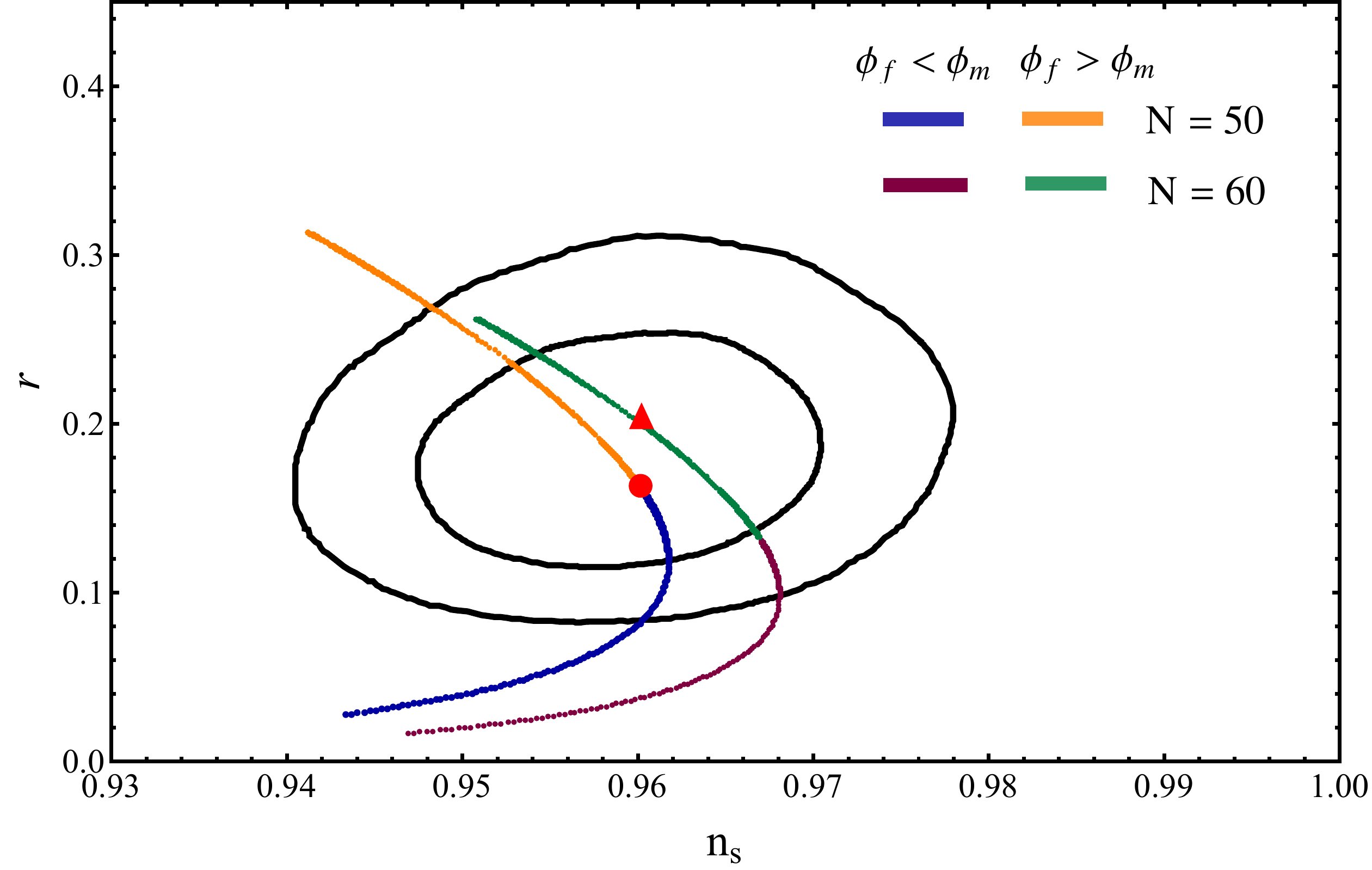}
\caption {
$r$ versus $n_s$ for the supersymmetric
inflaton potential with $(j,~k,~l)=(0, ~2, ~4)$.
}
\label{024(2)_1}
\end{figure}

Second, we consider the supersymmetric inflationary model with potential
 $V=|a+b\phi^2|^2=a^2+2ab\phi^2+b^2\phi^4$. For simplicity, we assume $a>0$ and $b<0$.
So the potential has two minima at $\phi=\phi_m=\pm \sqrt{-\frac{a}{b}}$. Without loss
of generality, we only consider the positive branch of the filed $\phi=\phi_m= \sqrt{-\frac{a}{b}}$.
The inflationary process can occur at either the left or right of the minimum.
When the slow-roll inflation occurs at the left of
the minimum, {\it i.e.},  $\phi_i<\phi_f<\phi_m$,
we present the numerical results for $r$ versus $n_s$ in Fig.~\ref{024(2)_1}.
For $n_s$ in its $1\sigma$ range $0.9603\pm0.0073$,
the range of $r$ is $[0.0254,~0.1585]$.
In addition, for the number of e-folding
$N_e=50$, $n_s$ and $r$ are within $1\sigma$ and $2\sigma$ regions
of the BICEP2 experiment for $a>-1650b$ and $a>-550b$, respectively.
Also, for $N_e=60$, $n_s$ and $r$ are within $2\sigma$ region for
$a>-1650b$, but no viable parameter space
for $1\sigma$ region.
Also, the best fit point with $n_s=0.96$ and $r=0.158$
for the BICEP2 data can be obtained for $N_e=50$ and $a>-3 \times 10^7 b$.
 For example,  $a=1$ and $b=-3 \times 10^{-7}$, and the corresponding
$\phi_i, ~\phi_e$, and $\phi_{m}$ are respectively $3148.08, ~3160.86$, and $3162.28$.

Furthermore, when the slow-roll inflation occurs at the right of
the minimum, {\it i.e.}, $\phi_m<\phi_f<\phi_i$,
the numerical results for $r$ versus $n_s$ are given in Fig.~\ref{024(2)_1}.
Interestingly, we will always get a larger $r$ than the above case
for any value of $a$ or $b$.
With  $n_s$ in its $1\sigma$ range $0.9603\pm0.0073$,
the range of $r$ is about $[0.1319,~0.2484]$.
In addition, for the number of e-folding
$N_e=50$, $n_s$ and $r$ are within $1\sigma$ and $2\sigma$ regions
of the BICEP2 experiment for $a>-165b$ and $a>-33b$, respectively.
Also, for $N_e=60$, $n_s$ and $r$ are within $1\sigma$ region
for $a>-17b$, and $2\sigma$ region for
the viable parameter space. Let us present two best fit points for the BICEP2 data.
The best fit point with $n_s=0.96$ and $r=0.16$ can be realized for
$N_e=50$ and $a\thickapprox -1\times10^6 b$.
For example, $a=1$ and $b=-1\times10^{-6}$,
and the corresponding  $\phi_i, ~\phi_f$, and $\phi_{m}$ are
respectively $1014.25, ~1001.42$, and $1000.0$.
Another best fit point with $n_s=0.96$ and $r=0.2$  can be obtained
for $a=165$ and $b=-1$. For example, $a=165$ and $b=-1$,
 and the corresponding  $\phi_i, ~\phi_f$, and $\phi_{m}$ are
$30.5877$, $14.3371$, and $12.8452$,  respectively.


\subsection{Inflaton Potential with $(j,~k,~l)=(0, ~3, ~4)$}

\begin{figure}[h]
\centering
\includegraphics[height=5cm]{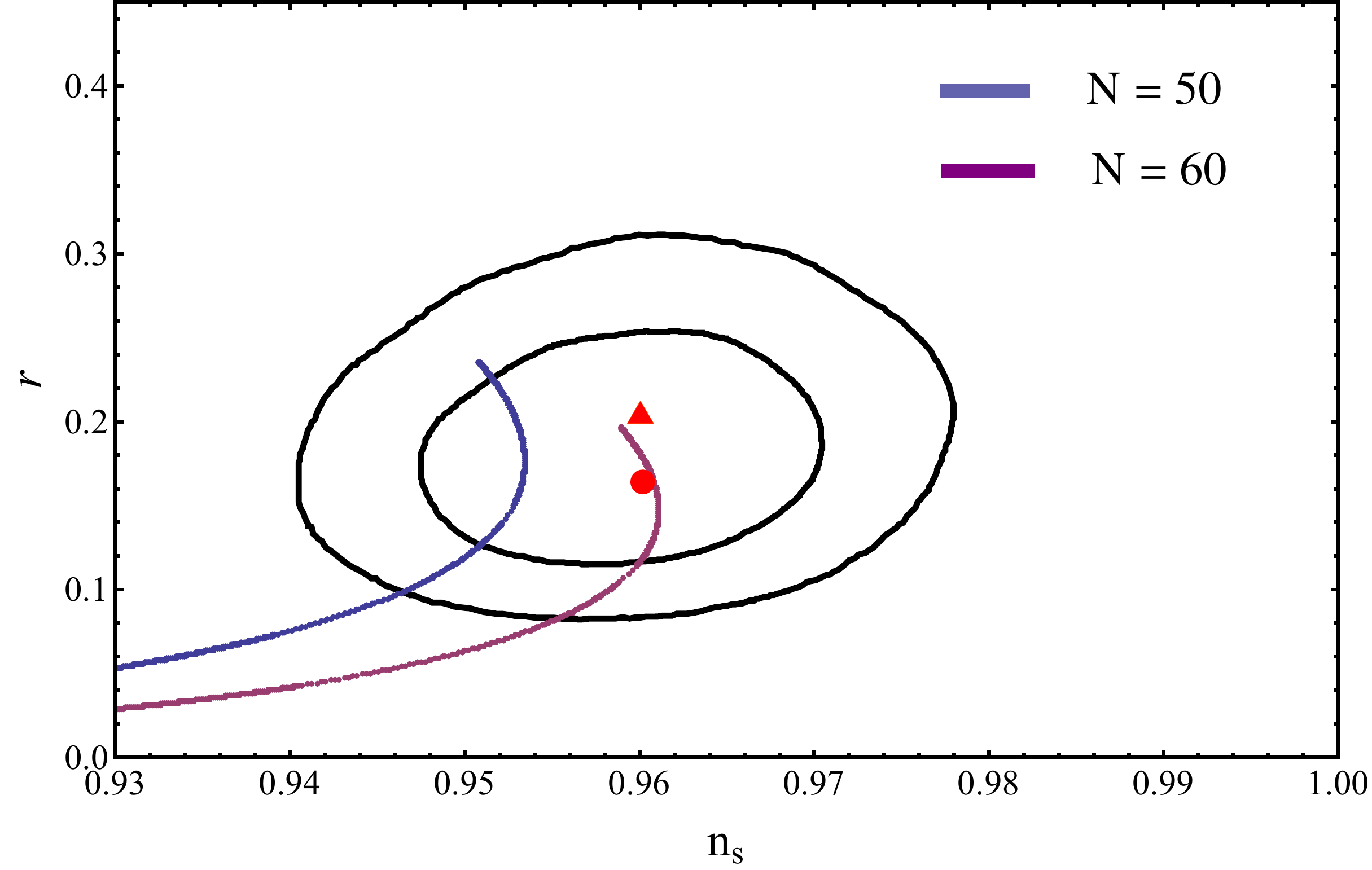}
\caption {
$r$ versus $n_s$ for the
inflaton potential with $(j,~k,~l)=(0, ~3, ~4)$ where the inflationary trajectory is
at the left of the maximum.
}
\label{034_1}
\end{figure}

 We consider the inflaton potential $V=a_0+a_3\phi^3+a_4\phi^4$.
First, we study
 the hill-top scenario with $a_0>0$, $a_3>0$, and $a_4<0$.
So there is a maximum at $\phi_M=-\frac{3 a_3}{4 a_4}$.
When the slow-roll inflation occurs at the left of
the maximum, {\it i.e.}, $\phi_f<\phi_i<\phi_M$,
we present the numerical results for $r$ versus $n_s$ in Fig.~\ref{034_1}.
The range of $r$ is about  $[0.0742,~0.1956]$ for $n_s$ within its
$1\sigma$ range $0.9603\pm0.0073$, which can be consistent with the BICEP2 results.
In the viable parameter space, we generically have $a_0 < a_3$.
For the number of e-folding $N_e=50$, $n_s$ and $r$ are within $1\sigma$ and $2\sigma$ regions
of the BICEP2 experiment for $a_3>-33a_4$ and $a_3>-26a_4$, respectively.
And for $N_e=60$, $n_s$ and $r$ are within $1\sigma$ and $2\sigma$ regions
of the BICEP2 experiment for $a_3>-40a_4$ and $a_3>-29a_4$, respectively.
Let us present two best fit points for the BICEP2 data.
The best fit point with $n_s=0.96$ and $r=0.16$ can be realized for
$N_e=59$, $a_3> 10a_0$, and $a_3\thickapprox-58.4a_4$, for example,
$a_0=10, ~ a_3=100$, and $a_4=-1.71$,
and the corresponding  $\phi_i, ~\phi_f$, and $\phi_{M}$ are
respectively $18.0429, ~2.07119$, and $43.8596$.
Another best fit point with $n_s=0.959$ and $r=0.196$ can be obtained
for $N_e=60$, $a_3\approx 5a_0$, and $a_3\thickapprox-1000a_4$,  for instance,
$a_0=20, a_3=100$, and $a_4=-0.1$,  and the corresponding  $\phi_i, ~\phi_f$, and
$\phi_{M}$ are $19.0411, ~2.07322$, and $750.0$, respectively.

\begin{figure}[h]
\centering
\includegraphics[height=5cm]{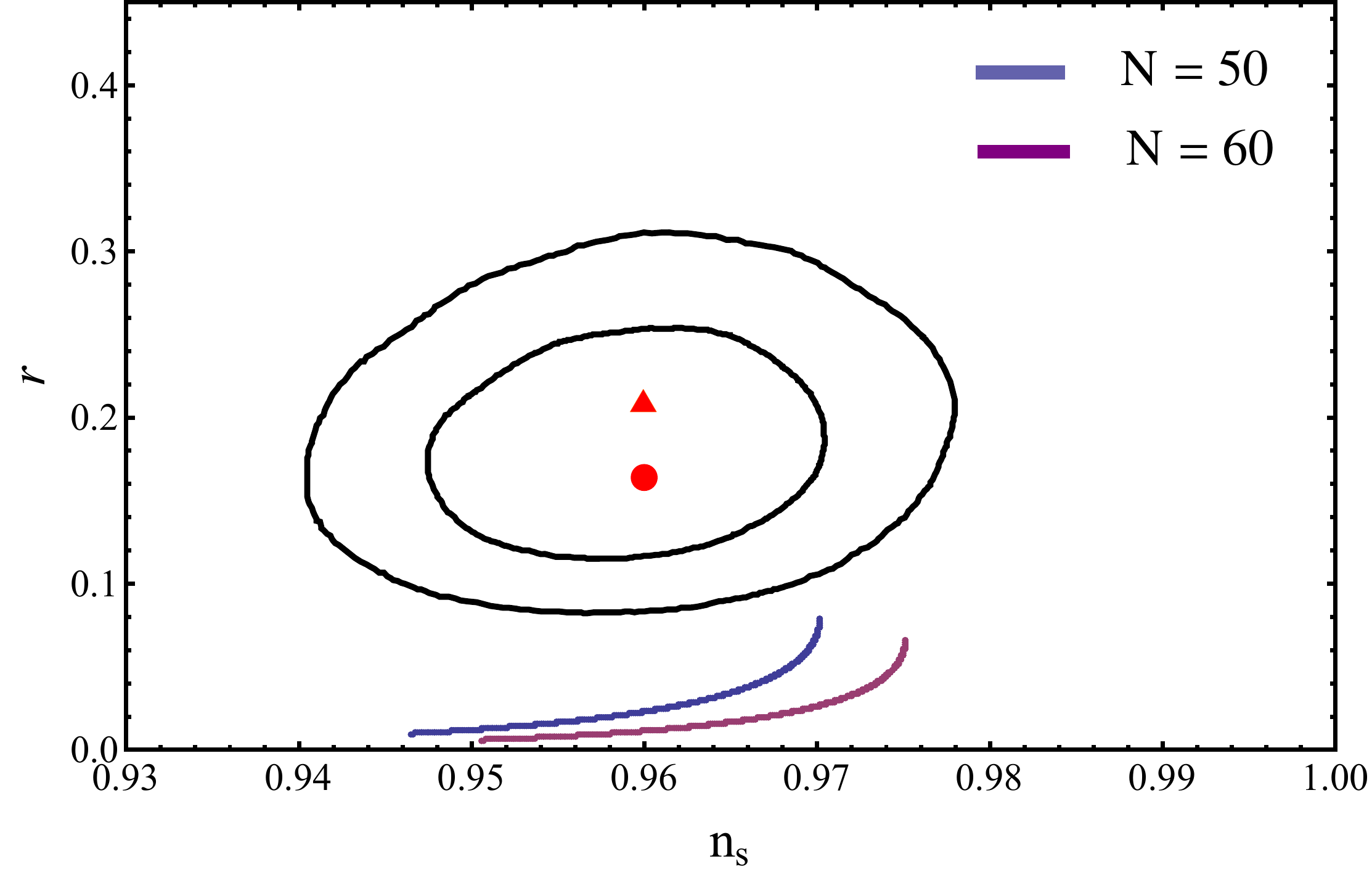}
\caption {
$r$ versus $n_s$ for the
inflaton potential with $(j,~k,~l)=(0, ~3, ~4)$ where the inflationary trajectory is
at the right of the maximum.
}
\label{034_2}
\end{figure}

Moreover, we consider the slow-roll inflation occurs at the right of
the maximum, {\it i.e.}, $\phi_M<\phi_i<\phi_f$.
The numerical results for $r$ versus $n_s$ is given in Fig.~\ref{034_2}.
For $n_s$ within the $1\sigma$ range $0.9603\pm0.0073$,
the range of $r$ is $[0.0067, ~0.0454]$, which is large enough
to be tested at the future Planck and QUBIT experiments.

\begin{figure}[h]
\centering
\includegraphics[height=5cm]{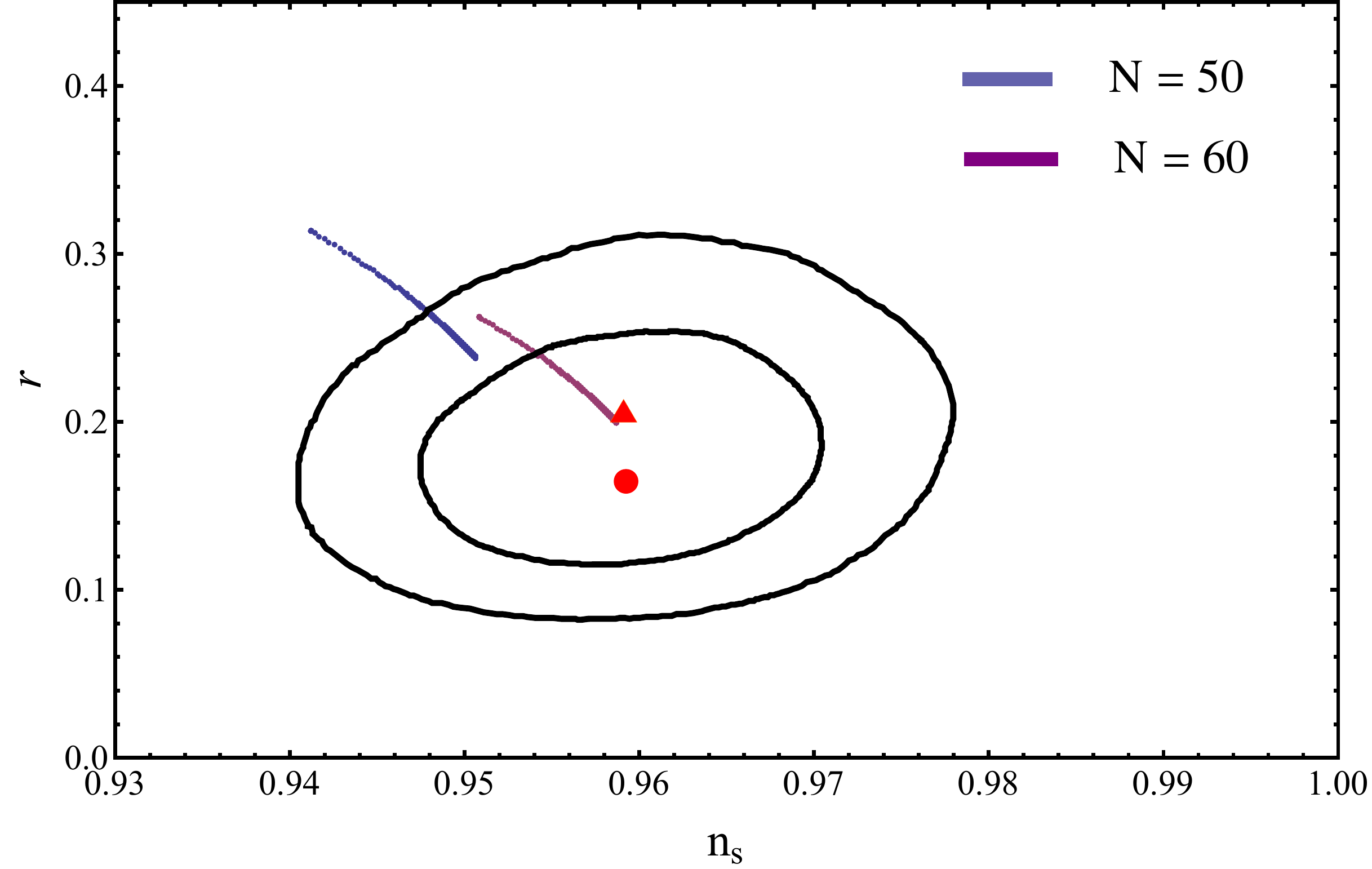}
\caption {
$r$ versus $n_s$ for the
inflaton potential with $(j,~k,~l)=(0, ~3, ~4)$ and $a_3<0$, where the inflationary trajectory is
at the left of the minimum.
}
\label{034_3}
\end{figure}

Second, we consider the other case with  $a_0>0$, $a_3<0$, and $a_4>0$,
which has a minimum at $\phi_m=-\frac{3 a_3}{4 a_4}$.
When the slow-roll inflation occurs at the left of
the minimum, {\it i.e.}, $\phi_i<\phi_f<\phi_m$,
we present the numerical results for $r$ versus $n_s$ in Fig.~\ref{034_3}.
The range of $r$ is about  $[0.1995,~0.2473]$ for $n_s$ within its
$1\sigma$ range $0.9603\pm0.0073$, which can be consistent with the BICEP2 results.
In the viable parameter space, we generically have $a_0 \thickapprox 1$.
For the number of e-folding $N_e=50$, $n_s$ and $r$ are within $2\sigma$ region
of the BICEP2 experiment for $-a_3>50a_4$ but no viable parameter space for $1\sigma$ region.
And for $N_e=60$, $n_s$ and $r$ are within $1\sigma$ region
of the BICEP2 experiment for $-a_3>15a_4$, and will always lie in $2\sigma$ region
for any values of $a_3$ and $a_4$.
The best fit point with $n_s=0.958$ and $r=0.199$ can be realized for
$N_e=60$, $a_0\thickapprox 1$, and $a_3\thickapprox-1000a_4$, for example,
$a_0=1, ~ a_3=-1000$, and $a_4=1$,
and the corresponding  $\phi_i, ~\phi_f$, and $\phi_{m}$ are
respectively $-19.1319, ~-2.1226$, and $750$.

\begin{figure}[h]
\centering
\includegraphics[height=5cm]{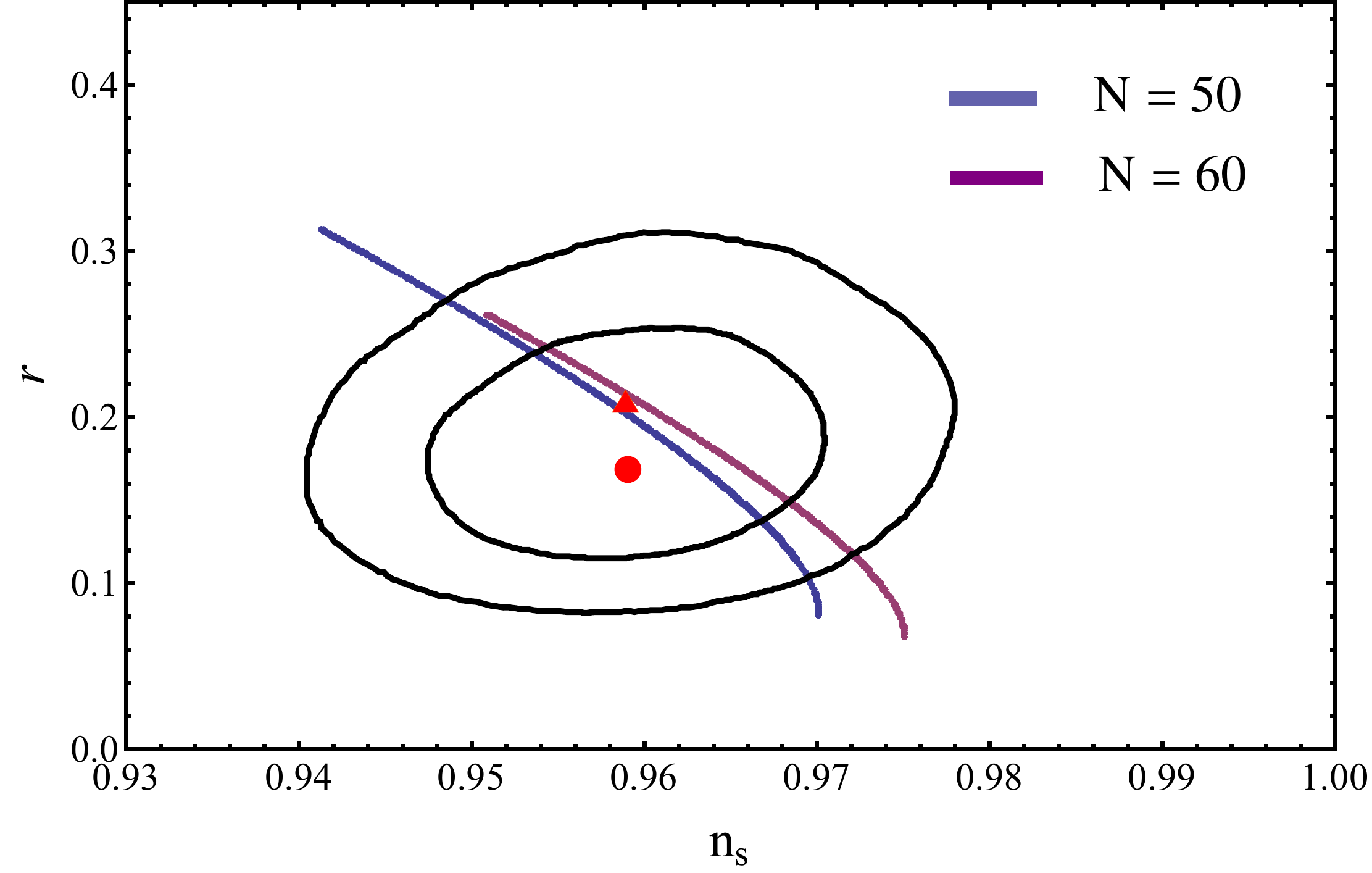}
\caption {
$r$ versus $n_s$ for the
inflaton potential with $(j,~k,~l)=(0, ~3, ~4)$ and $a_3<0$, where the inflationary trajectory is
at the right of the minimum.
}
\label{034_4}
\end{figure}

In addition, let us consider the slow-roll inflation, which occurs at the right of
the minimum, {\it i.e.}, $\phi_m<\phi_f<\phi_i$. We present the numerical results
for $r$ versus $n_s$ in Fig.~\ref{034_4}.
 The range of $r$ is about  $[0.1311,~0.2512]$ for $n_s$ within its
$1\sigma$ range $0.9603\pm0.0073$, which can be consistent with the BICEP2 results.
In the viable parameter space, we have $a_0 \thickapprox 1$ in general.
For the number of e-folding $N_e=50$, $n_s$ and $r$ are within $1\sigma$ and $2\sigma$ regions
of the BICEP2 experiment for $10a_4<-a_3<150a_4$ and $15a_4<-a_3<60a_4$, respectively.
And for $N_e=60$, $n_s$ and $r$ are within $1\sigma$ and $2\sigma$ regions
of the BICEP2 experiment for $-a_3<55a_4$ and $8a_4<-a_3<32a_4$, respectively.
The best fit point with $n_s=0.96$ and $r=0.2$ can be obtained
for $N_e=54$, $a_0=1$, and $-a_3\thickapprox-19a_4$,  for instance,
$a_0=1, a_3=-19$, and $a_4=1$,  and the corresponding  $\phi_i, ~\phi_f$, and
$\phi_{m}$ are $33.5051, ~19.7918$, and $14.25$, respectively.

\subsection{Inflaton Potential with $(j,~k,~l)=(1, ~2, ~3)$}

\begin{figure}[h]
\centering
\includegraphics[height=5cm]{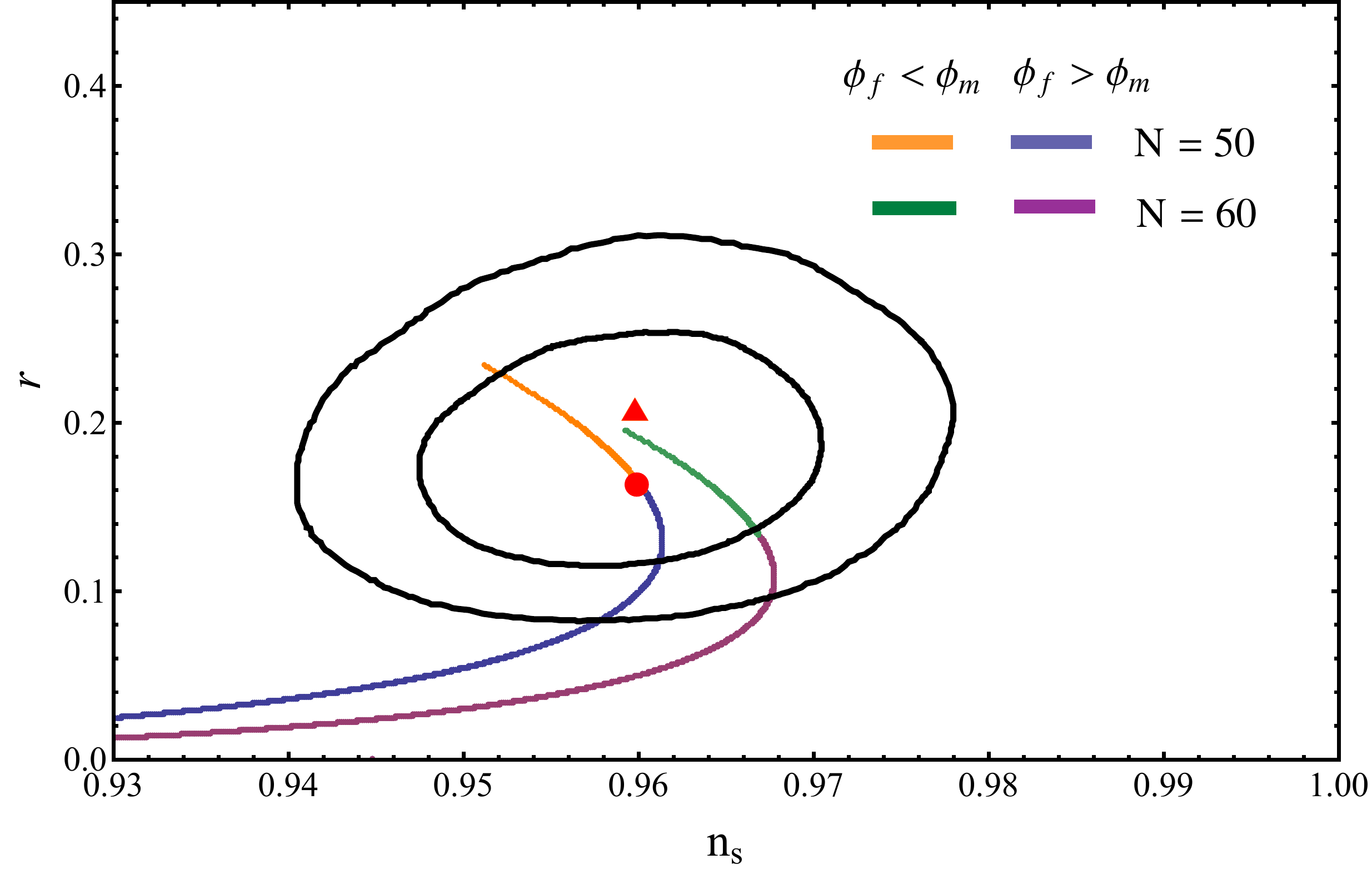}
\caption {
$r$ versus $n_s$ for the
inflaton potential with $(j,~k,~l)=(1, ~2, ~3)$ where the inflationary trajectories are
at the left and right of the minimum.
}\label{123_1}
\end{figure}

We consider the inflaton potential $V=a_1\phi+a_2\phi^2+a_3\phi^3$.
For simplicity, we only study
 the hill-top scenario with $a_1>0$, $a_2>0$, and $a_3<0$.
So, there exist a minimum at $\phi_m=-\frac{a_2}{3 a_3}-\frac{1}{3} \sqrt{\frac{a_2^2-3 a_1 a_3}{a_3^2}}$ and
a maximum at $\phi_M=-\frac{a_2}{3 a_3}+\frac{1}{3} \sqrt{\frac{a_2^2-3 a_1 a_3}{a_3^2}}$.
We find that only the inflationary processes near the minimum will give us a proper $r$.
First,  for the slow-roll inflation at the left of the minimum,
we present the numerical results for $r$ versus $n_s$ in Fig.~\ref{123_1}.
With $n_s$ in its $1\sigma$ range $0.9603\pm0.0073$,
the range of $r$ is about  $[0.1234, 0.2207]$, which can be consistent with the BICEP2 results.
Moreover, for the number of e-folding
$N_e=50$, $n_s$ and $r$ are within $1\sigma$ and $2\sigma$ regions
of the BICEP2 experiment for $-20a_3<a_1<-1000a_3$ and $a_1<1000 ~{\rm Max}(a_2,-a_3)$, respectively.
Also, for $N_e=60$, $n_s$ and $r$ are within $1\sigma$ and $2\sigma$ regions
for $a_1<-100a_3$ and $a_1<-1000a_3$, respectively.
Let us present two best fit points for the BICEP2 data.
The best fit point with $n_s=0.963$ and $r=0.16$ can be realized for
$N_e=50$, $a_2>10a_1$, and $a_2\thickapprox-10^3 a_3$, for example,
$a_1=0.1, ~a_2=1$, and $a_3=-10^{-3}$,
and the corresponding  $\phi_i, ~\phi_f, ~\phi_{m}$, and $\phi_{M}$ are
respectively $-14.2966, ~-1.46697, ~-0.0499963$, and $666.717$.
Another best fit point with $n_s=0.958$ and $r=0.2$ can be obtained
for $N_e=58$, $a_2>10a_1$, and $a_2\thickapprox-3.3 a_3$,  for example,
 $a_1=0.1, a_2=1$ and $a_3=-0.3$, and the corresponding  $\phi_i, ~\phi_f, ~\phi_{m}$, and
$\phi_{M}$ are $-17.9616, ~ -1.64821, ~-0.0000499989$, and 2.22227, respectively.

Second,  we consider the slow-roll inflation at the right of the minimum.
The numerical results for $r$ versus $n_s$ are given in Fig.~\ref{123_1} as well.
The range of $r$ is about  $[0.0337,~0.158]$ for $n_s$ in its
$1\sigma$ range $0.9603\pm0.0073$, which can be
 consistent with the BICEP2 results.
In addition, for the number of e-folding
$N_e=50$, $n_s$ and $r$ are within $1\sigma$ and $2\sigma$ regions
of the BICEP2 experiment for $a_2>-50a_3$ and $a_1< a_2 (1+\ln{ -\frac{a_2}{50a_3} })$
 and for  $a_2>-32a_3$ and $a_1<8 [a_2 (1+\ln{-\frac{a_2}{32a_3} })]$, respectively.
Also, for $N_e=60$, $n_s$ and $r$ are within $2\sigma$ region for
$a_2>-50a_3$ and $a_1<2 [a_2 (1+\ln{-\frac{a_2}{50a_3} })]$, but no viable parameter space
for $1\sigma$ region.
Especially, the best fit point with $n_s=0.96$ and $r=0.158$
for the BICEP2 data can be obtained for $N_e=50$, $a_2>10a_1$, and $a_2>-10^4 a_3$.
For example, $a_1=0.1, ~a_2=1$, and $a_3=-10^{-4}$, and the corresponding
$\phi_i, ~\phi_f, ~\phi_{m}$, and $\phi_{M}$ respectively are
$14.1599, ~1.36588, ~-0.0499996$, and  $6666.72$.

\begin{figure}[h]
\centering
\includegraphics[height=5cm]{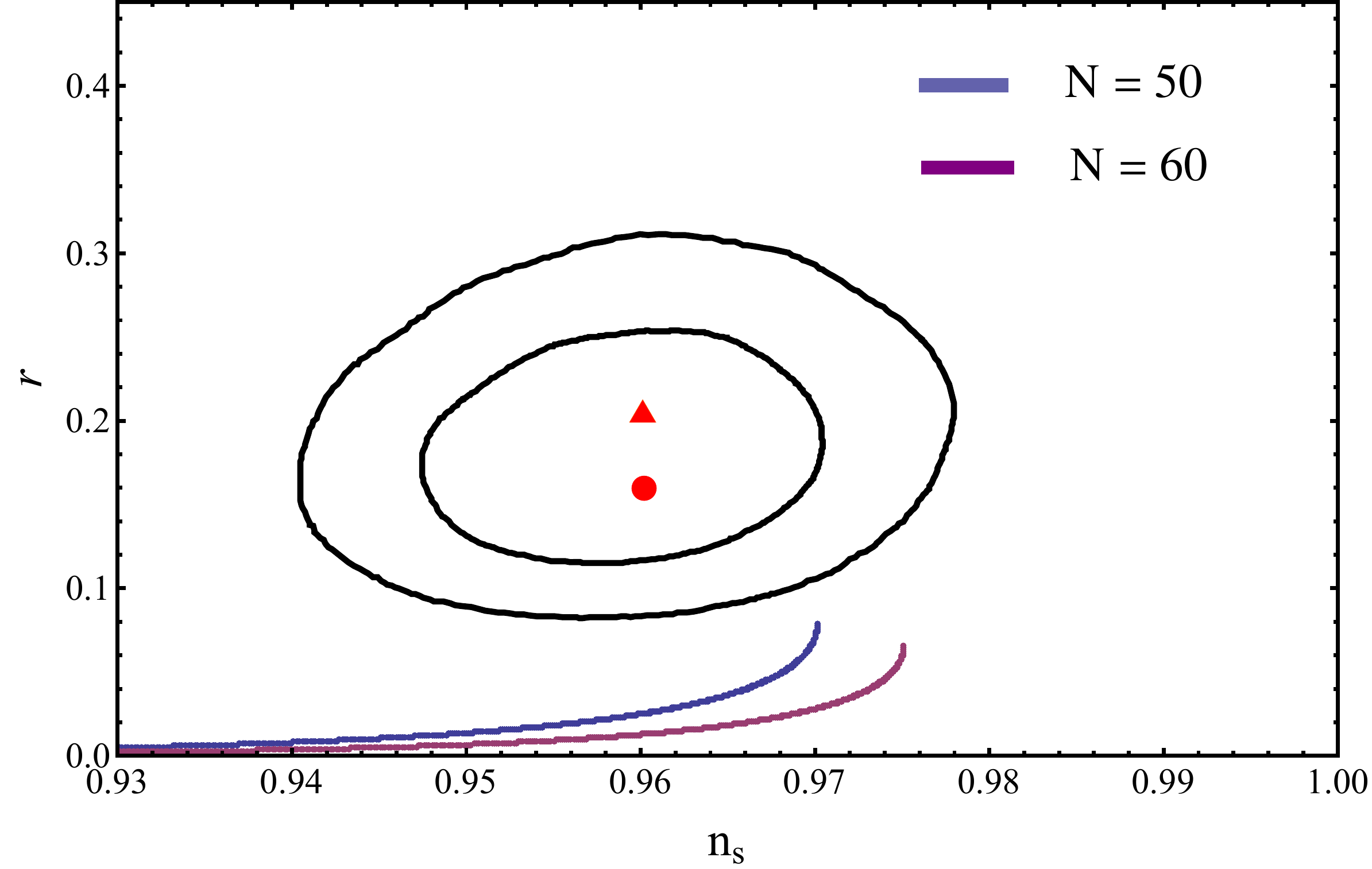}
\caption{
$r$ versus $n_s$ for the
inflaton potential with $(j,~k,~l)=(1, ~2, ~3)$ where the inflationary trajectory is
at the right of the maximum.
}
\label{123_2}
\end{figure}

Third, for the slow-roll inflation at the right of the maximum,
 we present the numerical results for $r$ versus $n_s$ in Fig.~\ref{123_2}.
So we cannot find the proper parameter space which can give a large enough $r$
in the $2\sigma$ region of the BICEP2 data.
For $n_s$ within the $1\sigma$ range $0.9603\pm0.0073$,
the range of $r$ is $[0.0083,~0.0471]$,
which can still be tested at the future Planck and QUBIT experiments.

\subsection{Inflaton Potential with $(j,~k,~l)=(1, ~2, ~4)$}

 For the inflaton potential $V=a_1 \phi+ a_2 \phi^2 +a_4 \phi^4$,
we consider the hill-top scenario with $a_4<0$.
Thus, either we have only one maximum at
\begin{equation}
\phi_m=\frac{\left(1-i \sqrt{3}\right) a_2}{2 \sqrt[3]{3} \sqrt[3]{\sqrt{3} \sqrt{27 a_1^2 a_4^4+8 a_2^3 a_4^3}-9 a_1 a_4^2}}-\frac{\left(1+i
   \sqrt{3}\right) \sqrt[3]{\sqrt{3}
\sqrt{27 a_1^2 a_4^4+8 a_2^3 a_4^3}-9 a_1 a_4^2}}{4\ 3^{2/3} a_4}~,
\label{Eq-MM}
\end{equation}
or we have one minimum given by the above Eq.~(\ref{Eq-MM}) and two maxima at
\begin{equation}
\phi_{M1}=\frac{1}{2} \left(\frac{\sqrt[3]{\sqrt{3} \sqrt{27 a_1^2 a_4^4+8 a_2^3 a_4^3}-9 a_1 a_4^2}}{3^{2/3} a_4}-\frac{2 a_2}{\sqrt[3]{3} \sqrt[3]{\sqrt{3}
   \sqrt{27 a_1^2 a_4^4+8 a_2^3 a_4^3}-9 a_1 a_4^2}}\right)~,
\end{equation}
 and
\begin{equation}
 \phi_{M2}=\frac{\left(1+i \sqrt{3}\right) a_2}{2 \sqrt[3]{3} \sqrt[3]{\sqrt{3} \sqrt{27 a_1^2 a_4^4+8 a_2^3 a_4^3}-9 a_1 a_4^2}}-\frac{\left(1-i
   \sqrt{3}\right) \sqrt[3]{\sqrt{3}
\sqrt{27 a_1^2 a_4^4+8 a_2^3 a_4^3}-9 a_1 a_4^2}}{4\ 3^{2/3} a_4}~,~\,
\end{equation}
with $\phi_{M1}<\phi_{M2}$. For the former case with $\phi_m$ as a maximum,
because the parameters can only
be considered in a very restricted way, we cannot get a proper $r$. Therefore,
we will consider the later case with $\phi_m$ a minimum.

\begin{figure}[h]
\centering
\includegraphics[height=5cm]{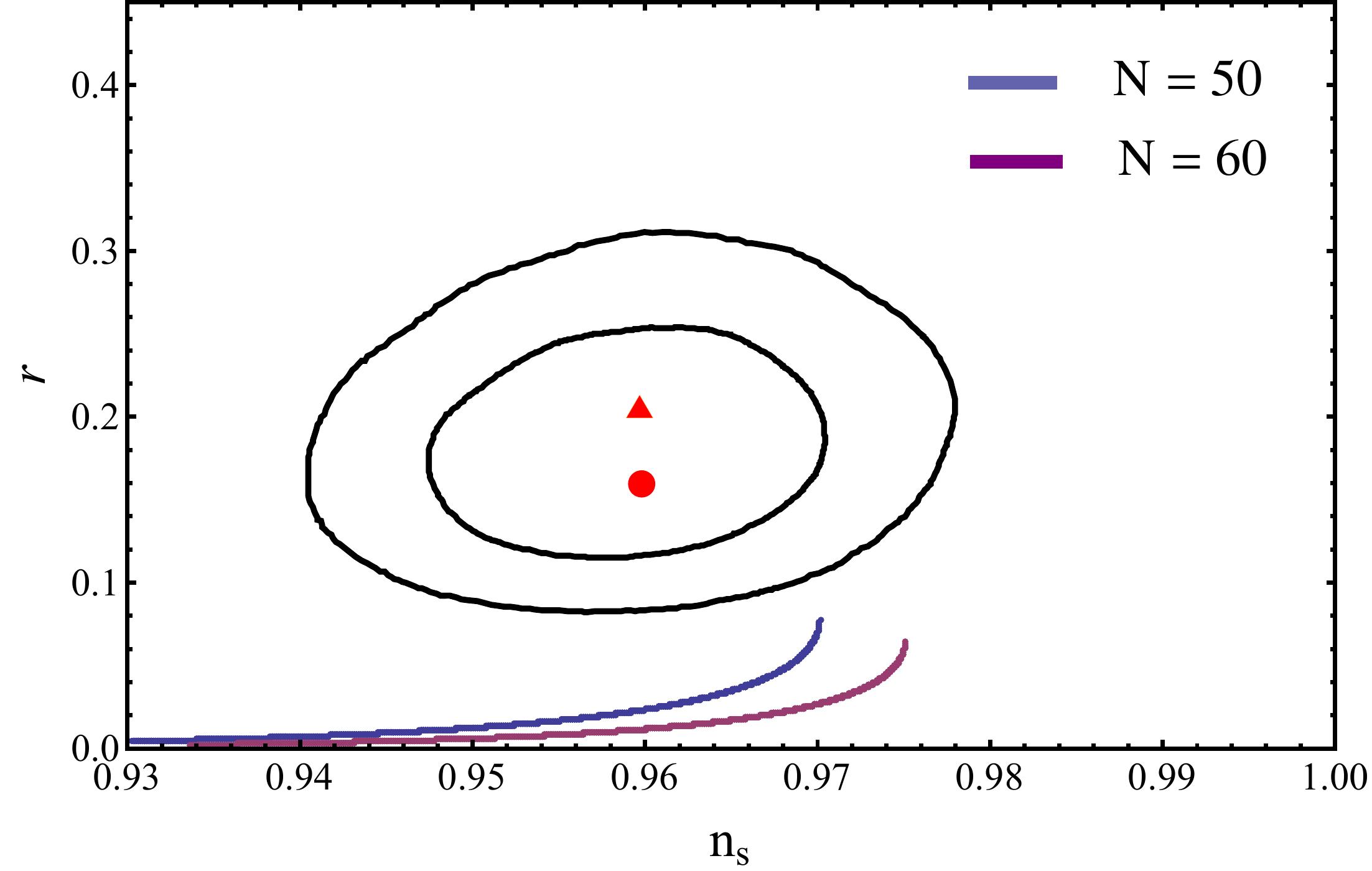}
\caption {
$r$ versus $n_s$ for the
inflaton potential with $(j,~k,~l)=(1, ~2, ~4)$ where
the inflationary trajectory is at the left of the maximum $\phi_{M1}$.
}
\label{124_1}
\end{figure}

First, we consider the inflation at the left of the maximum, {\it i.e.},
$\phi_f<\phi_i<\phi_{M1}$.
We present the numerical results for $r$ versus $n_s$ in Fig.~\ref{124_1}.
So we cannot find the viable parameter space which can generate a large enough $r$.
For $n_s$ in the $1\sigma$ range $0.9603\pm0.0073$,
the range of $r$ is $[0.0084,~0.0449]$. Interestingly,
such $r$ can still be within the reach of the future Planck and QUBIT experiments.

\begin{figure}[h]
\centering
\includegraphics[height=5cm]{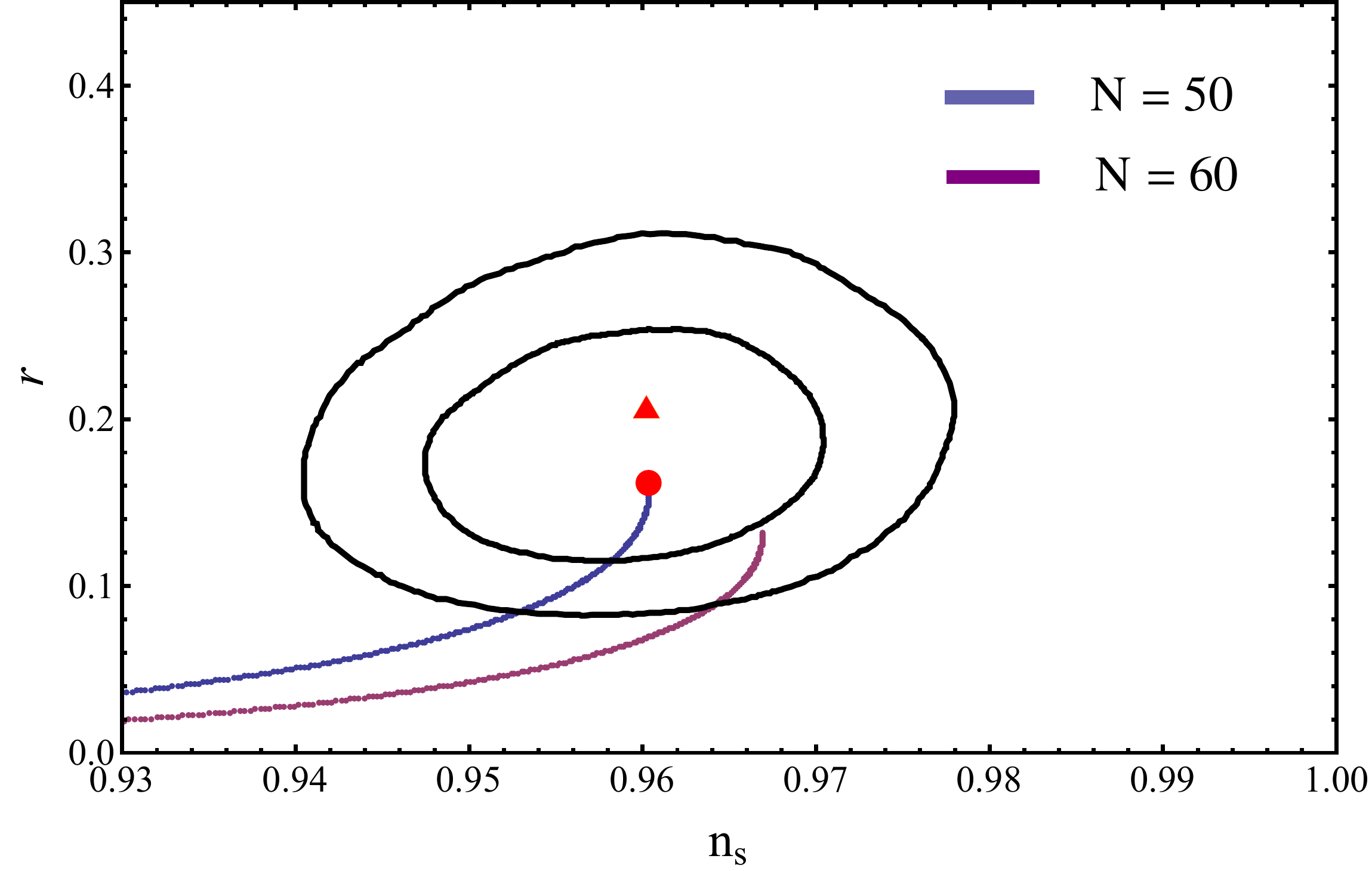}
\caption {
$r$ versus $n_s$ for the
inflaton potential with $(j,~k,~l)=(1, ~2, ~4)$ where
the inflationary trajectory is between $\phi_{M1}$ and $\phi_m$.
}
\label{124_2}
\end{figure}

Second, we consider the inflationary trajectory between $\phi_{M1}$ and $\phi_m$,
{\it i.e.}, $\phi_{M1}<\phi_i<\phi_f<\phi_m$.
We present the numerical results for $r$ versus $n_s$ in Fig.~\ref{124_2}.
For $n_s$ within the $1\sigma$ range $0.9603\pm0.0073$,
the range of $r$ is $[0.0487,~0.1585]$, which can be consistent with the BICEP2 experiment.
In addition, for the number of e-folding
$N_e=50$, $n_s$ and $r$ are within $1\sigma$ and $2\sigma$ regions
of the BICEP2 experiment for $a_2>-1250a_4$ and $a_1<[a_2 (1+20\ln{\frac{a_2}{-1250a_4}})]/10$
and for $a_2>-660a_4$ and $a_1<[a_2 (1+20\ln{\frac{a_2}{-660a_4}})]/10$, respectively.
Also, for $N_e=60$, $n_s$ and $r$ are within $2\sigma$ region for
$a_2>-1100a_4$ and $a_1<[a_2 (1+20\ln{ \frac{a_2}{-1100a_4} })]/10$,
but no viable parameter space for $1\sigma$ region.
The best fit point with $n_s=0.96$ and $r=0.158$
for the BICEP2 data can be obtained for $N_e=50$, $a_2>10a_1$, and $a_2>-10^6 a_4$.
For example, $a_1=0.1, ~a_2=1$, and $a_4=-10^{-6}$, and the corresponding
$\phi_i, ~\phi_f, ~\phi_{M1}$, and $\phi_{m}$ are respectively
$-14.2625, ~-1.46598, ~-707.082$, and  $-0.05$.

\begin{figure}[h]
\centering
\includegraphics[height=5cm]{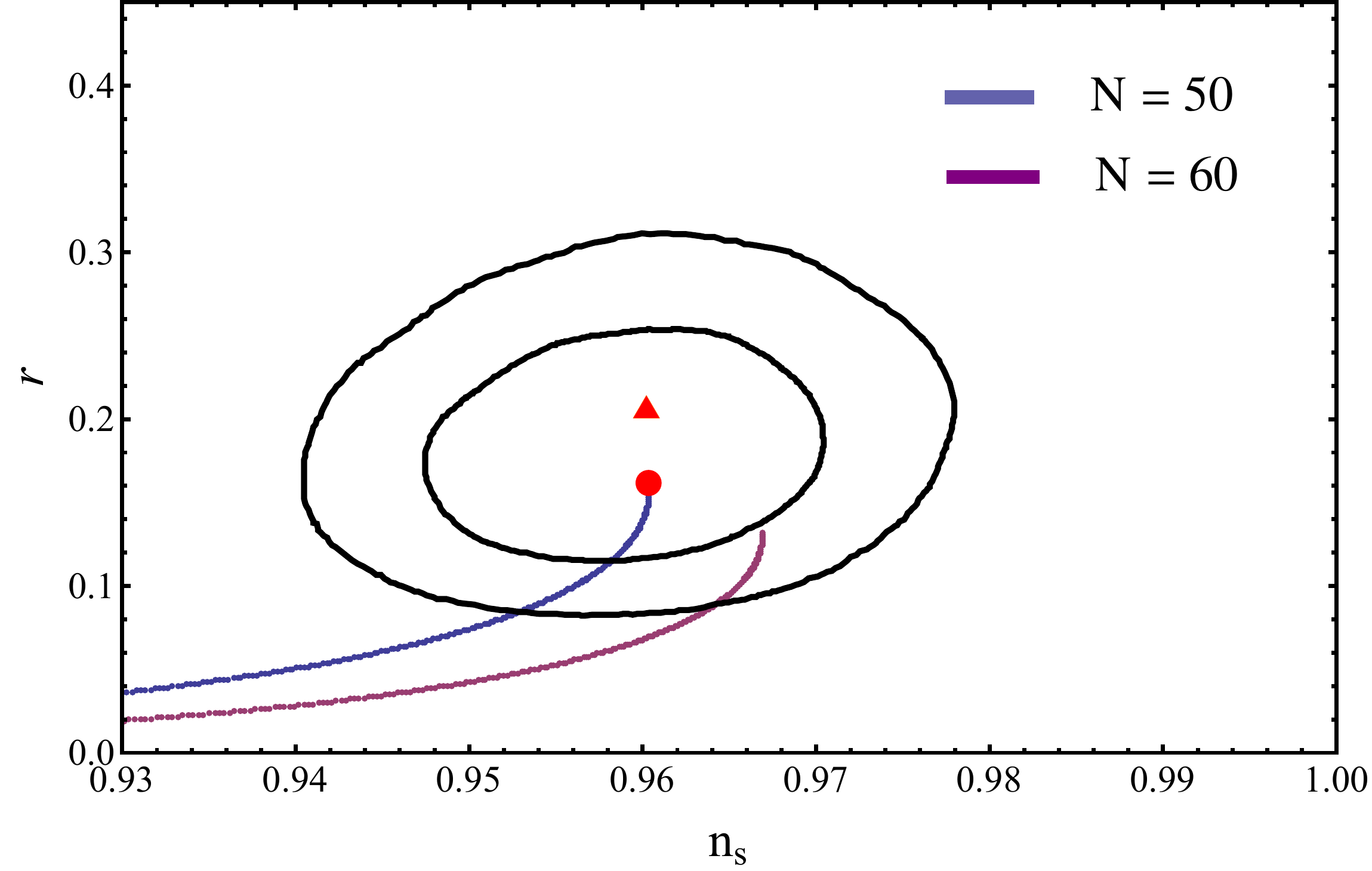}
\caption {
$r$ versus $n_s$ for the
inflaton potential with $(j,~k,~l)=(1, ~2, ~4)$ where
the inflationary trajectory is between $\phi_m$ and $\phi_{M2}$.
}
\label{124_3}
\end{figure}

Third, we consider the inflationary trajectory between $\phi_{m}$ and $\phi_{M2}$,
{\it i.e.}, $\phi_{m}<\phi_f<\phi_i<\phi_{M2}$.
We present the numerical results for $r$ versus $n_s$ in Fig.~\ref{124_3}.
For $n_s$ in the $1\sigma$ range $0.9603\pm0.0073$,
the range of $r$ is $[0.0487,~0.1585]$.
This case is similar to the above second case with
 $\phi_{M1}<\phi_i<\phi_f<\phi_m$, so we will not present benchmark point here.

\begin{figure}[h]
\centering
\includegraphics[height=5cm]{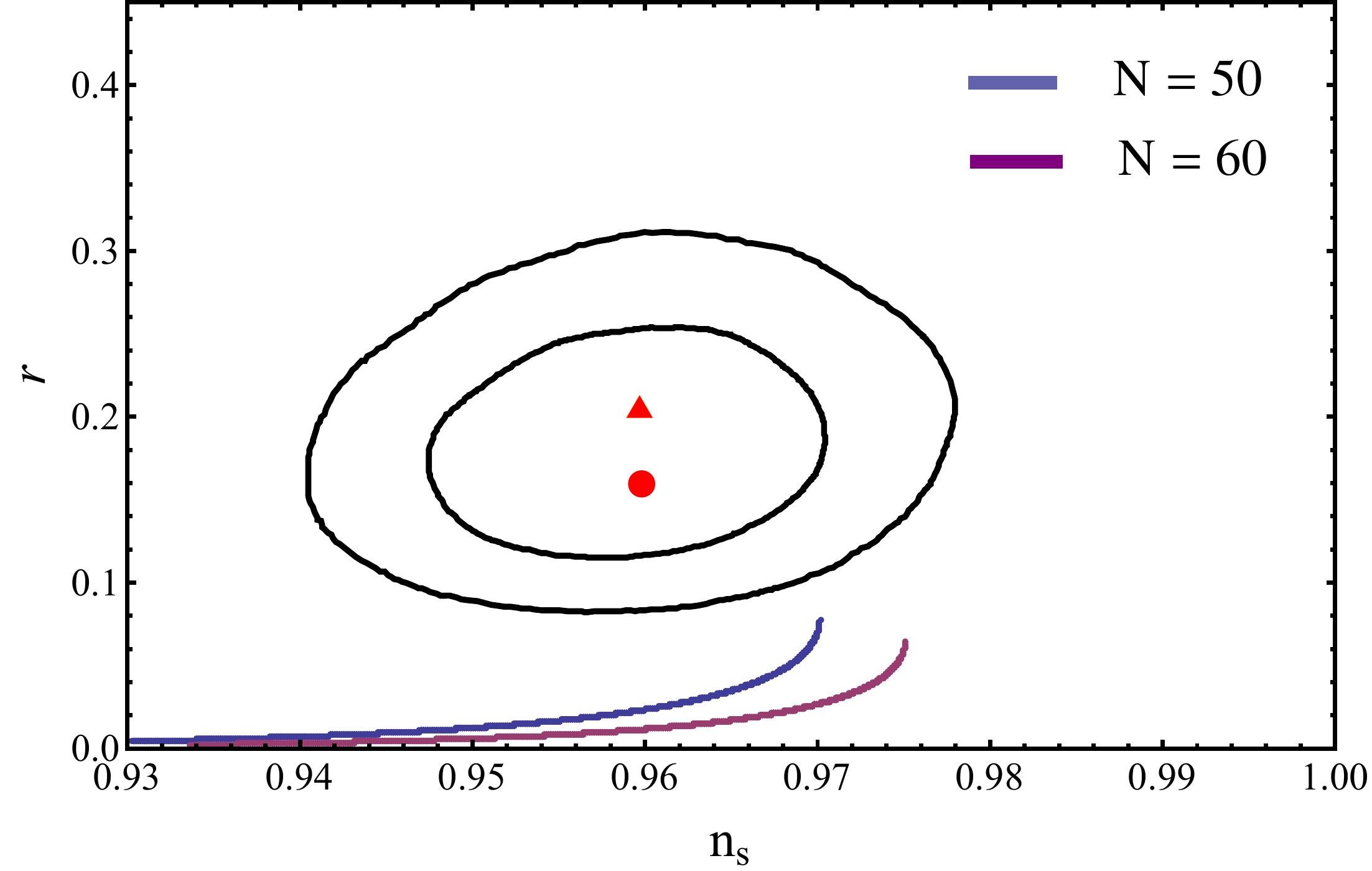}
\caption {
$r$ versus $n_s$ for the
inflaton potential with $(j,~k,~l)=(1, ~2, ~4)$ where
the inflationary trajectory is at the right of the maximum $\phi_{M2}$.
}
\label{124_4}
\end{figure}

Fourth, we consider the inflation at the right of the maximum $\phi_{M2}$,
{\it i.e.}, $\phi_{M2}<\phi_i<\phi_f$.
The numerical results for $r$ versus $n_s$ are given in Fig.~\ref{124_4}.
With $n_s$ in its $1\sigma$ range $0.9603\pm0.0073$,
the range of $r$ is about $[0.0084,~0.0449]$.
Similar to the first case, $r$ is not large enough, but
 can still be tested at the future Planck and QUBIT experiments.

\subsection{Inflaton Potential with $(j,~k,~l)=(1, ~3, ~4)$}

 We consider the inflaton potential $V=a_1\phi+a_3\phi^3+a_4\phi^4$.
For simplicity, we focus on the hill-top scenario with $a_1>0$ and $a_3>0$ while $a_4<0$.
So, there exists a maximum as follows
\begin{equation}
\phi=\phi_M=\frac{1}{4} \left(\frac{a_3^2}{a_4 \sqrt[3]{-a_3^3-8 a_1 a_4^2+4 \sqrt{4
   a_1^2 a_4^4+a_1 a_3^3 a_4^2}}}-\frac{a_3}{a_4}+\frac{\sqrt[3]{-a_3^3-8
   a_1 a_4^2+4 \sqrt{4 a_1^2 a_4^4+a_1 a_3^3 a_4^2}}}{a_4}\right)~.
\end{equation}

\begin{figure}[h]
\centering
\includegraphics[height=5cm]{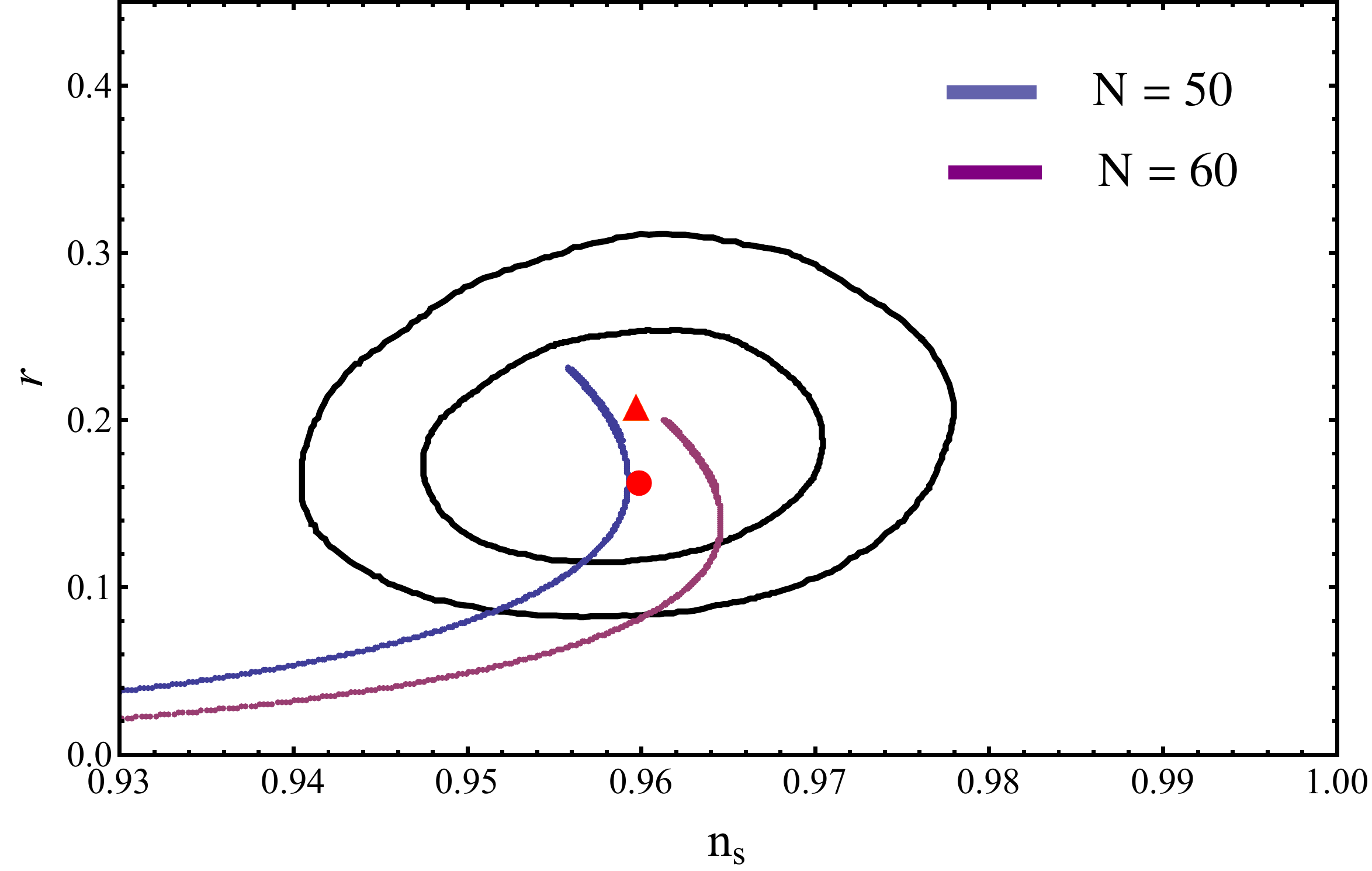}
\caption {
$r$ versus $n_s$ for the
inflaton potential with $(j,~k,~l)=(1, ~3, ~4)$ where the inflationary trajectory is
at the left of the maximum.
}
\label{134_1}
\end{figure}

First, for the inflation at the left of
the maximum with $\phi_f<\phi_i<\phi_M$,  we present the numerical results
for $r$ versus $n_s$ in Fig.~\ref{134_1}.
For $n_s$ within the $1\sigma$ range $0.9603\pm0.0073$,
the range of $r$ is $[0.0556,~0.2328]$, which can be consistent with the BICEP2 experiment.
In addition, for the number of e-folding
$N_e=50$, $n_s$ and $r$ are within $1\sigma$ and $2\sigma$ regions
of the BICEP2 experiment for $a_3>-30a_4$ and $100a_3<a_1<100\left[3\tan{\left(\sec^{-1}{\left(-a_3/10a_4+3\right)}-2.5\right)}+7\right] a_3$ and for  $a_3>-25a_4$ and $a_1<100\left[3\tan{\left(\sec^{-1}{\left(-a_3/10a_4+0.8\right)}-2.5\right)}+13\right] a_3$ , respectively.
Also, for $N_e=60$, $n_s$ and $r$ are within $1\sigma$ and $2\sigma$ regions for
$a_3>-35a_4$ and
$a_1<100\left[3\tan{\left(\sec^{-1}{\left(-a_3/10a_4+6.5\right)}-2.5\right)}+6\right] a_3$
and for $a_3>-30a_4$ and
$a_1<100\left[3\tan{\left(\sec^{-1}{\left(-a_3/10a_4+1.7\right)}-2.5\right)}+8\right] a_3$,
respectively.
Let us present two best fit points for the BICEP2 data.
The best fit point with $n_s=0.96$ and $r=0.16$ can be realized for
$N_e=50$, $a_1\thickapprox 100a_3$, and $a_3\thickapprox-36.5a_4$,
for example, $a_1=10000, a_3=100$, and $a_4=-2.74$,
and the corresponding  $\phi_i, ~\phi_f$, and $\phi_{M}$ are
respectively $12.3513, ~0.714071$, and $28.4959$.
Another best fit point with $n_s=0.96$ and $r=0.2$ can be obtained
for $N_e=60$, $a_1\thickapprox 90a_3$, and $a_3\thickapprox-1000a_4$, for example,
 $a_1=9000, ~a_3=100$, and $a_4=-0.1$,  and the corresponding  $\phi_i, ~\phi_f$, and
$\phi_{M}$ are $15.1757, ~0.715088$, and $750.04$, respectively.

\begin{figure}[h]
\centering
\includegraphics[height=5cm]{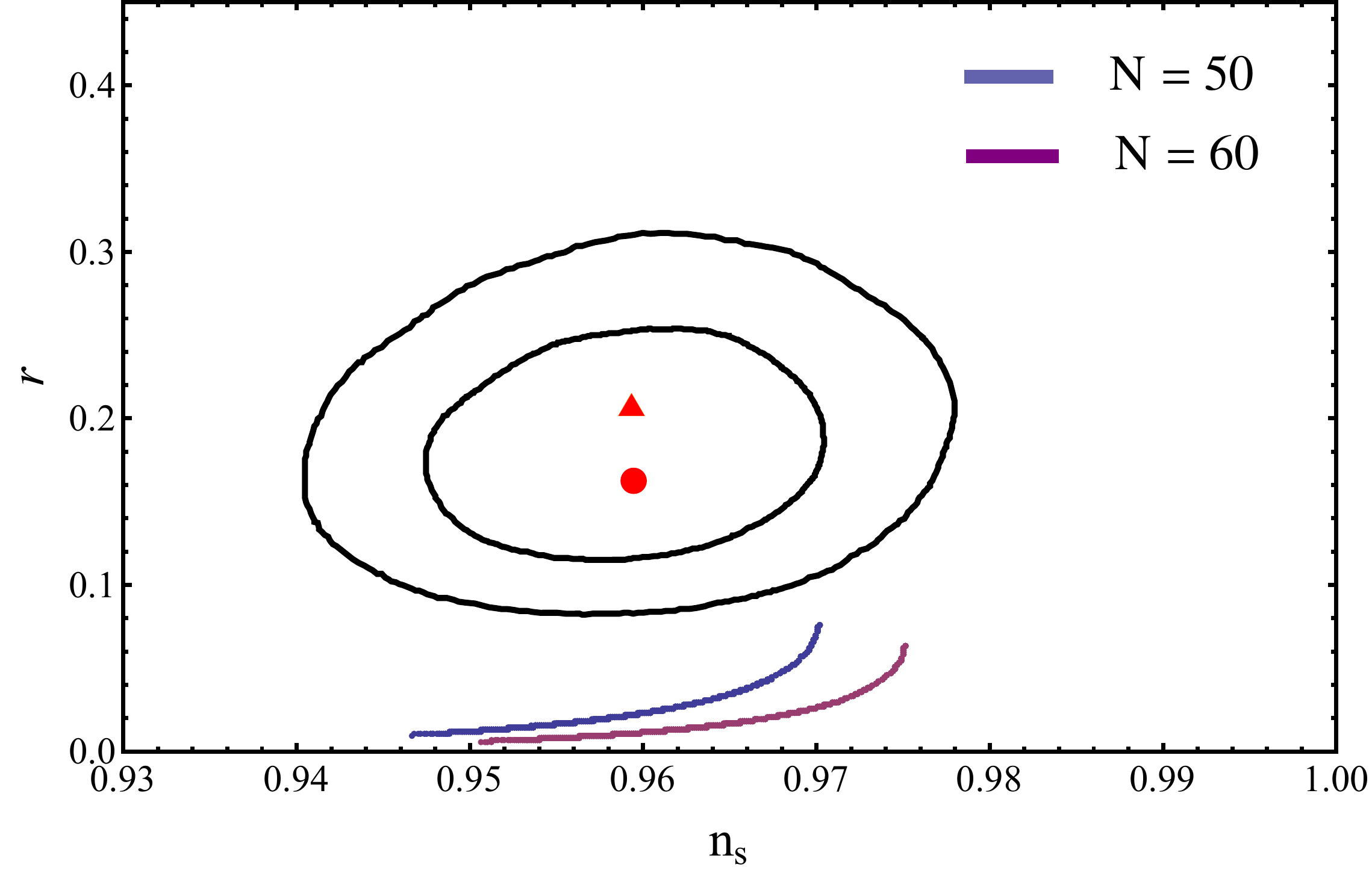}
\caption {
$r$ versus $n_s$ for the
inflaton potential with $(j,~k,~l)=(1, ~3, ~4)$ where the inflationary trajectory is
at the right of the maximum.
}
\label{134_2}
\end{figure}

Second, we consider the inflation at the right of
the maximum with $\phi_M<\phi_f<\phi_i$,  we present the numerical results
for $r$ versus $n_s$ in Fig.~\ref{134_2}.
For $n_s$ within the $1\sigma$ range $0.9603\pm0.0073$,
the range of $r$ is $[0.0081,~0.0458]$, which is still within
the reach of the future Planck and QUBIT experiments.

\subsection{Inflaton Potential with $(j,~k,~l)=(2, ~3, ~4)$}

\begin{figure}[h]
\centering
\includegraphics[height=5cm]{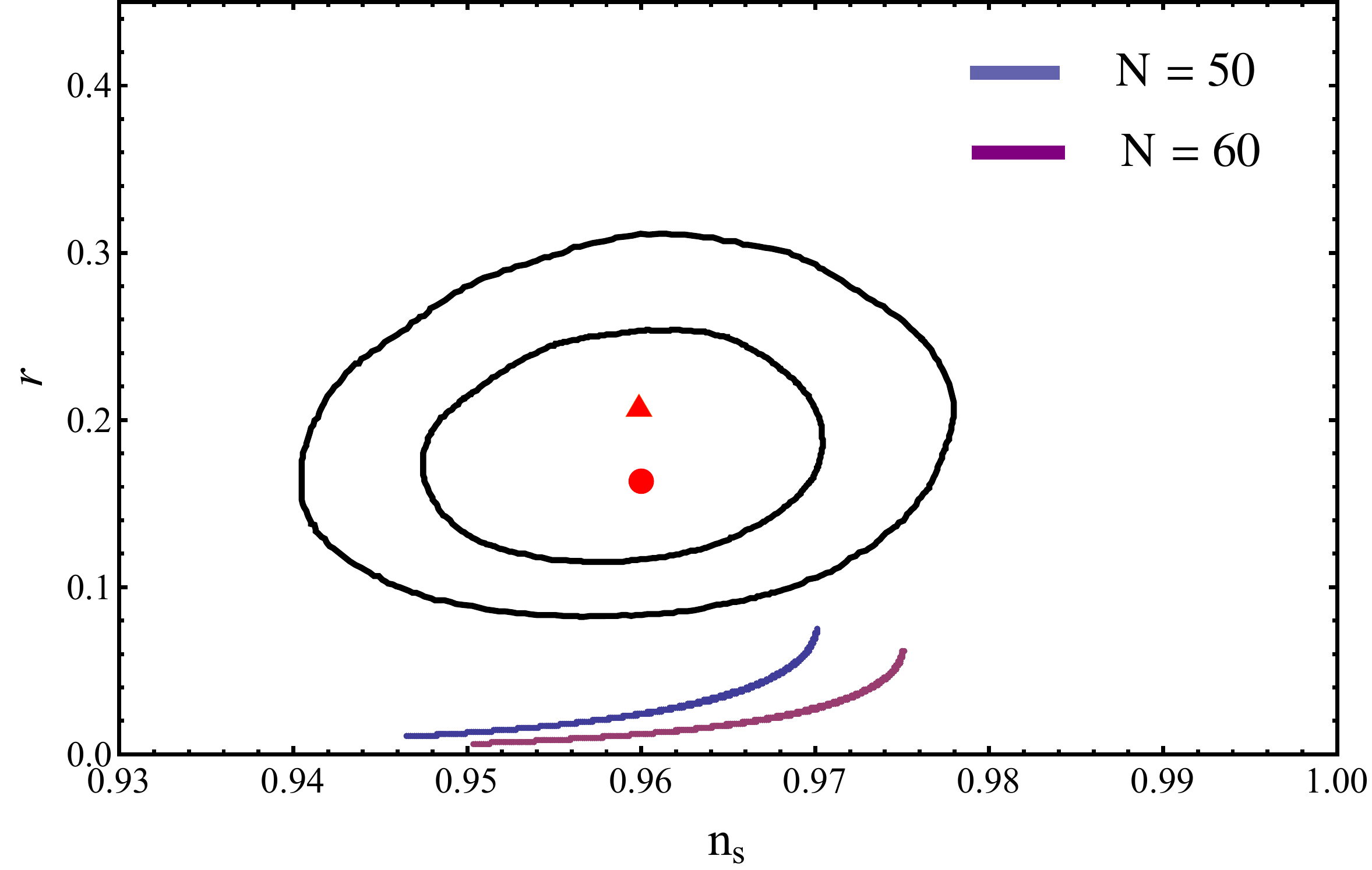}
\caption {
$r$ versus $n_s$ for the non-supersymmetric
inflaton potential with $(j,~k,~l)=(2, ~3, ~4)$
where the inflationary trajectory is
at the left of the maximum $\phi_{M1}$.
} \label{234_1}
\end{figure}

First, we consider the non-supersymmetric models with inflaton potential
$V=a_2\phi^2+a_3\phi^3+a_4\phi^4$. For simplicity, we assume $a_2>0$ and $a_3>0$,
while $a_4<0$. Thus, there exist a minimum at $\phi_m=0$ as well as two maxima at
$\phi_{M1}=\frac{-3 a_3-\sqrt{9 a_3^2-32 a_2 a_4}}{8 a_4}$
and $\phi_{M2}=\frac{-3 a_3+\sqrt{9 a_3^2-32 a_2 a_4}}{8 a_4}$.
Thus, we shall discuss four cases as follows:

(1) When the slow-roll inflation occurs at the left of
the maximum $\phi_{M1}$, {\it i.e.}, $\phi_f<\phi_i<\phi_{M1}$,
$a_2$ must be large enough to get a $\phi_f$ with a relatively large absolute value,
and we present the numerical results for $r$ versus $n_s$ in Fig.~\ref{234_1}.
For $n_s$ in the $1\sigma$ range $0.9603\pm0.0073$,
the range of $r$ is $[0.0073,~0.0472]$, which is out of the $2\sigma$ region
for the BICEP2 data. Interestingly, we still have large enough tensor-to-scalar ratio
within the reach of the future Planck and QUBIT experiments.

\begin{figure}[h]
\centering
\includegraphics[height=5cm]{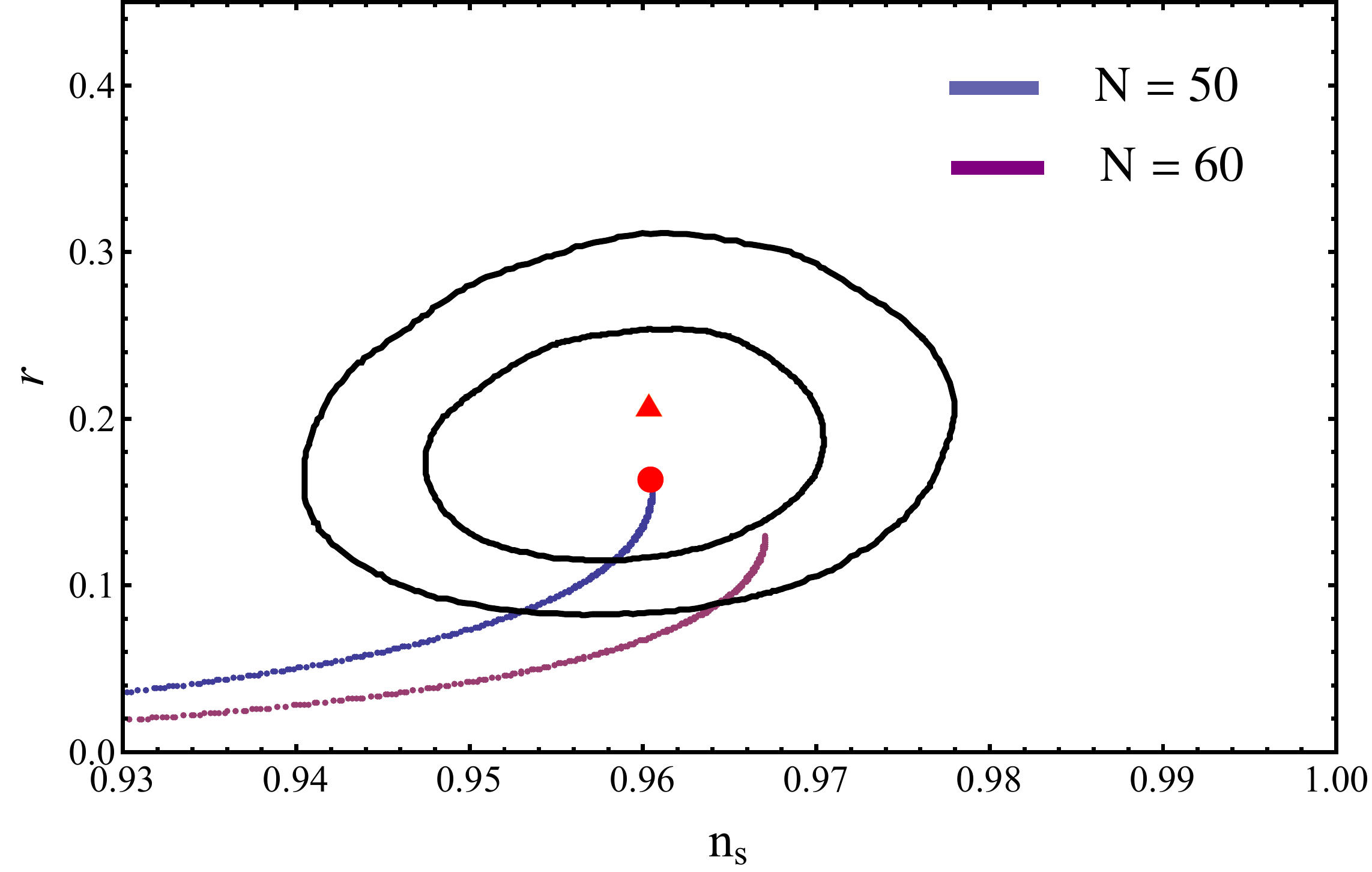}
\caption {
$r$ versus $n_s$ for the non-supersymmetric
inflaton potential with $(j,~k,~l)=(2, ~3, ~4)$
where the inflationary trajectory is
at the right of the maximum $\phi_{M1}$.
} \label{234_2}
\end{figure}

(2) For the slow-roll inflation occurs at the right of $\phi_{M1}$,
{\it i.e.}, $\phi_{M1}<\phi_i<\phi_f<\phi_{m}$, we can obtain large $r$ via chaotic inflation
by requiring $a_3\ll a_2$ and $a_4\ll a_2$.
The numerical results for $r$ versus $n_s$ are given in Fig.~\ref{234_2}.
With $n_s$ in the $1\sigma$ range $0.9603\pm0.0073$,
the range of $r$ is $[0.0496,~0.1585]$, which can be consistent with the BICEP2 experiment.
In the viable parameter space, we generically have $a_2>1000a_3$.
Moreover, for the number of e-folding
$N_e=50$, $n_s$ and $r$ are within $1\sigma$ and $2\sigma$ regions
of the BICEP2 experiment for $a_2>-1250a_4$ and $a_2>-660a_4$, respectively.
Also, for $N_e=60$, $n_s$ and $r$ are within $2\sigma$ region for
$a_2>-1000a_4$, but no viable parameter space for $1\sigma$ region.
The best fit point with $n_s=0.96$ and $r=0.158$
for the BICEP2 data can be obtained for $N_e=50$, $a_2>10^4 a_3$,
and $a_2>-10^6a_4$.
For instance, $a_2=10000, ~a_3=1$, and $a_4=-0.01$, and the corresponding
$\phi_i, ~\phi_e$, and $\phi_{M1}$ are  respectively
$-14.2086, ~-1.41411$, and $-670.6$.

\begin{figure}[h]
\centering
\includegraphics[height=5cm]{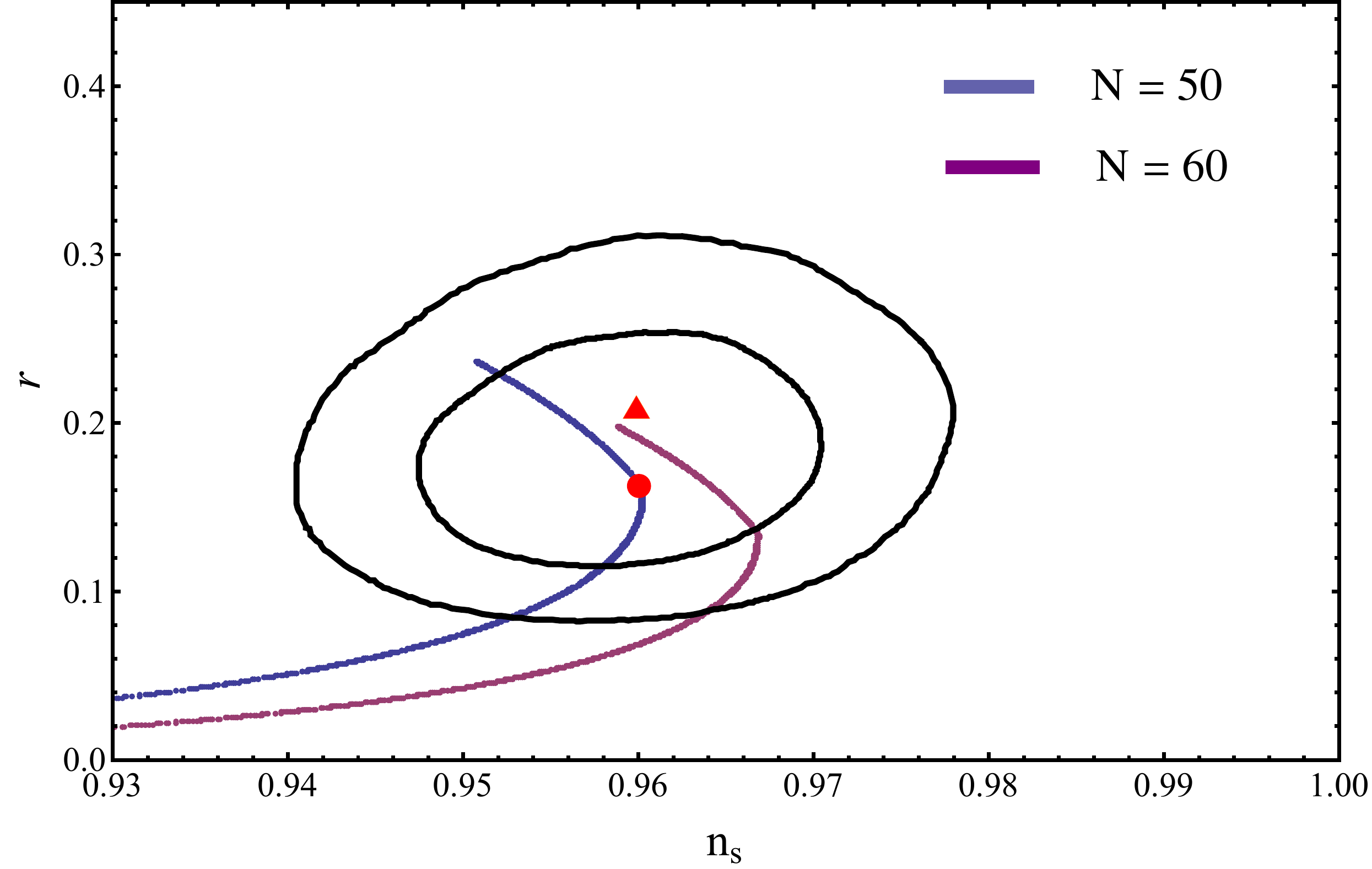}
\caption {
$r$ versus $n_s$ for the non-supersymmetric
inflaton potential with $(j,~k,~l)=(2, ~3, ~4)$
where the inflationary trajectory is
at the left of the maximum $\phi_{M2}$.
}
\label{234_3}
\end{figure}

(3) When the slow-roll inflation occurs at the left of
the maximum $\phi_{M2}$, {\it i.e.}, $\phi_{m}<\phi_f<\phi_i<\phi_{M2}$,
we present the numerical results for $r$ versus $n_s$ in Fig.~\ref{234_3}.
For $n_s$ in the $1\sigma$ range $0.9603\pm0.0073$,
the range of $r$ is $[0.0490,~0.2228]$, which can be consistent with the BICEP2 experiment.
For the number of e-folding $N_e=50$, $n_s$ and $r$ are within $1\sigma$ and $2\sigma$ regions
of the BICEP2 experiment for $a_2>-1000a_4$ and $a_3>-150a_4$
and for $a_2>-660a_4$ and $a_3>-100a_4$, respectively.
For the number of e-folding $N_e=60$, $n_s$ and $r$ are within $1\sigma$ and $2\sigma$ regions
of the BICEP2 experiment for $a_2>-1550a_4$ and $a_3>-150a_4$
and for $a_2>-1250a_4$ and $a_3>-100a_4$, respectively.
Let us give two best fit points for the BICEP2 data.
The best fit point with $n_s=0.96$ and $r=0.16$ can be realized for
$N_e=50$, $a_2\approx900 a_3$, and $a_3>-10^3a_4$, for example,
 $a_2=90, ~a_3=0.1$, and $a_4=-0.0001$, and the corresponding
$\phi_i, ~\phi_f$, and $\phi_{M2}$ are
respectively $14.249, ~1.41532$, and $1143.52$.
Another best fit point with $n_s=0.959$ and $r=0.1953$
can be obtained
for $N_e=60$, $a_2\approx a_3$, and $a_3\approx-10^3a_4$, for instance,
  $a_2=1, ~a_3=1$, and $a_4=-0.001$, and the corresponding
$\phi_i, ~\phi_f$, and $\phi_{M2}$ are $18.7512, ~1.87413$, and $750.666$,
respectively.

(4) When the slow-roll inflation occurs at the right of
the maximum $\phi_{M2}$, {\it i.e.}, $\phi_{M2}<\phi_i<\phi_{f}$,
we will not study it here since it is the same as the above case (1).

\begin{figure}[h]
\centering
\includegraphics[height=5cm]{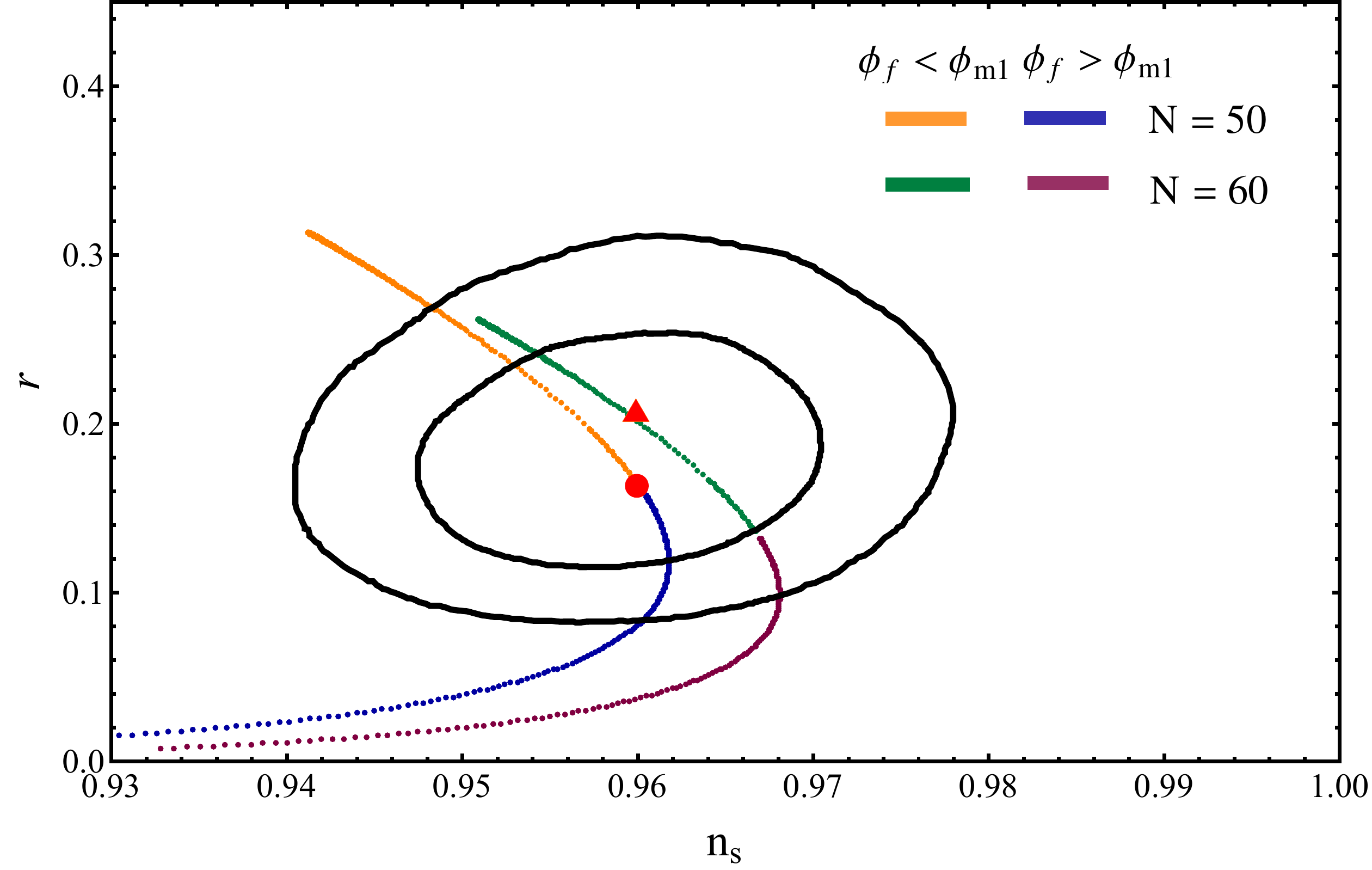}
\caption {
$r$ versus $n_s$ for the supersymmetric
inflaton potential with $(j,~k,~l)=(2, ~3, ~4)$
where the inflationary trajectories are
at the left and right of the minimum $\phi_{m1}$.
}
\label{024_2_1}
\end{figure}

Second, we study the supersymmetric models with inflaton potential
$V=|a\phi+b\phi^2|^2=a^2\phi^2+2ab\phi^3+b^2\phi^4$. For simplicity,
we assume $a>0$ while $b<0$. Thus, there exist a maximum at $\phi_M=-\frac{a}{2b}$
and two minima at $\phi_{m1}=0$ and $\phi_{m2}=-\frac{a}{b}$. And
we shall consider the following four cases:

(1) When the slow-roll inflation occurs at the left of
the minimum $\phi_{m1}$, {\it i.e.}, $\phi_i<\phi_f<\phi_{m1}$,
we present the numerical results for $r$ versus $n_s$ in Fig.~\ref{024_2_1}.
With $n_s$ in the $1\sigma$ range $0.9603\pm0.0073$,
the range of $r$ is $[0.1369,~0.2490]$, which can be consistent with the BICEP2 experiment.
Moreover, for the number of e-folding $N_e=50$, $n_s$ and $r$ are within $1\sigma$ and $2\sigma$ regions
of the BICEP2 experiment for $a>-30b$
and $a>-15b$, respectively.
For the number of e-folding $N_e=60$, $n_s$ and $r$ are within $1\sigma$ region
of the BICEP2 experiment for $a>-8b$ and are generically
in $2\sigma$ region.
Let us give two best fit points for the BICEP2 data.
The best fit point with $n_s=0.96$ and $r=0.16$ can be realized for
$N_e=50$, and $a\thickapprox -2000 b$, for example,
 $a=2000 $ and $ b=-1$, and the corresponding
$\phi_i, ~\phi_f$, and $\phi_{m1}$ are
respectively $-14.2462, ~-1.41521$, and $0$.
Another best fit point with $n_s=0.959$ and $r=0.20$ can be obtained
for $N_e=60$, and $a\thickapprox -26 b$, for instance,
  $a=26 $ and $ b=-1$, and the corresponding
$\phi_i, ~\phi_f$, and $\phi_{m1}$ are  $-17.7247, ~-1.49091$, and  $0$,
respectively.

(2) When the slow-roll inflation occurs at the right of the minimum $\phi_{m1}$ and the left of
the maximum $\phi_{M}$, {\it i.e.}, $\phi_{m1}<\phi_f<\phi_i<\phi_M$, to have relatively large $r$,
we find that $|b|$ cannot be equal to or larger than $a$.
 The numerical results for $r$ versus $n_s$ are also given in Fig.~\ref{024_2_1}.
For $n_s$ in the $1\sigma$ range $0.9603\pm0.0073$,
the range of $r$ is $[0.0254,~0.1584]$, which can be consistent with the BICEP2 experiment.
In addition, for the number of e-folding $N_e=50$, $n_s$ and $r$ are
within $1\sigma$ and $2\sigma$ regions
of the BICEP2 experiment for $a>-85b$ and $a>-47b$, respectively.
For the number of e-folding $N_e=60$, $n_s$ and $r$ are within $2\sigma$ region for
$a>-85b$, while no viable parameter space for $1\sigma$ region.
Especially, the best fit point with $n_s=0.96$ and $r=0.158$
for the BICEP2 data can be obtained for $N_e=50$, and $a>-10^4 b$.
For example,  $a=1$ and $b=-10^{-4}$, and the corresponding
$\phi_i, ~\phi_e$, and $\phi_{M}$ respectively are $14.2025, ~1.41391$, and $3333.33$.

(3) When the slow-roll inflation occurs at the right of
the maximum $\phi_{M}$, {\it i.e.}, $\phi_{M}<\phi_i<\phi_f<\phi_{m2}$,
it is the same as the above case (2) and then we will not
discuss it here.

(4) When the slow-roll inflation occurs at the right of
the minimum $\phi_{m2}$, {\it i.e.}, $\phi_{m2} < \phi_f <\phi_i$,
we will not study it here since it is the same as the above case (1).



\subsection{The Most General Renormalizable Supersymemtric Inflationary Models}

We briefly comment on the most general renormalizable
supersymmetric inflationary models with the following inflaton potential
\begin{eqnarray}
V&=& (a'+b'\phi'+c'\phi^{\prime 2})^2~,~\,
\end{eqnarray}
where $a'$, $b'$, and $c'$ are all non-zero.
Redefining the inflaton field and parameters as follows
\begin{eqnarray}
\phi ~\equiv~ \phi'+\frac{b'}{2c'}~,~~a~\equiv~a'-\frac{b^{\prime 2}}{4c'}~,~
b~\equiv~c'~.~\,
\end{eqnarray}
we obtain the inflaton potential
\begin{eqnarray}
V&=& (a+b\phi^{ 2})^2~.~\,
\end{eqnarray}
This is the same as the supersymmetric
inflaton potential, which is studied in the subsection E.
Thus, we will not repeat it here.

\subsection{Numerical Result Summary}

To summarize the above results for $n_s$ within its $1\sigma$ range $0.9603\pm0.0073$,
we present the ranges of $r$ for different signs of parameters
in the non-supersymmetric and supersymmetric models
respectively in Tables~\ref{Tab-Non-SUSY} and \ref{Tab-SUSY}.
Interestingly, we always have large enough tensor-to-scalar ratios, which
are within the reach of the future Planck and QUBIT experiments.

\begin{table}[htbp]
\centering
\caption{\label{table1} The ranges of $r$ for  different signs
 of parameters and $n_s$ within its $1\sigma$ range $0.9603\pm0.0073$
in the non-supersymmetric models.}
\begin{tabular}{c|c|c|c|c|c}
\hline
Model & Sign of the Parameters & Range \uppercase\expandafter{\romannumeral1} & Range \uppercase\expandafter{\romannumeral2} & Range \uppercase\expandafter{\romannumeral3} & Range \uppercase\expandafter{\romannumeral4} \\
\hline

\multirow{2}{*}{(0, 1, 2)} & (+, +, $-$) &  [0.0132, 0.0534] &  [0.0132, 0.0534] & & \\
\cline{2-6}
& (+, $-$, +) &  [0.0132, 0.1610] &  [0.0132, 0.1610] & & \\
\hline

\multirow{2}{*}{(0, 1, 3)} & (+, +, $-$) &  [0.1231, 0.2237] & [0.0337, 0.0669]  & [0.0085, 0.0482] & \\
\cline{2-6}
& (+, $-$, $-$) &  [0.1670, 0.2427] & & &\\
\hline

\multirow{2}{*}{(0, 1, 4)} & (+, +, $-$) &   [0.0250, 0.0732] &  [0.0077, 0.0459]  &  & \\
\cline{2-6}
& (+, $-$, +) & NO FIT &  [0.1288, 0.2498] & &\\
\hline

\multirow{3}{*}{(0, 2, 3)} & (+, +, $-$) &   [0.1363, 0.2206] &   [0.0645, 0.160]  & [0.0097, 0.0431] & \\
\cline{2-6}
& (+, $-$, $-$) & [0.1249, 0.2242] &   [0.0104, 0.0512] & [0.0099, 0.0505] &\\
\cline{2-6}
& (+, $-$, +) &  [0.0099, 0.0485] &   [0.0099, 0.0515] & [0.1232, 0.2253] &\\
\hline

(0, 2, 4) & (+, +, $-$) &    [0.0480, 0.1565] &   [0.0072, 0.0444]  &  & \\
\hline

\multirow{2}{*}{(0, 3, 4)} & (+, +, $-$) &   [0.0742, 0.1956] &    [0.0067, 0.0454]  & & \\
\cline{2-6}
& (+, $-$, +) &  [0.1995, 0.2473] &    [0.1311, 0.2512] & &\\
\hline

(1, 2, 3) & (+, +, $-$) &   [0.1234,0.2207] &    [0.0337, 0.158]  &  [0.0083, 0.0471] & \\
\hline

(1, 2, 4) & (+, +, $-$) &   [0.0084, 0.0449] &     [0.0487, 0.1585]  &  [0.0487, 0.1585] &  [0.0084, 0.0449]\\
\hline

(1, 3, 4) & (+, +, $-$) &     [0.0556, 0.2328] &       [0.0081, 0.0458]  &  & \\
\hline

(2, 3, 4) & (+, +, $-$) &    [0.0073, 0.0472] &    [0.0496, 0.1585]  &  [0.0490, 0.2228] &  [0.0073, 0.0472] \\
\hline

\end{tabular}
\label{Tab-Non-SUSY}
\end{table}

\begin{table}[htbp]
\centering
\caption{\label{table2} The ranges of $r$ for  different signs
 of parameters and $n_s$ within its $1\sigma$ range $0.9603\pm0.0073$
in the supersymmetric models.}
\begin{tabular}{c|c|c|c|c|c}
\hline
Model & Sign of the Parameters & Range \uppercase\expandafter{\romannumeral1} & Range \uppercase\expandafter{\romannumeral2} & Range \uppercase\expandafter{\romannumeral3} & Range \uppercase\expandafter{\romannumeral4} \\
\hline

$|a+b\phi|^2$ & (+, $-$) &    [0.1322, 0.1584] &   [0.1322, 0.1584]  &  & \\
\hline

$|a+b\phi^2|^2$ & (+, $-$) &     [0.1319, 0.2484] &   [0.0254, 0.1585]  &  [0.0254, 0.1585] &  [0.1319, 0.2484] \\
\hline

$|a\phi+b\phi^2|^2$ & (+, $-$) &     [0.1369, 0.2490] &    [0.0254, 0.1369]  &  [0.0254, 0.1369] &  [0.1369, 0.2490] \\
\hline

\end{tabular}
\label{Tab-SUSY}
\end{table}

\section{Conclusion}

We have systematically studied the renormalizable three-term polynomial inflation
in the supersymmetric and non-supersymmetric models. We can construct
the supersymmetric inflaton potentials via the supergravity theory, and
 we showed that the general renormalizable
supergravity model is equivalent to one kind of our supersymmetric models.
Although the running of the spectral index is out of the $2\sigma$ range for all
the models, we found
that the spectral index and tensor-to-scalar ratio can be consistent
with the Planck and BICEP2 results.
Even if we do not consider the BICEP2 experiment, our inflationary models can
not only highly agree with the Planck observations, but also
saturate its upper bound on the tensor-to-scalar ratio ($r \le 0.11$).
In short, our models can be tested at the future Planck and QUBIC experiments.



\begin{acknowledgments}

We would like to thank Xiao Liu very much for helpful discussions.
This research was supported in part by the Natural Science
Foundation of China under grant numbers 10821504, 11075194, 11135003, 11275246, 11305110, and by the National
Basic Research Program of China (973 Program) under grant number 2010CB833000 (TL).

\end{acknowledgments}

\end{document}